\documentclass[twocolumn]{aastex631}

\usepackage{color}
\usepackage{CJK}
\usepackage{graphicx}
\usepackage{amsmath}
\usepackage{amssymb}
\usepackage{appendix}
\usepackage[mathlines]{lineno}

\hypersetup{pdfauthor={Name}}

\def\ra#1#2#3{#1$^{\rm h}$#2$^{\rm m}$#3$^{\rm s}$}
\def\dec#1#2#3{#1$^{\rm d}$#2$^{\rm m}$#3$^{\rm s}$}

\graphicspath{{./}{figures/}}

\shorttitle{LFBOT Hosts}
\shortauthors{Nugent et al.}

\begin{document}

\title{The Environments of Luminous Fast Blue Optical Transients: Evidence for a Compact Object and Wolf-Rayet Star Merger Origin} 

\correspondingauthor{A. E. Nugent}
\email{anya.nugent@cfa.harvard.edu}

\newcommand{\NU}{\affiliation{Center for Interdisciplinary Exploration and Research in Astrophysics (CIERA) and Department of Physics and Astronomy, Northwestern University, Evanston, IL 60208, USA}}

\newcommand{\Purdue}{\affiliation{Purdue University, 
Department of Physics and Astronomy, 525 Northwestern Avenue, West Lafayette, IN 47907, USA}}

\newcommand{\CfA}{\affiliation{Center for Astrophysics\:$|$\:Harvard \& Smithsonian, 60 Garden St. Cambridge, MA 02138, USA}}

\newcommand{\UCSC}{\affiliation{Department of Astronomy and Astrophysics, University of California, Santa Cruz, CA 95064, USA}}

\newcommand{\IS}{\affiliation{Centre for Astrophysics and Cosmology, Science Institute, University of Iceland, Dunhagi 5, 107 Reykjav\'ik, Iceland}}

\newcommand{\DAWN}{\affiliation{Cosmic Dawn Center (DAWN), Niels Bohr Institute, University of Copenhagen, Jagtvej 128, 2100 Copenhagen \O, Denmark}}

\newcommand{\PUCV}{\affiliation{Instituto de F\'isica, Pontificia Universidad Cat\'olica de Valpara\'iso, Casilla 4059, Valpara\'iso, Chile}}

\newcommand{\IPMU}{\affiliation{Kavli Institute for the Physics and Mathematics of the Universe (Kavli IPMU), 5-1-5 Kashiwanoha, Kashiwa, 277-8583, Japan}}

\newcommand{\PSU}{\affiliation{Department of Astronomy \& Astrophysics, The Pennsylvania State University, University Park, PA 16802, USA}}

\newcommand{\ICDS}{\affiliation{Institute for Computational \& Data Sciences, The Pennsylvania State University, University Park, PA, USA}}

\newcommand{\IGC}{\affiliation{Institute for Gravitation and the Cosmos, The Pennsylvania State University, University Park, PA 16802, USA}}

\newcommand{\Swin}{\affiliation{ Centre for Astrophysics and Supercomputing, Swinburne University of Technology, Hawthorn, VIC, 3122, Australia}}

\newcommand{\Curtin}{\affiliation{ International Centre for Radio Astronomy Research, Curtin University, Bentley, WA 6102, Australia}}

\newcommand{\MQ}{\affiliation{Department of Physics \& Astronomy, Macquarie University, NSW 2109, Australia}}

\newcommand{\MQAAAstro}{\affiliation{Macquarie University Research Centre for Astronomy, Astrophysics \& Astrophotonics, Sydney, NSW 2109, Australia}}

\newcommand{\CSIRO}{\affiliation{CSIRO, Space and Astronomy, PO Box 76, Epping NSW 1710 Australia}}

\newcommand{\KICP}{\affiliation{Kavli Institute for Cosmological Physics, The University of Chicago, 5640 South Ellis Avenue, Chicago, IL 60637, USA}}

\newcommand{\UChicago}{\affiliation{Department of Astronomy \& Astrophysics, University of Chicago, 5640 S Ellis Avenue, Chicago, IL 60637, USA}}

\newcommand{\UA}{\affiliation{University of Arizona, Steward Observatory, 933~N.~Cherry~Ave., Tucson, AZ 85721, USA}}

\newcommand{\EFI}{\affiliation{Enrico Fermi Institute, The University of Chicago, 933 East 56th Street, Chicago, IL 60637, USA}}

\newcommand{\mpia}{\affiliation{Max-Planck-Institut f\"{u}r Astronomie (MPIA), K\"{o}nigstuhl 17, 69117 Heidelberg, Germany}}

\newcommand{\GWU}{\affiliation{Department of Physics, The George Washington University, Washington, DC 20052, USA}}

\newcommand{\UCB}{\affiliation{Department of Astronomy, University of California, Berkeley, CA 94720-3411, USA}}

\newcommand{\RU}{\affiliation{Department of Astrophysics/IMAPP, Radboud University, PO Box 9010,
6500 GL, The Netherlands}}

\newcommand{\LJMU}{\affiliation{Astrophysics Research Institute, Liverpool John Moores University, IC2, Liverpool Science Park, 146 Brownlow Hill, Liverpool L3 5RF, UK}}

\newcommand{\LU}{\affiliation{School of Physics and Astronomy, University of Leicester, University Road, Leicester. LE1 7RH, UK}}

\newcommand{\Adler}{\affiliation{The Adler Planetarium, 1300 South DuSable Lake Shore Drive, Chicago, IL 60605, USA}}

\newcommand{\ANU}{\affiliation{Research School of Astronomy and Astrophysics, Australian National University, Canberra, ACT 2611, Australia}}

\newcommand{\Car}{\affiliation{Cardiff Hub for Astrophysics Research and Technology, School of Physics \& Astronomy, Cardiff University, Queen's Buildings, Cardiff CF24 3AA, UK}}

\newcommand{\IAIFI}{\affiliation{The NSF AI Institute for Artificial Intelligence and Fundamental Interactions}}

\newcommand{\MIT}{\affiliation{Department of Physics and Kavli Institute for Astrophysics and Space Research, Massachusetts Institute of Technology, 77 Massachusetts Avenue, Cambridge, MA 02139, USA}}

\newcommand{\Hawaii}{\affiliation{Institute for Astronomy, University of Hawai‘i, 640 N. A‘ohoku Pl., Hilo, HI 96720, USA}}

\newcommand{\Weizmann}{\affiliation{Department of Particle Physics and Astrophysics, Weizmann Institute of Science, 234 Herzl St, 7610001 Rehovot, Israel}}

\newcommand{\Minnesota}{\affiliation{School of Physics and Astronomy, University of Minnesota, Minneapolis, MN 55455, USA}}

\newcommand{\CCA}{\affiliation{Center for Computational Astrophysics, Flatiron Institute, 162 W. 5th Avenue, New York, NY 10011, USA}}

\newcommand{\Columbia}{\affiliation{Department of Physics and Columbia Astrophysics Laboratory, Columbia University, New York, NY 10027, USA}}

\newcommand{\LSU}{\affiliation{Department of Physics \& Astronomy, Louisiana State University, Baton Rouge, LA 70803, USA}}

\newcommand{\LANL}{\affiliation{Center for Nonlinear Studies, Los Alamos National Laboratory, Los Alamos, NM 87545 USA}}

\author[0000-0002-2028-9329]{Anya E. Nugent}
\CfA

\author[0000-0002-5814-4061]{V.~Ashley Villar}
\CfA
\IAIFI

\author[0000-0002-4670-7509]{Brian D.~Metzger}
\CCA
\Columbia

\author[0000-0003-2624-0056]{Christopher~L.~Fryer}
\LANL

\author[0000-0002-2942-3379]{Eric~Burns}
\LSU

\author[0000-0002-5025-4645]{Alexa C. Gordon}
\NU

\author[0000-0002-7197-9004]{Danielle Frostig}
\CfA

\author[0000-0002-9363-8606]{Yuxin Dong}
\NU

\begin{abstract}
We present a comprehensive analysis of the host galaxies of 11 luminous fast blue optical transients (LFBOTs). We model new and archival host photometry and spectroscopy with \texttt{Prospector}. We determine that all LFBOT hosts are actively star-forming with recent bursts of star formation and have a median stellar mass of $\log(M_*/M_\odot)=9.61^{+0.74}_{-1.61}$, present-day star formation rate SFR=$0.99^{+14.85}_{-0.95}$~$M_\odot$~yr$^{-1}$, and gas-phase oxygen abundance metallicity 12+log(O/H)=$8.59^{+0.18}_{-0.22}$. To contextualize these results, we compare them to the host properties of Hydrogen-poor superluminous supernovae (SLSNe-I), several core-collapse supernova subtypes (CCSN; SNe Ibc, II, and Ibn) and long gamma-ray bursts (LGRBs). We find that LFBOT hosts are more star-forming than CCSN hosts, but less star-forming than SLSN-I hosts. We further show that LFBOT hosts are more metal-poor than SN Ibc and II hosts, but more metal-rich than SLSN-I and LGRB hosts. Finally, we find that, similar to SLSNe-I and unlike CCSNe and LGRBs, a large fraction of LFBOTs occur in their hosts' faintest pixel or outside their host galaxy's light. Our results indicate that LFBOTs have a massive stellar origin that do not trace active star-forming regions within their hosts and have a weaker metallicity-dependence than other extreme transients. For these reasons, we favor a compact-object and Wolf-Rayet star merger progenitor scenario over other previously proposed models, such as tidal disruption events and failed or successful CCSN. Future discoveries of LFBOTs with the Rubin observatory will help to increase their sample size and place firmer constraints on their environments and progenitors.
\end{abstract}


\section{Introduction}
\label{sec:intro}
Luminous Fast Blue Optical Transients (LFBOTs), or ``cow-like" events after the prototypical member AT2018cow \citep{prentice2018, ho2019, margutti2019, perley2019}, represent a unique class of extreme astrophysical transient phenomena. LFBOTs are characterized by their rapid lightcurve evolution, reaching peak brightness in $\lesssim1$~week and decaying to half-peak brightness in a similar time, and high peak optical luminosities ($L_\textrm{peak} > 10^{43}$~erg~s$^{-1}$) that rival those of superluminous supernovae (SLSNe). Importantly, their lightcurves cannot be explained by the radioactive decay of $^{56}$Ni, thereby distinguishing LFBOTs from ``normal" supernovae (SNe) and core-collapse (CC) events \citep{perley2019}. LFBOT spectra are dominated by a hot, blue, featureless continuum (blackbody temperatures $>$10$^4$~K) for weeks after they reach peak brightness, hinting that they are powered by a central heating source. The luminous and variable X-ray emission observed for several events further corroborates this idea and signifies that the central engine is likely a compact object  \citep{margutti2019, coppejans2020, perley2021, ho2022_xnd, yao2022, ho2023, chrimes2024, nayana2025, omand2026}. In addition, a handful of LFBOTs appear to have bright radio and millimeter emission, deduced to come from a dense, extended circumstellar medium (CSM) surrounding the central engine  \citep{ho2019, ho2020_koala, ho2022_xnd, margutti2019, coppejans2020, bright2022, yao2022, chrimes2024, chrimes2024_offset, nayana2025, sevilla2026}.

While there are now 11 \textit{bona-fide} events that comprise the LFBOT class, the nature of their progenitor and central engine still remains highly uncertain. Given that LFBOTs are intrinsically rare transients ($<1\%$ the CCSN volumetric rate; \citealt{coppejans2020, ho2023_fbot}), they likely have unusual astrophysical origins. For example, it has been theorized that LFBOTs could represent rare end-states to massive star evolution, such as failed SNe that create accreting black holes (BH; \citealt{margutti2019, perley2019, qlc2019, antoni2022, chrimes2024}), pulsational-pair instability SNe (PPISNe; \citealt{leung2020}), or magnetar-powered SNe \citep{prentice2018, vurm2021}. In each of these scenarios, a compact object forms and produces hot, blue, LFBOT-like emission from either rapid rotation or the accretion on the compact object. Interestingly, these theories also directly connect LFBOTs to other transient phenomena from stellar deaths, including: CCSNe, long gamma-ray bursts (LGRBs; arise from collapsars), and hydrogen-poor SLSNe (SLSNe-I; thought to originate from magnetar-powered core-collapse events). On the other hand, \citet{metzger2022} and \citet{klencki2025} hypothesize that LFBOTs (and possibly Type Ibn and Icn SNe) originate from mergers of stellar-mass BHs or neutron stars (NSs) with Wolf-Rayet (WR) stars. \citet{metzger2022} highlights that this model can reproduce all consistently observed properties of LFBOTs, including their low $^{56}$Ni masses, featureless spectra, extended CSM, and aspherical ejecta of LFBOTs, which many other progenitor models cannot do. Given the observational similarities of LFBOTs to TDEs of supermassive BHs (SMBH), including their early, blue featureless spectra and late-time UV and He emission, it has also been proposed they may derive from TDEs of intermediate-mass BHs (IMBHs; \citealt{kwo+2019, perley2019, ho2023}) or stellar-mass BHs or NSs \citep{tsuna2025}. Several studies have also debated whether LFBOTs simply represent shock interactions with dense CSM material \citep{margalit2022, pellegrino2022, gutierrez2026, khatami2024, gnh+2026}, but these models fail to explain the variable X-ray emission observed for several LFBOTs \citep{margutti2019, yao2022}. Finally, it is completely possible that the LFBOT class comprises multiple of these proposed progenitor models, as several studies have shown the difficulties in fitting all LFBOTs under a single progenitor model \citep{chrimes2026, sevilla2026}.

The lack of consensus on LFBOT origins from their observed characteristics motivates consideration of properties independent of their transient emission to place meaningful constraints on their progenitors. Indeed, their host galaxies may be vital in this effort, as the connection of transients to their environments gives critical insight into their progenitor formation pathways and timescales. These studies have proven fruitful in discerning the progenitors of a wide variety of transients, including several types of CCSNe and LGRBs \citep{prieto2008, svensson2010, anderson2009, arcavi2010, li2011, schulze2021, taggart2021, qin2024, frankenblast}. Several studies on individual (or a smaller subset of) LFBOTs have found that they typically occur in actively star-forming, low-mass galaxies with low metallicities \citep{perley2019, coppejans2020, ho2020_koala, yao2022, perley2019, sevilla2026}. At face value, the presence of LFBOTs in this particular environment may hint that they originate from young massive star progenitors, as is the case for other transient populations observed in similar hosts, like LGRBs and SLSNe-I \citep{svensson2010,perley2013,chen2013, lunnan2014, lunnan2015, angus2016, perley2016, schulze2018, schulze2021, taggart2021}. This was further inferred in \citet{sun2023}, where they determined that AT2018cow was spatially coincident with a star-bursting region within its host, where it is presumed that a large, massive star population may exist. On the contrary, \citet{chrimes2024_offset} found that LFBOT AT2023fhn was $\approx17$~kpc from its host galaxy's center, far removed from any star-formation. This offset is quite large in comparison to those observed for CCSNe, LGRBs, and SLSNe-I, thereby challenging progenitor scenarios tied to young stellar populations. Thus, a uniform analysis of LFBOT environments is critical to characterize their host properties, galactocentric offsets, and connection to recent star formation, which, ultimately, may clarify the nature of their progenitors.

Here, we present a uniform analysis of the host galaxy and environmental properties of 11 LFBOTs: AT2018cow \citep{prentice2018, ho2019, margutti2019, perley2019}, CSS161010 \citep{coppejans2020, gutierrez2026}, ZTF18abvkwl \citep{ho2020_koala}, AT2020mrf \citep{yao2022}, AT2020xnd \citep{perley2021, bright2022, ho2022_xnd}, AT2022tsd \citep{ho2023}, AT2023fhn \citep{chrimes2024, chrimes2024_offset, sevilla2026}, AT2023hkw \citep{sevilla2026}, AT2023vth \citep{sevilla2026}, AT2024qfm \citep{sevilla2026}, and AT2024wpp \citep{pkk+2025, lebaron2026, nayana2025, omand2026, perley2026}. This represents the largest sample of LFBOT host galaxy analysis thus far in the literature. We take advantage of previously published host associations in the aforementioned literature to conduct our environmental analysis. We note that there are two other claimed LFBOTs that we do not include in this study: AT2022abfc and AT2024aehp \citep{sevilla2026}. We exclude AT2022abfc as there is a lack of optical lightcurve information that hinders our confidence in its characterization as an LFBOT, and we exclude AT2024aehp given that its optical lightcurve at late times deviates from other well-studied LFBOTs. In Section~\ref{sec:obs}, we detail photometric and spectroscopic observations of the host galaxies. We discuss our host galaxy stellar population modeling techniques and present our main findings on LFBOT host properties in Section~\ref{sec:prosp}. We compare the global host properties to host properties of other well-studied transient populations in Section~\ref{sec:compare}. We furthermore perform an analysis on the LFBOT galactocentric offsets and fractional fluxes in Section~\ref{sec:offsets_res}. We postulate on the LFBOT progenitor in Section~\ref{sec:disc}, under the assumption that they originate from a single progenitor scenario, even if this is not necessarily true. Finally, we discuss our main conclusions in Section~\ref{sec:conc}.

We employ a standard WMAP9 cosmology of $H_{0}$ = 69.6~km~s$^{-1}$~Mpc$^{-1}$, $\Omega_\textrm{m}$ = 0.286, $\Omega_\textrm{vac}$ = 0.714 \citep{Hinshaw2013, blw+14}. 

\section{Observations}
\label{sec:obs}
\subsection{Photometry}
\label{sec:phot}
We collect photometric observations of LFBOT host galaxies using a combination of previously published photometry, data from our own observing programs, and publicly available data from archival surveys. We collect available host galaxy photometry for AT2018cow, ZTF18abvkwl ($u$-band), AT2020mrf, and AT2020xnd from \citet{perley2019}, \citet{ho2020_koala}, \citet{yao2022}, and \citet{perley2021}, respectively. For the hosts of AT2020xnd ($r$-band), AT2022tsd ($r$ and $z$-band), and AT2024wpp ($giz$-bands), we obtain 15$\times$120~sec exposures ($20\times$90~sec for the $z$-band observations) with the 6.5-m MMT Observatory's Binospec optical imager and spectrograph (PI: Nugent). We furthermore obtain 7$\times$200~sec exposures in $V$-band for the host of AT2023fhn and  9$\times$200~sec exposures in $G$-band for the host of AT2022tsd with the 10-m Keck Observatory's Low Resolution Imaging Spectrometer (LRIS; PI: Gordon). For the hosts of AT2024qfm and AT2024wpp, we collect near-IR $YJHK$ imaging (30~min total exposure time) with the MMT's MMT and Magellan Infrared Spectrograph (MMIRS; PI: Frostig). 

We reduce all photometric images from our programs using \texttt{POTPyRI}\footnote{https://github.com/CIERA-Transients/POTPyRI} (Pipeline for Optical/infrared Telescopes in Python for Reducing Images; \citealt{POTPyRI}), which performs bias subtraction, flat fielding, astrometry (via Astrometry.net; \citealt{lang2010}), image alignment and stacking, PSF photometry, and flux calibration. We perform aperture photometry on the hosts using \texttt{SEP} \citep{Barbary2016}, a Python-based version of \texttt{Source Extractor} \citep{Bertin1996}. Within \texttt{SEP}, we measure Kron radii of the hosts and extract a photometric measurement and uncertainty within these radii, using their standard \texttt{sep.sum$\_$ellipse} routine.

For the hosts of CSS161010, ZTF18abvkwl, AT2022tsd, AT2023fhn, A2023hkw, AT2023vth, AT2024qfm, and AT2024wpp, we use \texttt{FrankenBlast} \citep{frankenblast}, a customized version of the \texttt{Blast} web-application \citep{blast}, to obtain additional photometry of the hosts from public archival surveys. \texttt{FrankenBlast} collects images of the hosts from the \textit{Galaxy Evolution Explorer} (GALEX; \citealt{Galex}), the Panoramic Survey Telescope and Rapid Response System (Pan-STARRS; \citealt{PanSTARRS}), the Dark Energy Camera Legacy Survey (DECaLS) Data Release 10 \citep{DECaLS}, the \textit{Two Micron All-Sky Survey} (2MASS; \citealt{2MASS}), and the \textit{Wide-field Infrared Survey Explorer} (WISE; \citealt{WISE}). In each of the obtained images, \texttt{FrankenBlast} constructs an elliptical aperture around the host and extracts a flux measurement and uncertainty via \texttt{Astropy photutils} \citep{photutils}. We refer the readers to \citet{frankenblast} for more details on \texttt{FrankenBlast} photometric technique used in this work. We list all host galaxy photometric observations used in the stellar population modeling fits (Section \ref{sec:sp_model}) and magnitudes (not corrected for Galactic extinction) in Table \ref{tab:phot}. Prior to modeling the host galaxy SEDs (Section~\ref{sec:sp_model}), we correct all photometry for Galactic extinction in the direction of each LFBOT using the \citet{sf11} extinction maps.

\subsection{Spectroscopy}
We supplement our photometric observations with host galaxy spectra from both our own programs and archival data that were previously-published. We obtain MMT/Binospec spectra (PI: Nugent) for the hosts of AT2018cow, AT2020xnd, AT2020tsd, AT2023fhn, AT2023hkw, AT2023vth, AT2024qfm, and AT2024wpp. We further collect archival spectroscopic observations from Keck/LRIS for the hosts of CSS161010, ZTF18abvkwl, and AT2020mrf, which were originally published in \citet{coppejans2020}, \citet{ho2020_koala}, and \citet{yao2022}, respectively. We list all spectroscopic observations and exposure times in Table~\ref{tab:spec}.

We use \texttt{PypeIt} \citep{prochaska2020} to reduce the MMT/Binospec and Keck/LRIS observations and extract the 1D spectra.  We do not re-reduce the host spectrum of ZTF18abvkwl as \citet{ho2020_koala} published a flux-calibrated 1D host spectrum available for download. We employ the standard \texttt{PypeIt} pipeline to perform bias subtraction, flat-field correction, wavelength calibration, and spectral extraction on all other spectroscopic observations. We apply a flux calibration to each spectrum using spectrophotometric standard spectra and co-add the flux-calibrated 1D spectra. We use the \texttt{PypeIt}-measured spectral variance to determine spectral noise.

We correct all spectra for Galactic extinction using the \citet{MilkyWay} extinction law and the $V$-band extinction in the direction of each host \citep{sf11}. We identify high signal-to-noise (S/N$>5$) spectral features in each host spectrum and list the lines identified in Table~\ref{tab:spec}. We confirm the previously published redshifts for all LFBOTs using the host galaxy lines, except for AT2020xnd, where we do not detect any spectral lines. We note that the redshift for AT2020xnd was determined from the detection of weak H$\alpha$ emission (assumed to be from the host) in the transient spectrum \citep{perley2021}, observed with Keck/LRIS. As Keck/LRIS has a larger aperture than MMT/Binospec, and is thus more capable of detecting weaker emission, it is possible that this line was simply too low S/N to be detected in the Binospec spectrum.

\begin{deluxetable*}{lccccccccc}
\tabletypesize{\footnotesize}
\tablecolumns{10}
\tablewidth{0pc}
\tablecaption{\texttt{Prospector} Stellar Population Modeling Results}
\label{tab:host_prop}
\tablehead{
\colhead{LFBOT} &
\colhead{$z$} &
\colhead{log($M_*/M_\odot$)} &
\colhead{SFR [$M_\odot$~yr$^{-1}$]} &
\colhead{log(sSFR) [yr$^{-1}$]} &
\colhead{Age [Gyr]} &
\colhead{log($Z_*/Z_\odot$)} &
\colhead{log($Z_\textrm{gas}/Z_\odot$)} &
\colhead{$A_V$ [mag]}  &
\colhead{$\log(U_\textrm{gas})$} }
\startdata
AT2018cow & 0.0141 & $9.41^{+0.02}_{-0.01}$ & $0.27^{+0.12}_{-0.15}$ & $-9.98^{+0.16}_{-0.34}$ & $4.92^{+2.67}_{-0.95}$ & $-0.28^{+0.02}_{-0.03}$ & $-0.17^{+0.15}_{-0.19}$ & $0.17^{+0.02}_{-0.02}$ & $-2.41^{+0.32}_{-0.38}$ \\
CSS161010 & 0.033 & $7.6^{+0.11}_{-0.12}$ & $0.05^{+0.02}_{-0.03}$ & $-8.9^{+0.22}_{-0.36}$ & $4.22^{+1.69}_{-2.08}$ & $-1.55^{+0.21}_{-0.26}$ & $-0.6^{+0.09}_{-0.08}$ & $1.65^{+0.33}_{-0.29}$ & $-3.97^{+0.04}_{-0.02}$ \\
ZTF18abvkwl & 0.2715 & $9.85^{+0.17}_{-0.16}$ & $23.42^{+11.0}_{-15.72}$ & $-8.5^{+0.25}_{-0.49}$ & $4.28^{+0.86}_{-0.7}$ & $0.07^{+0.07}_{-0.07}$ & $-0.6^{+0.05}_{-0.05}$ & $3.76^{+0.23}_{-0.2}$ & $-2.75^{+0.32}_{-0.19}$ \\
AT2020mrf & 0.1353 & $8.24^{+0.05}_{-0.06}$ & $0.05^{+0.03}_{-0.03}$ & $-9.54^{+0.2}_{-0.44}$ & $5.69^{+0.64}_{-0.82}$ & $-0.88^{+0.12}_{-0.19}$ & $-0.56^{+0.23}_{-0.13}$ & $0.55^{+0.32}_{-0.32}$ & $-3.69^{+0.23}_{-0.19}$ \\
AT2020xnd & 0.2433 & $7.74^{+0.34}_{-0.4}$ & $0.06^{+0.05}_{-0.03}$ & $-8.99^{+0.51}_{-0.49}$ & $3.41^{+1.95}_{-2.1}$ & $-0.91^{+0.73}_{-0.69}$ & $-0.6^{+0.24}_{-0.21}$ & $0.36^{+0.55}_{-0.25}$ & $-1.77^{+0.54}_{-0.66}$ \\
AT2022tsd & 0.2555 & $9.64^{+0.15}_{-0.15}$ & $1.74^{+3.15}_{-1.01}$ & $-9.4^{+0.45}_{-0.41}$ & $2.91^{+1.98}_{-1.59}$ & $-0.88^{+0.2}_{-0.31}$ & $-0.28^{+0.16}_{-0.15}$ & $0.47^{+0.49}_{-0.29}$ & $-3.06^{+0.18}_{-0.16}$ \\
AT2023fhn & 0.2375 & $10.24^{+0.1}_{-0.15}$ & $11.26^{+8.82}_{-5.38}$ & $-9.17^{+0.29}_{-0.32}$ & $4.85^{+0.89}_{-1.33}$ & $-0.64^{+0.21}_{-0.19}$ & $-0.81^{+0.04}_{-0.02}$ & $1.03^{+0.46}_{-0.34}$ & $-1.94^{+0.37}_{-0.29}$ \\
AT2023hkw & 0.335 & $11.28^{+0.03}_{-0.04}$ & $18.77^{+6.31}_{-7.47}$ & $-10.01^{+0.13}_{-0.22}$ & $5.37^{+0.48}_{-0.43}$ & $0.15^{+0.03}_{-0.04}$ & $-0.53^{+0.13}_{-0.11}$ & $3.43^{+0.68}_{-0.34}$ & $-2.43^{+0.49}_{-0.93}$ \\
AT2023vth & 0.0747 & $10.03^{+0.09}_{-0.07}$ & $11.11^{+3.18}_{-8.45}$ & $-9.02^{+0.16}_{-0.6}$ & $7.85^{+2.61}_{-2.33}$ & $-0.6^{+0.04}_{-0.03}$ & $-0.4^{+0.06}_{-0.08}$ & $5.6^{+0.15}_{-0.12}$ & $-3.49^{+0.07}_{-0.07}$ \\
AT2024qfm & 0.227 & $10.37^{+0.05}_{-0.04}$ & $3.82^{+1.69}_{-1.26}$ & $-9.79^{+0.17}_{-0.18}$ & $5.98^{+0.53}_{-0.55}$ & $0.17^{+0.01}_{-0.02}$ & $-0.66^{+0.09}_{-0.05}$ & $0.8^{+0.36}_{-0.22}$ & $-2.08^{+0.28}_{-0.26}$ \\
AT2024wpp & 0.0866 & $8.86^{+0.06}_{-0.05}$ & $0.03^{+0.09}_{-0.01}$ & $-10.4^{+0.62}_{-0.28}$ & $4.07^{+1.57}_{-1.71}$ & $-1.78^{+0.05}_{-0.07}$ & $-0.43^{+0.13}_{-0.19}$ & $0.15^{+0.11}_{-0.12}$ & $-3.44^{+0.13}_{-0.14}$ \\ \hline 
All & $0.23^{+0.03}_{-0.17}$ & $9.61^{+0.74}_{-1.61}$ & $0.99^{+14.85}_{-0.95}$ & $-9.51^{+0.71}_{-0.64}$ & $4.99^{+1.44}_{-1.96}$ & $-0.61^{+0.74}_{-0.94}$ & $-0.55^{+0.26}_{-0.16}$ & $0.9^{+2.86}_{-0.72}$  & $-2.89^{+0.91}_{-0.78}$
\enddata
\tablecomments{The redshifts and \texttt{Prospector}-determined stellar masses ($M_*$), SFRs, specific SFRs (sSFR), mass-weighted stellar population age, stellar metallicities ($Z_*$), gas-phase metallicities ($Z_\textrm{gas}$), dust extinction ($A_V$), and gas ionization parameter ($U_\textrm{gas}$) for our LFBOT host population.}
\end{deluxetable*}
\section{Global Host Stellar Population Properties}
\label{sec:prosp}

In this section, we detail our stellar population modeling methods, an overview of LFBOT host stellar population property results, and comparisons to previous LFBOT literature that include global host property analysis.

\subsection{Modeling}
\label{sec:sp_model}

To determine the stellar population properties of our LFBOT host galaxy population, we fit their photometric and spectroscopic (if available) observations with stellar population modeling code, \texttt{Prospector} \citep{Leja2019, jlc+2021}. We prefer \texttt{Prospector} over other SED fitting tools given its ability to jointly fit both photometric and spectroscopic observations (a rare and computationally expensive feature), as well as the greater flexibility in model prescriptions, especially in regards to the galaxy star formation history. Moreover, a handful of our comparison samples were modeled with \texttt{Prospector} (see Section~\ref{sec:compare}), thereby making our results more aligned previous studies. Within \texttt{Prospector}, we employ the nested sampling fitting routine \texttt{dynesty} to sample various combinations of stellar population properties of interest and produce their posterior distributions. \texttt{Prospector} creates model photometry and spectroscopy through \texttt{FSPS} (Flexible Stellar Population Synthesis) and \texttt{python-FSPS} \citep{FSPS_2009, FSPS_2010}. Internally,  \texttt{Prospector} uses the \texttt{MILES} spectral libraries \citep{MILES}, \texttt{MIST} models \citep{MIST}, and WMAP9 cosmology.

For all \texttt{Prospector} fits, we use the \citet{KriekandConroy13} dust attenuation model, in which we sample an offset from the \citet{calzetti2000} attenuation curve and the ratio of light attenuated from young ($\tau_{V,1}$; in units of optical depth) to old  ($\tau_{V,2}$) stellar light and a \citet{Chabrier2003} initial mass function (IMF). We also include two templates that probe the mid-IR, which will ensure that we are modeling the SED properly in that wavelength regime: the \citet{DraineandLi07} IR dust emission model and the \citet{nenkova2008} Active Galactic Nuclei (AGN) model. The \citet{DraineandLi07} model comprises three-components: polycyclic aromatic hydrocarbon mass fraction ($q_\textrm{pah}$; essentially specifies the grain size distribution), the minimum radiation field strength ($U_\textrm{min}$), and the relative contribution of dust heated at $U_\textrm{min}$ ($\gamma_e$). We choose to only sample $q_\textrm{pah}$ to ensure we are not over-fitting the mid-IR data. The AGN model contains two sampled parameters: $\tau_\textrm{AGN}$ (the mid-IR optical depth) and $f_\textrm{AGN}$ (the total AGN luminosity, expressed as a fraction of the total bolometric luminosity from the host). We note that only one host (AT2018cow) has well-sampled mid-IR data; thus for the rest of the hosts, the posterior distribution will likely match the prior distribution. To probe realistic combinations of total mass formed in the host ($M_F$) and stellar metallicity ($Z_*$), we impose the \citet{gcb+05} mass-metallicity relation on their priors. This most complete relation for mass and \textit{stellar} metallicity beyond $z=0$, making it favorable to use in our study (see~\citealt{Leja2019} for further details on this prior).  Finally, we model the host star formation history (SFH) with a seven-bin non-parametric \texttt{continuity} SFH model. The bins represent lookback time, with the first two bins spaced at 0-30~Myr and 30-100~Myr to probe recent star formation. The next five bins are equally space in logarithmic time from 100 Myr to the age of the Universe at each hosts' redshift. We assume a constant star formation rate (SFR) in each bin and sample the ratio of the SFR between two adjacent bins. We choose to only model each host with seven bins as this will recover the overall shape of the SFH well without overfitting the data.

For the ten LFBOT hosts with spectral line detections in their spectra, we model their spectral continua with a 12$^\textrm{th}$-order Chebyshev polynomial, given that the observed spectra cover a wide wavelength range ($>$4000\AA) and this polynomial order best captures fluctuations in the spectral continua without overfitting them. We additionally apply a spectral smoothing model to match the resolution of the model spectra to that of the observed spectra and a model to normalize each host spectral continuum to its photometry. Given that the majority of our spectra are high S/N, we employ a noise inflation model that inflates the noise in all spectroscopic pixels to ensure that the spectra are not over-weighted in comparison to the photometry in the fit. We furthermore include a a pixel outlier model, which marginalizes over poorly modeled noise in the spectra, like residual sky lines or missing absorption lines (see Appendix D in \citealt{jlc+2021}). We fit the spectral emission lines with a nebular marginalization template, which marginalizes over the emission lines to measure their fluxes. We also sample a gas-phase metallicity ($Z_\textrm{gas}$) and a dimensionless gas ionization parameter that measures the ratio of hydrogen ionizing photons density to hydrogen density ($U_\textrm{gas}$). We note that we also include this nebular emission model to fit the host of AT2020xnd (in which we do not include the observed spectrum, as there are no identifiable features that would be useful in constraining the host stellar population properties), but we caution that this will likely result in poorly-constrained gas-phase metallicities, given that there are no spectral lines to fit. In all fits, we impose a 5\% error floor on the photometric fluxes to avoid overfitting any single photometric measurement.

We finally convert several sampled parameters into more commonly-used physical galaxy properties. We calculate a mass-weighted age ($t_m$) and present-day ($0-100~\textrm{Myr}$) SFR using the SFR ratios and $M_F$. We additionally determine $M_*$ from $M_F$, the SFH, and $Z_*$. We lastly convert $\tau_{V,1}$ and $\tau_{V,2}$ to a total $V$-band dust extinction magnitude ($A_V$) by multiplying their sum by 1.086. 

\subsection{Results}
\label{sec:sp_res}
We present the LFBOT host galaxy stellar population properties in Table \ref{tab:host_prop} and present the model fit to the observational data in Figure~\ref{fig:appsed}. We report all values as the population median and 68\% confidence interval, estimated from resampling with replacement the weighted raw samples (typically $\approx30,000$ samples) into 100,000 uniformly weighted samples from each LFBOT host's \texttt{Prospector}-derived stellar population property posterior distributions. This number ensures that we fully sample the posterior distributions of each host stellar population property.  We find that LFBOT hosts are typically lower mass galaxies with moderate to high amounts of star formation, with $\log(M_*/M_\odot) = 9.61^{+0.74}_{-1.61}$, present-day SFR = $0.99^{+14.85}_{-0.95}$~$M_\odot$~yr$^{-1}$, and sSFR = $-9.51^{+0.71}_{-0.64}$~yr$^{-1}$. We furthermore characterize all hosts as star-forming, using Equation (2) in \citet{tacchella2022}, which depends on the galaxy's sSFR and age of the Universe at the galaxy's redshift.

\begin{deluxetable*}{lcccccccc}
\tabletypesize{\footnotesize}
\tablecolumns{7}
\tablewidth{0pc}
\tablecaption{LFBOT Host Metallicities}
\label{tab:met}
\tablehead{
\colhead{LFBOT} &
\colhead{12 + $\log(\textrm{O}/\textrm{H})$} &
\colhead{12 + $\log(\textrm{O}/\textrm{H})$} &
\colhead{12 + $\log(\textrm{O}/\textrm{H})$} &
\colhead{log([OIII]/H$\beta$)} &
\colhead{log([NII]/H$\alpha$)} &
\colhead{log([SII]/H$\alpha$)} &
\colhead{log([OI]/H$\alpha$)} & \\
\colhead{} &
\colhead{R23} &
\colhead{[NII]/H$\alpha$} &
\colhead{[NII]/[OII]} &
\colhead{} &
\colhead{} &
\colhead{} &
\colhead{}
}
\startdata
AT2018cow & $8.67^{+0.09}_{-0.09}$ & $\cdots$ & $8.73^{+0.08}_{-0.13}$ & $0.38^{+0.15}_{-0.15}$ &  $-0.87^{+0.14}_{-0.14}$ & $-0.84^{+0.16}_{-0.16}$ & $-1.84^{+0.21}_{-0.21}$ \\
CSS161010 & $\cdots$ & $8.28^{+0.09}_{-0.1}$ & $\cdots$ & $0.48^{+0.13}_{-0.12}$ & $-0.82^{+0.14}_{-0.14}$ & $-0.17^{+0.07}_{-0.07}$ & $-1.07^{+0.08}_{-0.08}$ \\
ZTF18abvkwl & $8.55^{+0.11}_{-0.13}$ & $\cdots$ & $8.89^{+0.03}_{-0.04}$ & $0.59^{+0.23}_{-0.23}$ & $-1.18^{+0.12}_{-0.12}$ & $-0.75^{+0.1}_{-0.1}$ & $-1.78^{+0.12}_{-0.12}$ \\
AT2020mrf & $8.66^{+0.1}_{-0.12}$ &  $\cdots$ & $8.61^{+0.16}_{-0.28}$ & $-0.54^{+0.37}_{-0.37}$ & $-0.79^{+0.34}_{-0.34}$ & $-0.28^{+0.21}_{-0.21}$ & $-1.2^{+0.27}_{-0.27}$ \\
AT2022tsd & $8.65^{+0.08}_{-0.09}$ & $\cdots$ & $8.78^{+0.07}_{-0.14}$ & $0.19^{+0.18}_{-0.18}$ & $-0.63^{+0.2}_{-0.2}$ & $-0.47^{+0.12}_{-0.12}$ &  $-1.49^{+0.15}_{-0.15}$ \\
AT2023fhn & $8.52^{+0.1}_{-0.11}$ & $\cdots$ & $8.69^{+0.07}_{-0.09}$ & $0.67^{+0.14}_{-0.14}$ & $-1.29^{+0.15}_{-0.15}$ & $-1.06^{+0.12}_{-0.12}$ & $-2.03^{+0.13}_{-0.13}$ \\
AT2023hkw & $8.51^{+0.21}_{-0.39}$ & $\cdots$ & $8.86^{+0.05}_{-0.06}$ & $0.7^{+0.21}_{-0.21}$ & $-1.27^{+0.16}_{-0.16}$ & $-0.86^{+0.15}_{-0.15}$ & $-1.87^{+0.18}_{-0.18}$ \\
AT2023vth & $8.92^{+0.04}_{-0.04}$ & $\cdots$ & $9.03^{+0.02}_{-0.02}$ & $-0.17^{+0.12}_{-0.12}$ & $-0.58^{+0.11}_{-0.11}$ & $-0.22^{+0.07}_{-0.07}$ & $-1.29^{+0.08}_{-0.08}$ \\
AT2024qfm &  $8.49^{+0.09}_{-0.12}$ & $\cdots$ & $8.52^{+0.1}_{-0.14}$ & $0.74^{+0.13}_{-0.13}$ & $-1.59^{+0.16}_{-0.16}$ & $-1.09^{+0.1}_{-0.1}$ & $-2.04^{+0.12}_{-0.12}$ \\
AT2024wpp & $8.62^{+0.11}_{-0.13}$ & $\cdots$ & $8.58^{+0.14}_{-0.27}$ & $-0.19^{+0.2}_{-0.2}$ & $-0.67^{+0.22}_{-0.22}$ & $-0.32^{+0.14}_{-0.14}$ &  $-1.3^{+0.16}_{-0.16}$ \\
\enddata
\tablecomments{Different emission-line measured metallicity estimates for the LFBOT hosts with observed spectra used in their \texttt{Prospector} fits. For LFBOT hosts that have a median $R_{23}$-determined 12 + $\log(\textrm{O}/\textrm{H})$ $<8.53$, we use their [NII]/H$\alpha$-determined metallicity. We do not calculate an R23 or [NII]/[OII]-based metallicity for the host of CSS161010, as the emission line fluxes blueward of [OIII] were not well-constrained in the \texttt{Prospector} fit.}
\end{deluxetable*}

\begin{figure*}[t]
\centering
\includegraphics[width=0.32\textwidth]{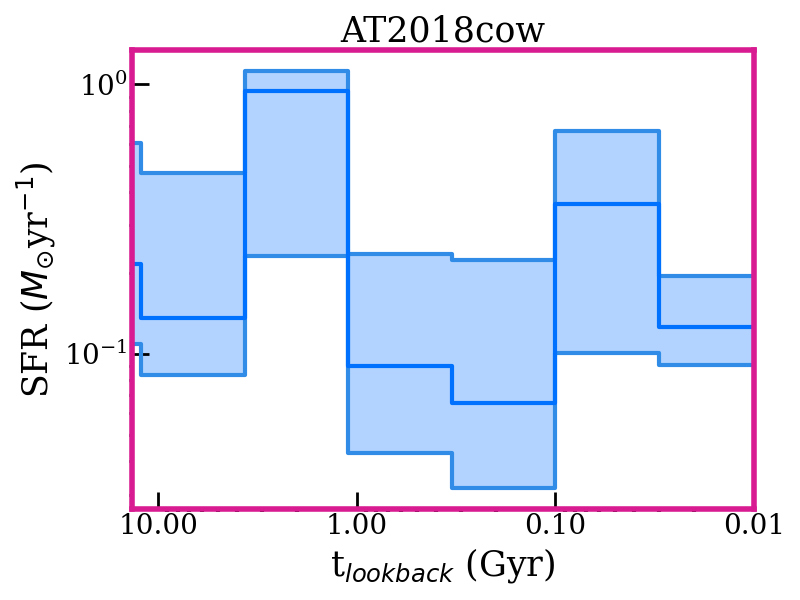}
\includegraphics[width=0.32\textwidth]{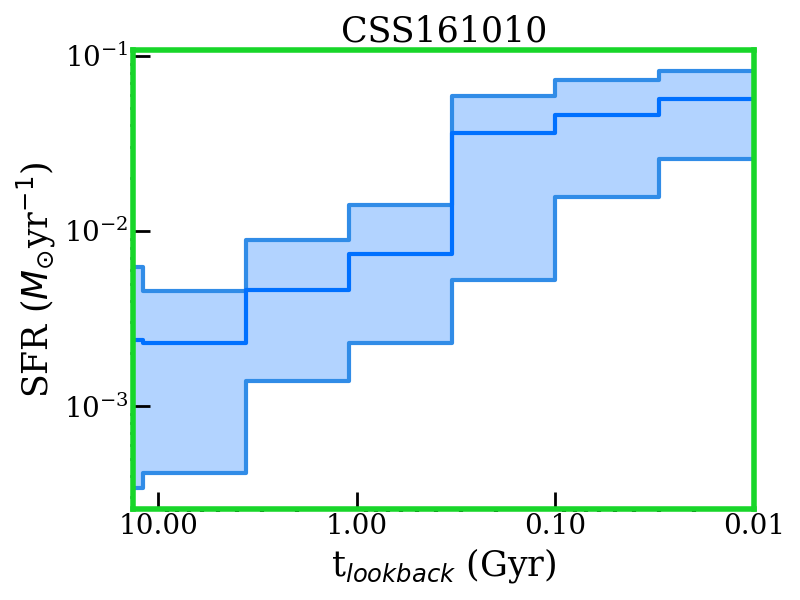}
\includegraphics[width=0.32\textwidth]{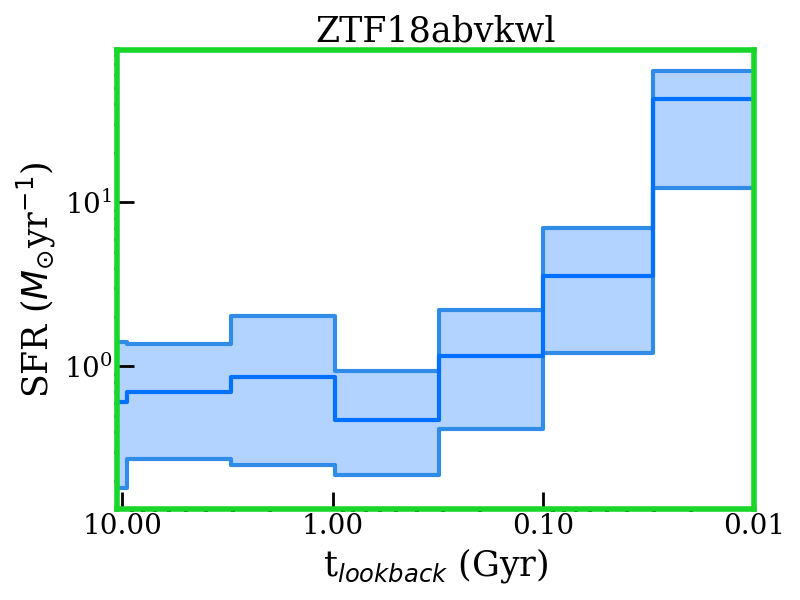}
\includegraphics[width=0.32\textwidth]{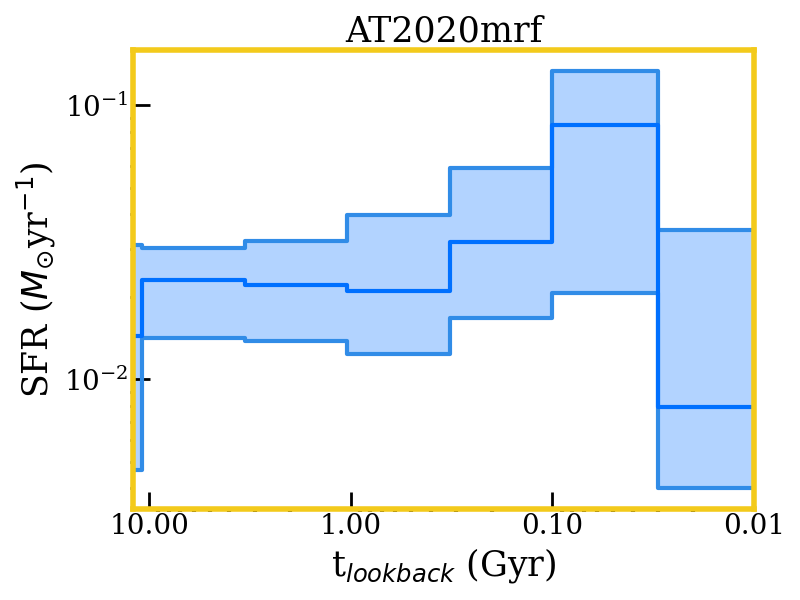}
\includegraphics[width=0.32\textwidth]{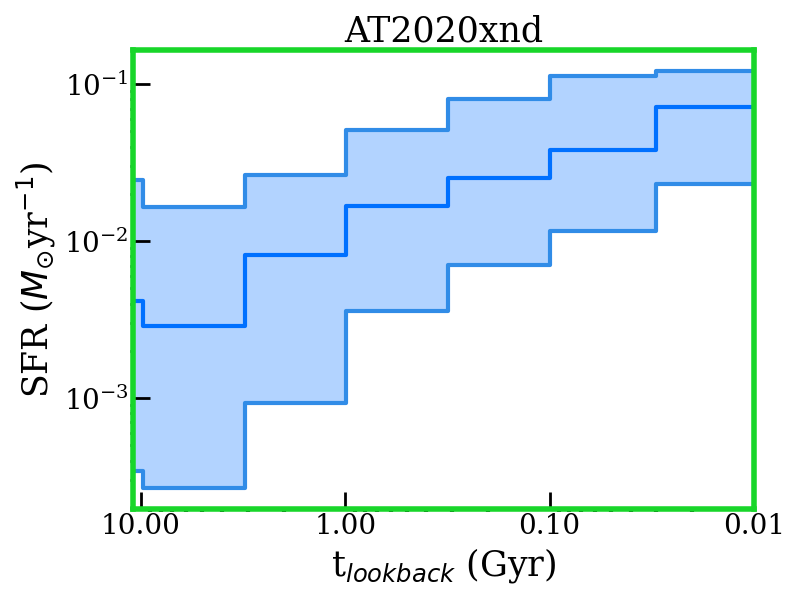}
\includegraphics[width=0.32\textwidth]{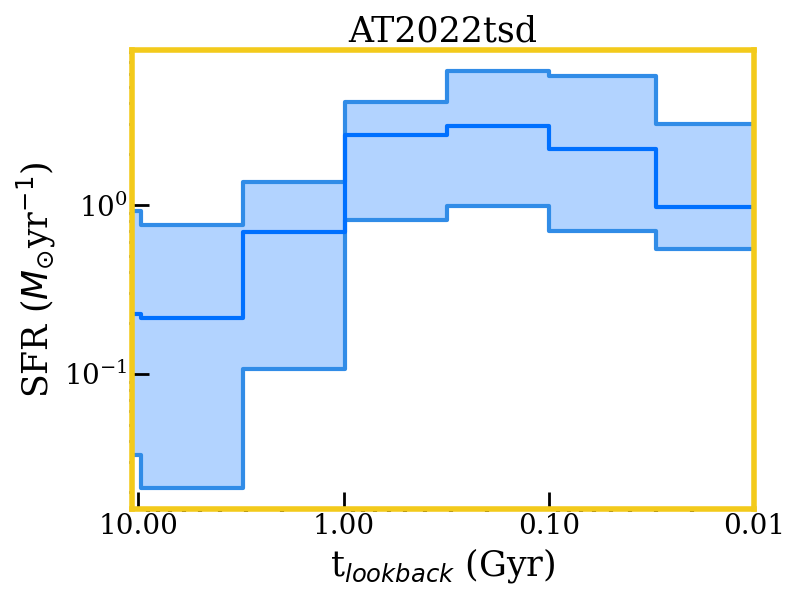}
\includegraphics[width=0.32\textwidth]{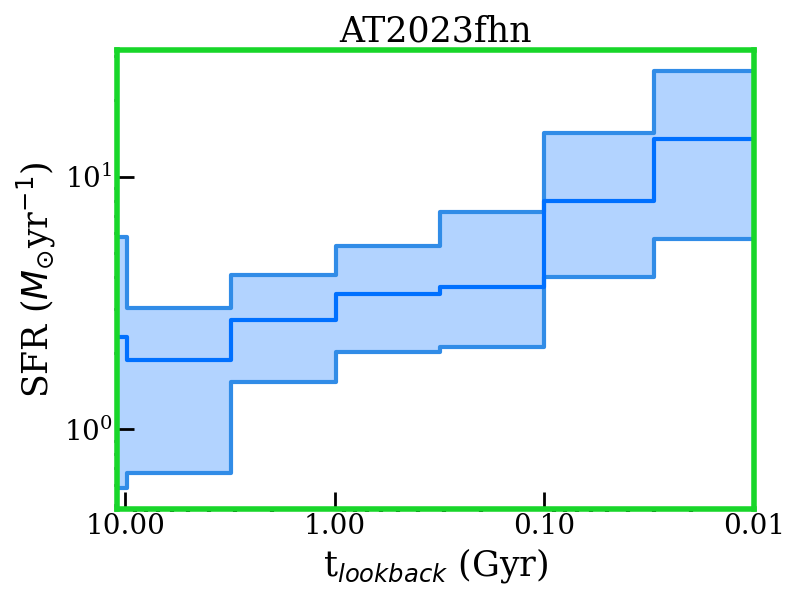}
\includegraphics[width=0.32\textwidth]{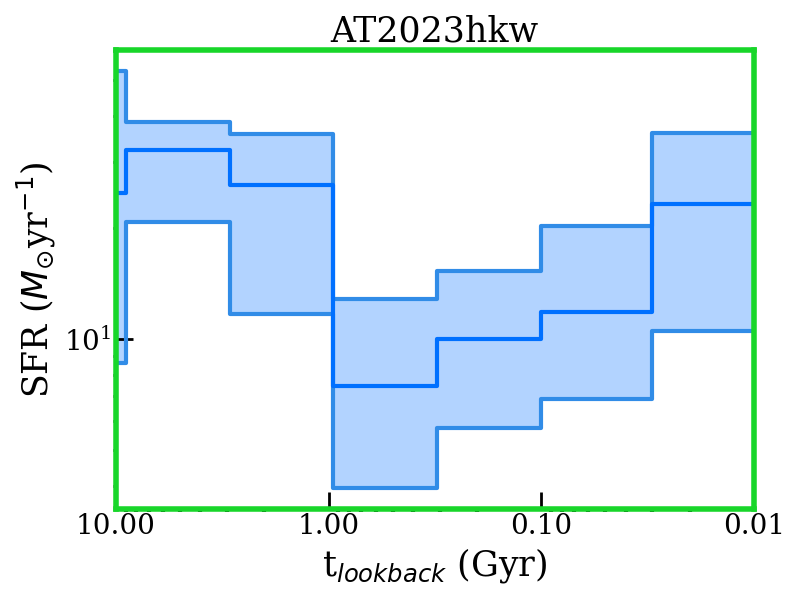}
\includegraphics[width=0.32\textwidth]{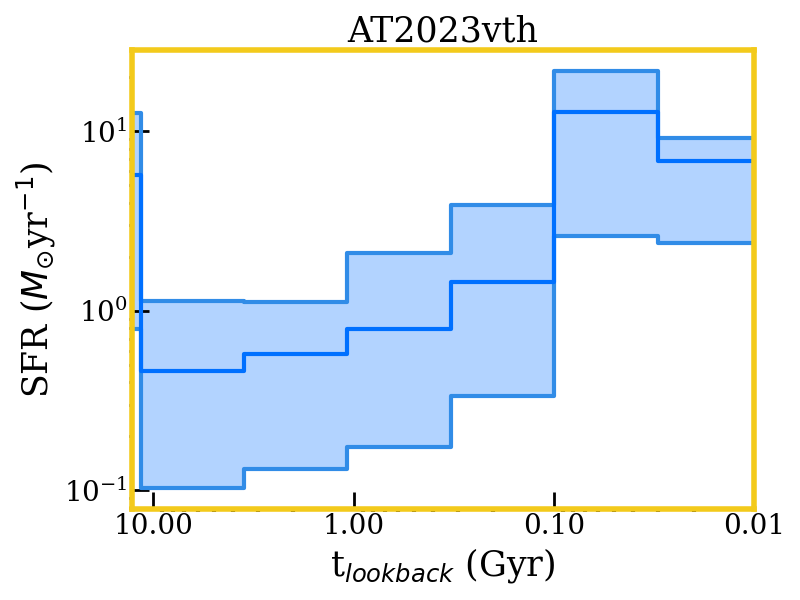}
\includegraphics[width=0.32\textwidth]{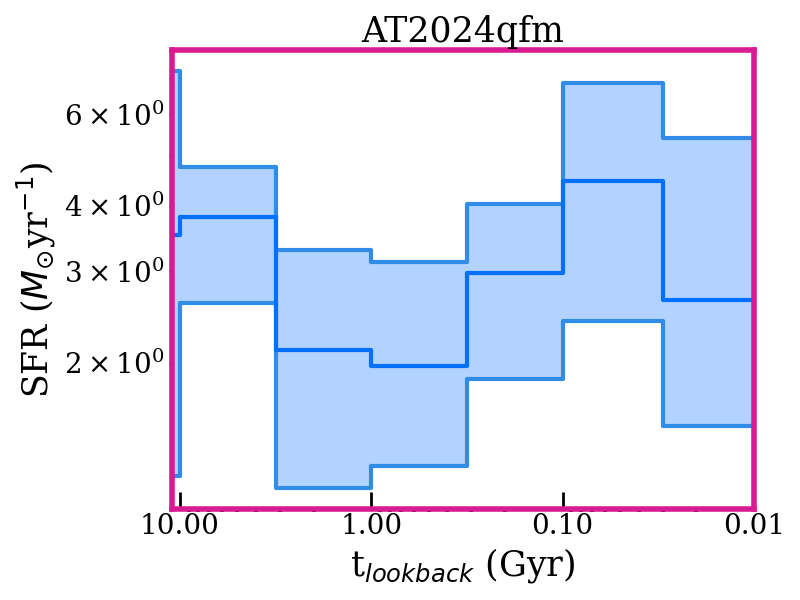}
\includegraphics[width=0.32\textwidth]{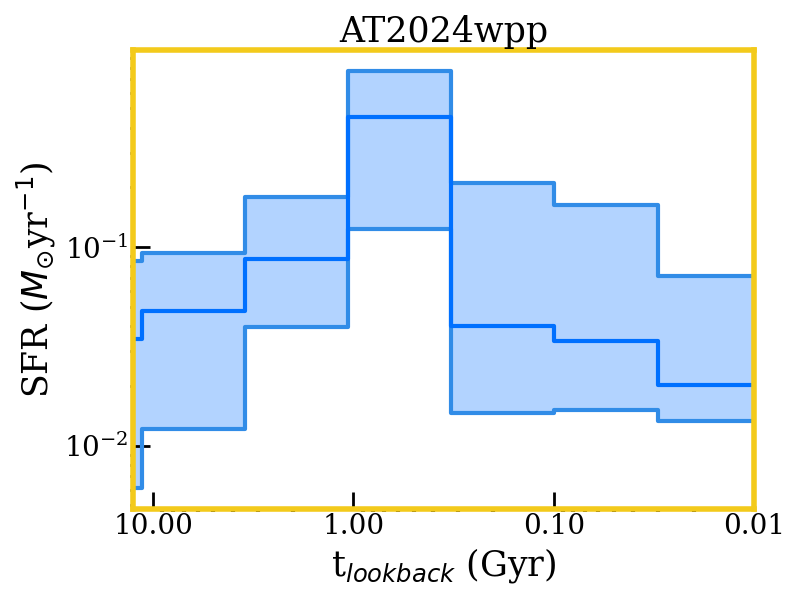}
\caption{The \texttt{Prospector}-derived SFHs of the 11 LFBOT hosts in our sample. We see a diversity of SFH shapes: five events have clearly rising SFHs (green border) and the rest appear to have had one (yellow border) or several (pink border) star formation bursts in their past have since declined or plateaued in SFR. All hosts are classified as currently star-forming, given the metric discussed in \citet{tacchella2022}.}
\label{fig:sfh}
\end{figure*}

Despite the apparent uniformity in present-day star-formation classification, it appears that LFBOT hosts have a diverse array of SFH shapes, which we highlight in Figure~\ref{fig:sfh}. We see that five hosts are still actively rising in SFR until present-day (hosts of CSS161010, ZTF18abvkwl, AT2020xnd, AT2023fhn, and AT2023hkw), and the other hosts have had past burst(s) of star formation, and have since declined or plateaued in SFR (hosts of AT2018cow, AT2020mrf, AT2022tsd, AT2023vth, AT2024qfm, and AT2024wpp). We note that the majority of hosts have had a star-formation burst or have reached their highest SFR in the past $\approx$100~Myr (with the exception of AT2024wpp, which appears to have had a single star formation burst  $\approx$1~Gyr ago). If LFBOTs are of stellar origin (see Section~\ref{sec:disc} for further discussion), and assuming their progenitor has the highest probability of forming in the most recent burst of star formation, this likely implies that that their progenitor formation timescale (delay time) is relatively short.

We generally find that LFBOT host stellar and gas-phase metallicities lean towards low-metallicity values, with $\log(Z_*/Z_\odot) = -0.61^{+0.74}_{-0.94}$ and $\log(Z_\textrm{gas}/Z_\odot) = -0.55^{+0.26}_{-0.16}$. As a many studies on general field galaxy populations and other transient host populations use gas-phase oxygen abundance metallicities ($12+\log(\textrm{O}/\textrm{H})$), we additionally determine this metallicity measurement for our LFBOT host population. We compute $12+\log(\textrm{O}/\textrm{H})$ for LFBOT hosts with emission line flux measurements determined through our \texttt{Prospector} fitting using three different calibration methods: $R_{23}$, [NII$\lambda6584$]/H$\alpha$, and [NII$\lambda6584$]/[OII]\footnote{We exclude the host of AT2020xnd from this analysis, as we did not use a spectrum in its \texttt{Prospector} fit.}. Each calibration works well over a different range of metallicities, with $R_{23}$, given by:
\begin{equation}
\label{eq:r23}
R_{23} = \log_{10}\Big( \frac{\textrm{OII}\lambda\lambda3727,3729 + \textrm{OIII}\lambda\lambda4959,5007} {\textrm{H}_\beta} \Big),
\end{equation}
valid over $8.53 < 12+\log(\textrm{O}/\textrm{H}) < 9.23$, the [NII]/H$\alpha$ method valid over $7.63 < 12+\log(\textrm{O}/\textrm{H}) < 8.53$, and [NII]/[OII] valid over the full metallicity range \citep{kewley2019}. As both $R_{23}$ and [NII]/H$\alpha$ both have low root-mean-square (RMS) errors in comparison to other calibrations and are more commonly used in transient host literature (see Section~\ref{sec:metallicity}), we prefer to use these two methods in our analysis \citep{kewley2019}. However, we use the [NII]/[OII] method to break the degeneracy between low and high metallicities, determine which calibration method should be reported for a host, and validate our metallicity measurements. \citet{kewley2008} show that when log([NII/[OII])$>-1.2$ ($12+\log(\textrm{O/H}) \gtrsim 8.4$), the $R_{23}$ method determines metallicities well. Below this limit, we must use better calibrators for low-metallicity galaxies, e.g., [NII]/H$\alpha$. We determine that all hosts in our sample have log([NII/[OII])$>-1.2$, thus we use the $R_{23}$ method to determine their metallicities. For the host of CSS161010, the emission line fluxes blueward of [OIII] were not measured well in the \texttt{Prospector} fit, thus we only determine an [NII]/H$\alpha$ metallicity. We note that as this host is a low-mass dwarf galaxy, we would assume from standard mass-metallicity relations that it should have a fairly low metallicity \citep{kewley2008}, and the [NII]/H$\alpha$ method will constrain it better than R23.

We follow the methodology in \citet{kewley2019} (see Tables 2 and 3) to determine $12+\log(\textrm{O}/\textrm{H})$, which is dependent on the specified calibration ($R_{23}$, [NII]/H$\alpha$, or [NII]/[OII]) and $\log(U_\textrm{gas})$. To correctly capture the uncertainty on each $12+\log(\textrm{O}/\textrm{H})$ measurement, we sample 1000 emission line fluxes for each line used in the respective calibration method from an asymmetric Gaussian distribution, using the 68\% interval on the flux provided by \texttt{Prospector}, and thus collect 1000 $12+\log(\textrm{O}/\textrm{H})$ values. We only use the median $\log(U_\textrm{gas})$, as provided in Table~\ref{tab:host_prop}. We present the metallicity measurements in Table~\ref{tab:met} and determine that LFBOT hosts have a $12+\log(\textrm{O}/\textrm{H}) = 8.59^{+0.18}_{-0.22}$. In general, we also find that the $R_{23}$ and [NII]/[OII] metallicities are the same within error.

We finally make a brief comment about potential biases in our stellar population modeling results. Typically, jointly fitting photometry and spectroscopy leads to accurate constraints on stellar population properties, as shown in \citet{jlc+2021}, thus we do strongly trust the accuracy of our results. However, \citet{wang2023} and \citet{frankenblast} both highlight how the nested sampling within \texttt{Prospector} can systematically lead to underestimated uncertainties specifically in fits with just photometry. Moreover, well-known degeneracies between stellar population age, stellar metallicity, and dust attenuation become stronger when only fitting photometry (see \citealt{conroy2013} for a review). Given that this only affects one host in our sample, we do not believe that this will have a strong impact on our conclusions.

\subsection{Comparisons to Previous LFBOT Literature}
\label{ref:lit}

We next briefly compare our main stellar population modeling results to those previously derived in other literature. We first emphasize that our study takes advantage of all available host galaxy photometry and spectroscopy and sophisticated, uniform stellar population modeling techniques to place the best constraints on LFBOT environments. We find estimates of host galaxy stellar mass and SFR for 9 LFBOTs in our sample: AT2018cow \citep{perley2019}, CSS161010 \citep{coppejans2020}, ZTF18abvkwl \citep{ho2020_koala}, AT2020mrf \citep{yao2022}, AT2024wpp \citep{perley2026}, AT2023fhn, AT2023hkw, AT2023vth, and AT2024qfm \citep{sevilla2026}. We note that the stellar population modeling techniques employed across these papers is inconsistent and generally focus on fitting host photometry exclusively, which will give poorer constraints on SFR, and larger uncertainties on all properties, than achieved here \citep{jlc+2021}. Moreover, the majority of these studies rely on parametric SFH models. As shown in \citet{Leja2019}, non-parametric SFH models (used here) tend to result in stellar masses that are 25–100\% larger than those returned by parametric SFH models and may lead to subtle differences in present-day SFRs. Indeed, we determine higher stellar masses for 8 LFBOTs in this sample (all except AT2024wpp). On average, the increase is $\approx0.3$~dex. The most extreme differences are for ZTF18abvkwl and AT2023vth, where we find stellar masses that are $\approx1$~dex higher than those published in \citet{ho2020_koala} and \citet{sevilla2026}, respectively. This is almost certainly due to our usage of a non-parametric SFH model and jointly fitting both the host photometry and spectrum in the SED fit.

The majority of our SFR measurements are consistent with those that were previously published, and we find only minimally higher SFRs for CSS161010, AT2023hkw, and AT2023vth. As with the stellar mass differences, this also due to our implementation of a non-parametric SFH. Given these subtle differences in host stellar population properties, we finally wish to emphasize the importance of using stellar population models with the same underlying physics to derive host properties for a given sample, as achieved here.

\begin{figure*}[t]
\centering
\includegraphics[width=0.9\textwidth]{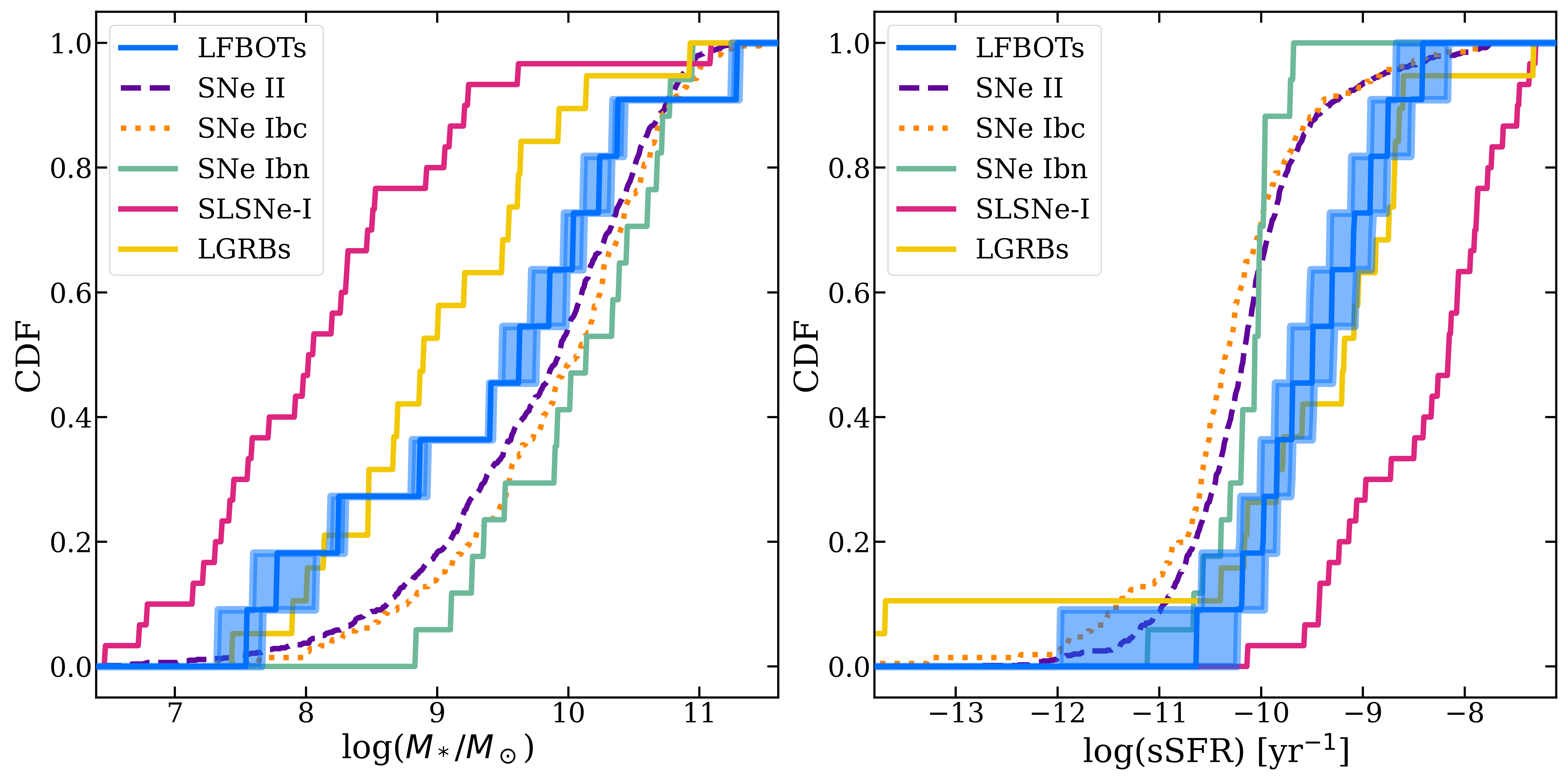}
\vspace{-0.1in}
\caption{\textit{Left:} The stellar mass CDFs for the hosts of LFBOTs (blue; median and 68\% interval), SNe II (purple dashed line), SNe Ibc (orange dotted line), SNe Ibn (green), SLSNe-I (pink), and LGRBs (yellow). LFBOT hosts are more massive than those of SLSNe-I and have statistically similar stellar mass distributions to SNe Ibc, Ibn, II and LGRB hosts. \textit{Right:} The sSFR CDFs for the same transient populations. LFBOT hosts are less star-forming than SLSN-I hosts, and more star-forming than CCSNe (SNe II, Ibc, and Ibn) hosts. LFBOT hosts' sSFR distribution is not statistically distinct from that of LGRB hosts.}
\label{fig:sfrmass}
\end{figure*}

\section{Comparison of LFBOT Hosts to Field Galaxies and Other Transient Hosts}
\label{sec:compare}
In this section, we contextualize the LFBOT host stellar population property results (Section~\ref{sec:sp_res}) by comparing them against field galaxy populations and host populations of transients that LFBOTs have been frequently compared to in the literature: SLSNe-I, Type II SNe (SNe II), Type Ibc SNe (SNe Ibc), Type Ibn SNe (SNe Ibn), and LGRBs. In Section~\ref{sec:mass_sf}, we compare the 1D $M_*$ and present-day star formation properties of these transient populations, and explore how they fall in the 2D SFR-$M_*$ plane in Section~\ref{sec:sfms}. In Section~\ref{sec:metallicity}, we cover the metallicities of the hosts and where they lie on the Baldwin, Phillips and Telervich (BPT) diagram.

\subsection{Mass and Star Formation}
\label{sec:mass_sf}
We begin by analyzing LFBOT host stellar masses and sSFRs against those of SLSN-I, LGRBs, SN II, Ibc, and Ibn hosts. We show cumulative distributions (CDFs) of LFBOT host $M_*$ and present-day specific SFR (sSFR  = SFR/$M_*$) in Figure~\ref{fig:sfrmass}. To build the distributions, we sample (with replacement) 5000 values from each LFBOT host's \texttt{Prospector}-derived stellar population property posterior distributions and create 5000 realizations on the CDF. We choose 5000 draws as there is little change in the CDFs beyond this point, and this number is not so large that it will artificially decrease the uncertainty in the CDF. We obtain host galaxy sSFRs and $M_*$ for SNe II, Ibc, and Ibn in \citet{frankenblast}, LGRBs in \citet{svensson2010, perley2013, wang2014, niino2017}, and SLSNe-I in \citet{schulze2021} and show their respective CDFs in Figure~\ref{fig:sfrmass}, as well. All comparison samples used here and elsewhere in the text are listed in Table~\ref{tab:lit}, for reference. We note that \citet{frankenblast} employed the same \texttt{Prospector} model as used in this work to derive host stellar population properties (minus the parameters used to fit the spectrum as they only employed host photometry in their study), thus their results are complementary to the ones derived here. \citet{schulze2021} also adopted a \texttt{Prospector} model to determine SLSN-I host $M_*$ and sSFR, however, they modeled SFH with a parametric delayed-$\tau$ model. As previously explained, this tends to lead to lower stellar mass estimates and differing SFR constraints than the non-parametric SFH model. In turn, there may be slight systematic differences between the LFBOT and SLSN-I $M_*$ and sSFR distributions caused by these varying methods. This also may be the case for the LGRB host sample, as \citet{svensson2010}, \citet{perley2013}, \citet{wang2014}, and \citet{niino2017} use a variety of stellar population modeling techniques to derive LGRB host properties. Finally, we note that redshift may also bias comparisons between LFBOT host properties and the other samples. For example, at higher redshift, we expect that galaxy populations will have lower $M_*$ and higher sSFR. To mitigate this redshift effect, we limit all host samples to $z<0.5$\footnote{This redshift limit only affects the SLSN-I and LGRB samples, which contain many more events at $z>0.5$. The CCSNe populations exclusively lie at $z<0.2$.}, where we expect that redshift will have a subtler effect on the overall $M_*$ and sSFR distributions. This is also the redshift range in which many field galaxy populations are analyzed \citep{speagle2014, whitaker2014, leja2022}, and ensures that we have statistically robust populations of LGRBs and SLSNe-I at low redshifts.

\begin{deluxetable*}{l|cccccc}
\tabletypesize{\footnotesize}
\tablecolumns{7}
\tablewidth{0pc}
\tablecaption{Results of Anderson-Darling Tests}
\label{tab:pvalue}
\tablehead{
\colhead{Transient} &
\colhead{$\log(M_*/M_\odot)$} &
\colhead{sSFR} &
\colhead{12+log(O/H)} &
\colhead{$R_\textrm{phys}$} &
\colhead{$R_\textrm{hnorm}$} &
\colhead{Fractional Flux}
}
\startdata
SLSNe & 100\% & 99\% & 99\% & 100\% & 0\% &  0\%\\
LGRBs & 0\% & 0\% & 62\% & 100\% & 74\% & 96\%\\
SNe Ibc & 4\% & 98\% & 100\% & 0\% & 94\% &  98\%\\
SNe II & 0\% & 96\% & 100\% & 0\% & 0\% & 54\% \\
SNe Ibn & 0\% & 97\% & 0\% & 0\% & 94\% &  $\cdots$ \\
\enddata
\tablecomments{Results of the AD tests, represented in the percentage of tests that reject the null hypothesis ($P_{AD}<0.05$), between LFBOT environmental properties and those of different transient classes. If $>50\%$ of tests can be rejected, we reject the null hypothesis that LFBOTs and the respective transient population have the same underlying environmental property distribution.}
\end{deluxetable*}

To quantitatively compare LFBOT $M_*$ and sSFR distributions to those of these selected transient populations, we perform Anderson-Darling (AD) rejection tests, with the null hypothesis that LFBOTs and the transient-of-interest derive from the same underlying stellar mass or sSFR distribution. We prefer to use AD tests as they are more sensitive in finding deviations in the tails of the distributions, while traditional Kolmogorov–Smirnov (KS) tests solely probe the centers of distributions. For each LFBOT-transient pair, we perform 5000 AD tests, with each test between one randomly-selected realization on the LFBOT $M_*$ or sSFR distribution and the respective transient $M_*$ or sSFR distribution. We then determine the percentage of tests that fall below probability $P_{AD} < 0.05$, the point at which we can reject the null hypothesis. If $>50\%$ of tests result in $P_{AD} < 0.05$, we reject the null hypothesis, as we assume that if the majority of tests fail, there is sufficient evidence that the distributions are distinct. We are more confident that the distributions indeed deviate if the fraction of failed tests is significantly $>50\%$. Finally, we note that $<50\%$ of tests rejected does not imply that there are strong correlations between the distributions, as we only measure the fraction of rejection. We list the results of all AD tests, here and elsewhere in the paper, in Table~\ref{tab:pvalue}.

We find that we cannot reject the null-hypothesis that LFBOTs trace the same $M_*$ or sSFR distributions as LGRBs, given that 0\% of $P_{AD} < 0.05$ when testing both properties. We furthermore determine that LFBOT $M_*$ and sSFRs are statistically distinct from those of SLSNe (100\% of $P_{AD} < 0.05$ for both $M_*$ and sSFR). Given the stark difference in the LFBOT and SLSN-I host $M_*$ and sSFR distributions, we do believe that this distinction is real and not heavily  biased by subtle differences in the stellar population modeling technique. While we also find that LFBOT hosts trace higher sSFRs than SNe II, Ibc, and Ibn ($>96\%$ of $P_{AD} < 0.05$ for all three comparisons), their stellar mass distributions are not statistically different ($<4\%$ of $P_{AD} < 0.05$ for all three comparisons). Thus, it appears that LFBOT hosts may be more star-forming than CCSN hosts but are not distinct in stellar mass. 

The similarity of LFBOT hosts to those of transients with young, massive stellar progenitors (CCSNe, LGRBs, SLSNe-I), with respect to star formation and stellar mass, strongly suggests that they too arise from young stellar progenitors. In particular, the lack of high-mass, quiescent LFBOT hosts, which are more common for transients with larger formation timescales (e.g., short GRBs and Type Ia SNe; \citealt{mannucci2005, leibler2010, mm2012, wang2013, maoz2014, Pan2014, fong2013, chen2021, wiseman2021, nugent2022, jeong2024}) or non-stellar origins (e.g., TDEs; \citealt{french2016, lawsmith2017, hammerstein2021}), informs that the LFBOT progenitor is much more strongly linked to recent star-formation in the host. We explore this dependency on star formation and stellar mass more in the following subsection, and their possible progenitors in Section~\ref{sec:disc}.

\subsection{Star-Forming Main-Sequence}
\label{sec:sfms}

\begin{figure*}[t]
\centering
\includegraphics[width=\textwidth]{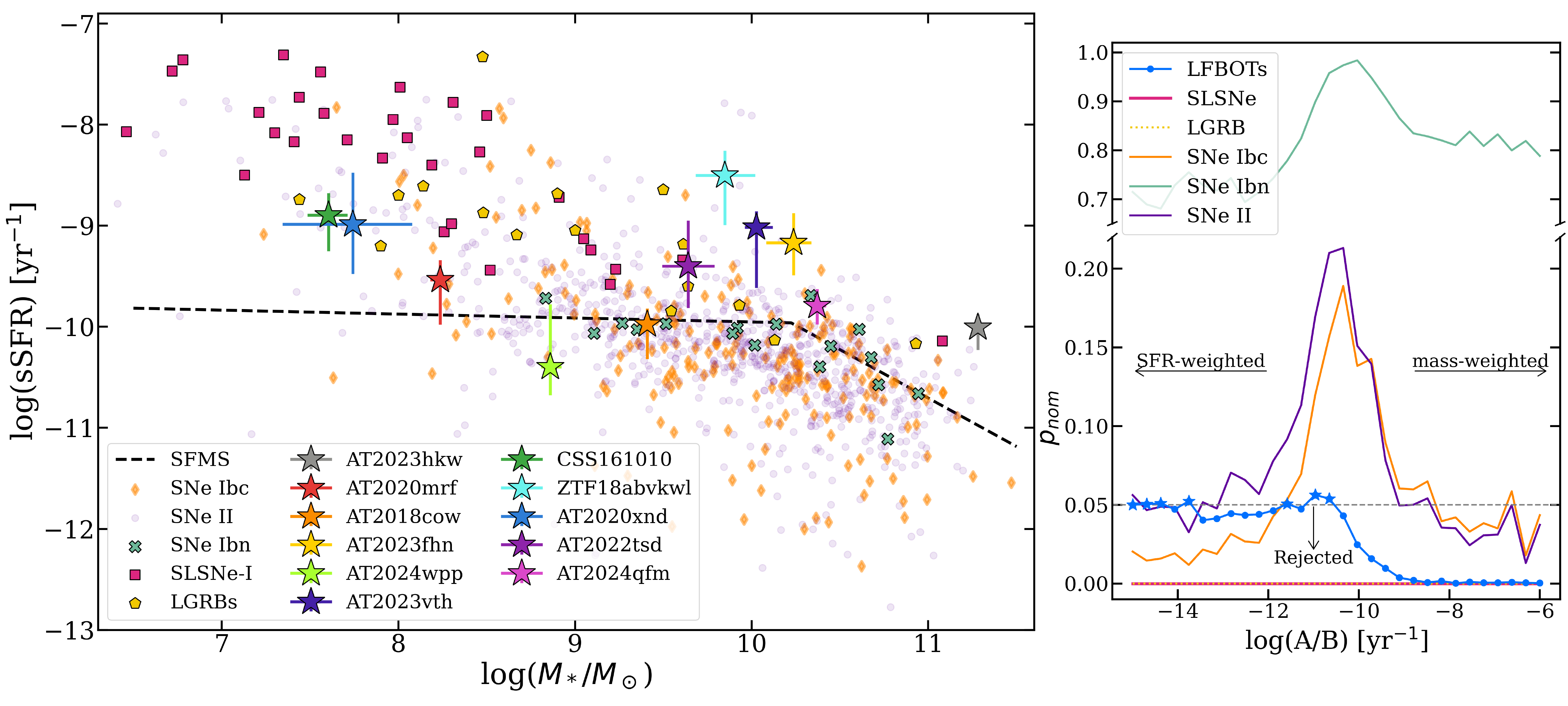}
\caption{\textit{Left:} The hosts of LFBOTs (stars; lines indicate median and 68\% interval), SNe II (purple circles), SNe Ibc (orange diamonds), SNe Ibn (green crosses), SLSNe-I (pink squares), and LGRBs (yellow pentagons) along the SFMS derived in \citet{leja2022} at $z<0.5$. LFBOTs hosts lie slightly above the SFMS and the CCSNe host populations, and are clearly distinct from LGRB and SLSN-I hosts. \textit{Right:} The results of the \citet{hm2026} test, which determines if a host galaxy population is drawn from a weighted SFR-$M_*$ distribution of a mock field galaxy population. We determine weights from $\log(A/B)$, where lower values indicate more SFR-weighted distributions, and higher values indicate more $M_*$-weighted distributions. If $p_\textrm{nom} < 0.05$, we can reject the null hypothesis that the host galaxy sample is drawn from the weighted distribution. We find that LFBOT hosts prefer more SFR-weighted distributions, especially in comparison the SNe II and Ibc hosts. The stars mark the solutions where the LFBOT 1D SFR and $M_*$ distributions were not statistically distinct from those of the weighted distribution. We cannot reject any weighted distributions for SNe Ibn hosts and we do not find any solution that fits the SFR-$M_*$ distributions of LGRB and SLSN-I hosts well.}
\label{fig:sfms}
\end{figure*}

As another test of how different LFBOT hosts are from the selected transient hosts, we showcase them along the star-forming main-sequence (SFMS) in Figure~\ref{fig:sfms}: a well-studied galaxy relation followed by star-forming galaxies as they gain in stellar mass \citep{speagle2014, whitaker2014, leja2022}. How transient hosts track the SFMS informs how dependent their progenitors are on star-formation or stellar mass for formation (with older progenitors more dependent on stellar mass, given their longer formation timescales). Visually, it appears that LFBOT lie slightly above the SFMS and the bulk of the CCSN host populations, seemingly tracing more star-forming galaxies. However, LFBOTs are not as high above the SFMS as SLSN-I and LGRB hosts \citep{schulze2018, schulze2021}. 

As a more quantitative assessment of how LFBOT hosts track the SFMS, we next determine how LFBOTs track both SFR and $M_*$ in the 2D SFR-$M_*$ plane, given the multivariate statistical test presented in \citet{hm2026}. \citet{hm2026} formulate a rejection test to determine whether a mock galaxy population, weighted by various combinations of SFR and $M_*$, can be ruled out to describe the SFR-$M_*$ distribution of a transient host population. This essentially tests the relative rates of transients tracking SFR or $M_*$ to determine how dependent their progenitor may be on either of these properties for formation. Similar to the AD test, \citet{hm2026} determine a nominal $p$-value ($p_\textrm{nom}$), where $p_\textrm{nom} < 0.05$ suggests the null hypothesis that the transient hosts derive from a specified weighted distribution can be rejected. Because this test was previously applied by \citet{frankenblast} to evaluate several SN host populations (including SNe II and Ibc), it provides a baseline to compare against our LFBOT host sample. For more details on this method, we refer the readers to \citet{hm2026}.

To build mock galaxy populations to compare to our LFBOT host sample, we sample galaxies from weighted probability density distributions with the same redshifts as the LFBOTs, given by $W \times \rho$($M_*$, SFR$|z)$. The ``unweighted" galaxy sample comes from \citet{leja2022}, in which they derived $\rho$($M_*$, SFR$|z)$ for galaxies in the COSMOS-2015 \citep{Laigle2016} and 3D-HST \citep{Skelton2014} surveys. All galaxies were modeled with a similar \texttt{Prospector} model as the one used here, making the results comparable to our work. We apply weights to this sample with the function:
\begin{equation}
\label{eq:weights}
    W_i = A \Big( \frac{M_{*,i}}{M_\odot}\Big) + B \Big( \frac{\textrm{SFR}_i}{M_\odot \textrm{yr}^{-1}}\Big),
\end{equation}
where $W_i$ represents the weight on the  $i^\textrm{th}$ galaxy in the sample. If $A \rightarrow 0$, the solution reduces to a SFR-weighted distribution, and if $B \rightarrow 0$, the solution reduces to a $M_*$-weighted distribution. To test a variety of weighted distributions against the LFBOT hosts, spanning very SFR to very $M_*$-weighted galaxy populations, we build 30 mock galaxy populations with $-15 \leq \log(A/B) \leq -6$, where $A+B=1$. For each weighted distribution, we draw 32,000 mock galaxies. Employing the \citet{hm2026} rejection test, we determine a single $p_\textrm{nom}$ for each weighted distribution, using only the median SFRs and $M_*$ of our LFBOT host population. In addition, we do the same tests for LGRB, SLSN-I, and SN Ibn (all at $z<0.5$) host populations as these do not exist in the literature\footnote{For the SN Ibn host population, we employ the same methods as described in \citet{frankenblast}, where they used a weighted distribution with an optical bias to exclude any mock galaxies with optical magnitude $>$23.5 mag (the same bias as their SN host sample). We do not use this bias for LFBOTs, LGRBs, and SLSNe as their host populations include galaxies fainter than this limit.}. 

We show the results of the rejection test in Figure~\ref{fig:sfms}. We find that we can reject the most mass-weighted SFR-$M_*$ distributions for LFBOTs, with their hosts seemingly preferring more SFR-weighted solutions. As another test of whether the 2D weighted distribution fits the LFBOT host galaxy SFR-$M_*$ distribution well, we perform KS tests between the 1D LFBOT and weighted SFR and $M_*$ distributions\footnote{We choose KS over AD tests in this scenario as the speed of the KS test allows us to more quickly compare the LFBOT hosts to the large simulated galaxy population and the sensitivity of the AD test is not needed in this case.}. To encapsulate uncertainty on the LFBOT SFR and $M_*$ distributions in the KS tests, we randomly sample 10,000 realizations on their SFR and $M_*$ CDFs. We reject the null hypothesis in the same manner as the AD tests described previously. We find that we cannot reject the null hypothesis that LFBOTs trace the same 1D $M_*$ and SFR distributions as the weighted mock galaxy sample with $-15 \lesssim \log(A/B) \lesssim -10.6$, implying LFBOTs do indeed trace more SFR-weighted distributions in both the 1D and 2D tests. 

The LFBOT host preference for more SFR-weighted SFR-$M_*$ distributions is distinct from the other transient host populations. In Figure~\ref{fig:sfms}, it is apparent that SNe II and Ibc hosts prefer more $M_*$-weighted solutions than LFBOTs (however, these solutions are still relatively SFR-weighted), likely owing to their larger populations of more massive hosts with lower SFRs. Interestingly, we cannot find any SFR or $M_*$-weighted distribution that LGRBs or SLSNe-I prefer. This is almost certainly due their hosts lying well above the SFMS and being highly star-forming galaxies. Moreover, SLSN-I and LGRB progenitors are likely more dependent on metallicity than either SFR or $M_*$ (see Section \ref{sec:metallicity}). Importantly, this highlights that LFBOTs trace a separate low-mass, star-forming galaxy population than either of these transients, implying that their progenitor is dependent on different environmental factors for formation. We further note that we cannot rule out any solutions for SNe Ibn, as their hosts span a range of $M_*$ and SFR, with no obvious high-density region in the SFR-$M_*$ plane. This mitigates any reasonable comparison we can achieve between LFBOT and SNe Ibn hosts with this 2D test. 

Overall, we conclude that LFBOTs do indeed prefer more star-forming hosts than most CCSN sub-types, but are less star-forming and not as low mass as SLSN-I hosts. Reiterating the conclusions from the previous subsection, we find this strong preference of LFBOT hosts on star-formation compelling evidence that their progenitor contains a young, massive stellar component.

Finally, we briefly comment on how our results on star formation and stellar mass may be influenced by the small sample size of LFBOTs. Currently the sample of LFBOTs is so small that the 1D distributions of sSFR and $M_*$, as well as their 2D SFR-$M_*$ distribution, will be heavily biased by any outliers in the sample. As a test of this, we remove the highest mass host in the sample, the host of AT2023hkw, which is $>1$ dex more massive than any other host in the sample and, thus, may be representative of an outlier. Without this host, we would be able to reject the null hypothesis that LFBOTs trace the same 1D $M_*$ distribution as SNe Ibc and Ibn, which we could not do with it added in. Thus, it is imperative to continue to build the population of LFBOTs and their hosts to determine with more certainty how their stellar mass and star formation distributions compare to those of other transients. Nevertheless, because our study uniformly models the host galaxies of all currently known, \textit{bona-fide} LFBOTs, it establishes a rigorous baseline of our current understanding of these environments.

\begin{figure}[t]
\centering
\includegraphics[width=0.47\textwidth]{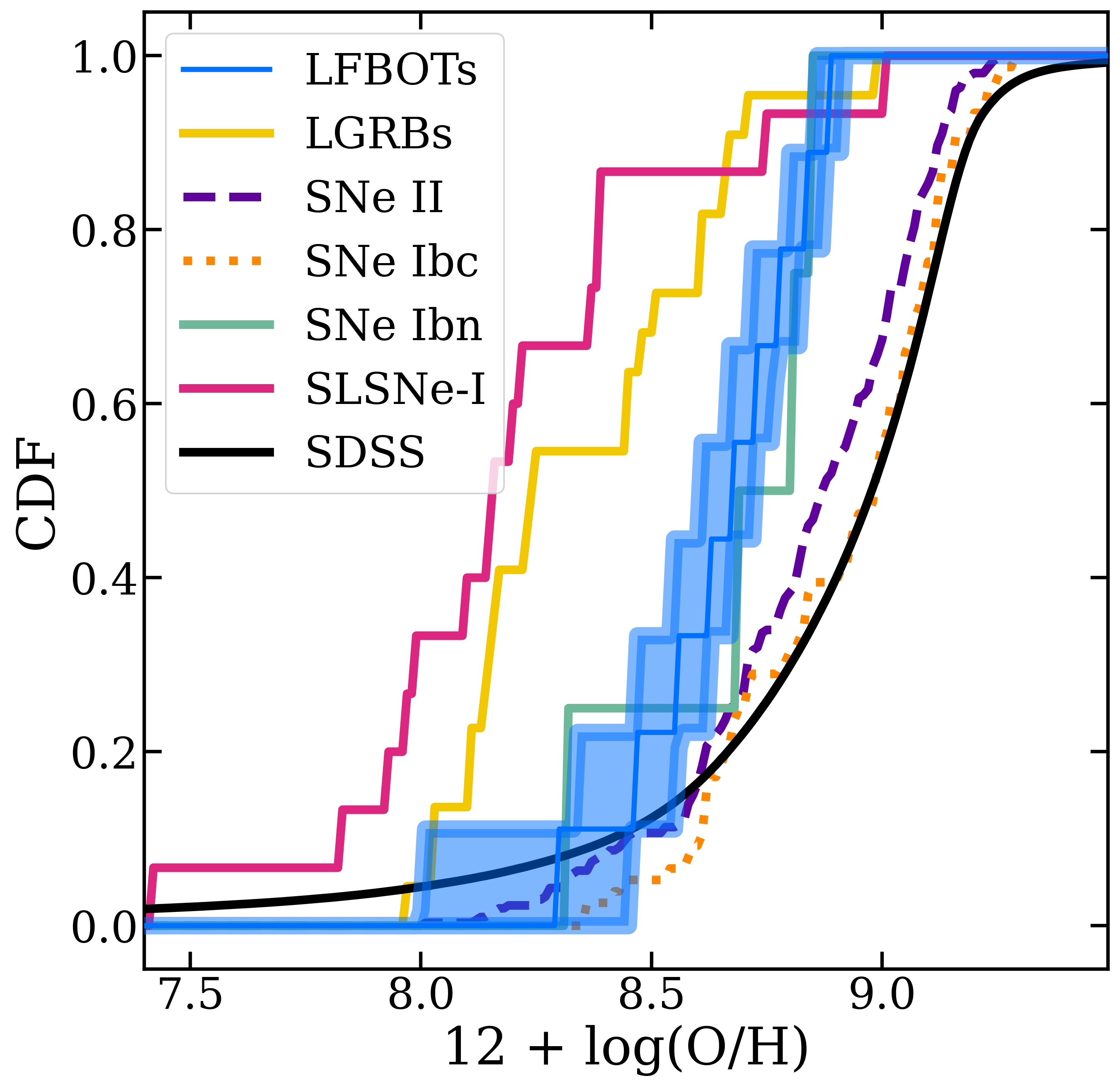}
\vspace{-0.1in}
\caption{CDFs of the gas-phase oxygen abundances (12+$\log(\textrm{O}/\textrm{H})$) for LFBOT hosts (blue), LGRBs (yellow), SNe II (purple dashed line), SNe Ibc (orange dotted line), SNe Ibn (green), SLSNe-I (pink), and a field galaxy population from SDSS (black). LFBOT hosts are clearly more metal-rich than SLSNe-I and LGRB hosts and more metal-poor than SNe Ibc and II hosts. In addition, LFBOTs trace lower metallicities than the field galaxy population.} 
\label{fig:OH}
\end{figure}

\subsection{Metallicity}
\label{sec:metallicity}
We next turn our attention to the metallicities of our LFBOT host population. In Figure~\ref{fig:OH}, we highlight the LFBOT $12+\log(\textrm{O}/\textrm{H})$ CDF (the median and 68\% interval are built from 1000 realizations on the CDF).  We additionally show CDFs of a general field galaxy population from the Sloan Digital Sky Survey (SDSS) Data Release 7 \citep{tremonti2004, sdssdr7}, a population of SN II, Ibc, and Ibn hosts from \citet{qin2024}, LGRB hosts from \citet{wang2014}, and SLSN-I hosts from \citet{perley2016}. All of these galaxy populations, except for the SLSN-I hosts, have $R_{23}$-based 12+log(O/H) metallicities, making the results comparable to our work. \citet{perley2016}, on the other hand, uses a variety of calibrations to determine their metallicities, with the majority determined through [NII]/H$\alpha$. As SLSN-I hosts are typically dwarf galaxies \citep{schulze2018, schulze2021}, which are assumed to have low metallicities (e.g., \citealt{kewley2008}), this metric is likely the best to use for this population. However, we note that there are subtle differences between the 12+log(O/H) metallicities determined through different calibrations (see \citealt{kk2004} and \citealt{kewley2019} for reviews), thus it is best practice to compare samples using the same metrics. Luckily, this has been achieved here for the most part. 

To determine if there are statistically significant differences in these distributions, we perform AD tests in the same way as described in in Section~\ref{sec:mass_sf}. We find that we can reject the null hypothesis that LFBOTs trace the same $12+\log(\textrm{O}/\textrm{H})$ distribution as the hosts of SLSNe-I (99\% of $P_{AD} < 0.05$), SNe Ibc (100\% of $P_{AD} < 0.05$), and SNe II (100\% of $P_{AD} < 0.05$). We find slight evidence that LFBOTs trace higher metallicities than LGRBs, as 62\% of $P_{AD} < 0.05$. We do not find any statistical distinction between the distributions of LFBOTs and SNe Ibn (0\% of $P_{AD} < 0.05$). We, do however, find that we cannot reject the null hypothesis that SNe Ibn trace the same metallicity distributions as the other CCSN hosts; thus, it is possible that small number statistics is playing a role in the outcome of our tests. We furthermore find statistical support that LFBOT hosts are more metal-poor than the general field galaxy population, as 100\% of $P_{AD} < 0.05$. Metallicity can largely influence the IMF in galaxies, where at low metallicities, galaxies are observed to have more top-heavy IMFs \citep{marks2012, li2023_imf}. Thus, this discrepancy between the field galaxies and LFBOTs may imply that LFBOT hosts contain a higher fraction of massive stars than typical in the field, which may indeed be a key component of the LFBOT progenitor.

\begin{figure*}[t]
\centering
\includegraphics[width=\textwidth]{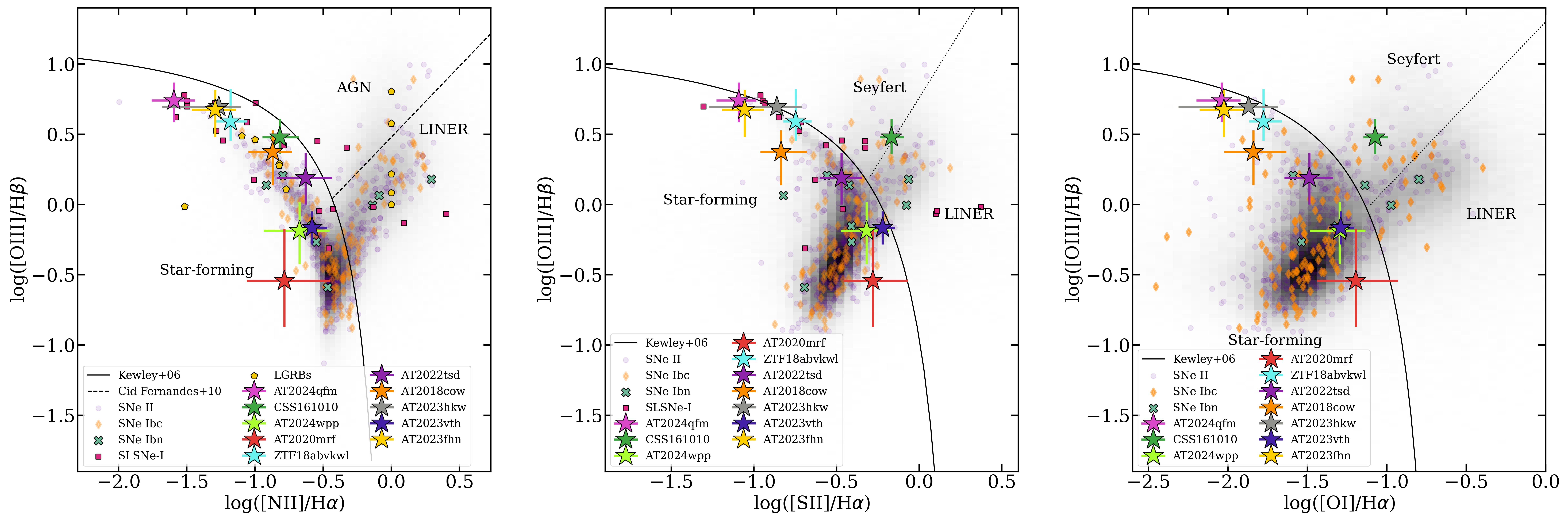}
\caption{Three variations of the BPT diagram: [OIII]/H$\beta$ versus [NII]/H$\alpha$ (\textit{left}), [SII]/H$\alpha$ (\textit{center}), [OI]/H$\alpha$ (\textit{right}), with the \citet{kewley2006} and \citet{cf2010} star-forming galaxy-AGN demarcation line over-plotted. LFBOT hosts are indicated with stars, SLSNe-I are represented by pink squares, LGRB hosts by yellow pentagons, SNe II hosts with purple circles, SNe Ibc hosts with orange diamonds, and SNe Ibn hosts with green crosses. We also show a field galaxy population from SDSS (grey histogram). In all three BPT diagnostics, LFBOTs appear to lie at higher [OIII]/H$\beta$ ratios than the field galaxy population and CCSN hosts, and occur in a more similar phase-space as SLSNe-I and LGRB hosts. This is likely caused by increased ionization in the LFBOT, SLSNe-I, and LGRB hosts due to higher amounts of star-formation, lower metallicities, and/or a larger massive star population.}
\label{fig:BPT}
\end{figure*}

As a further test of the metallicity dependence of LFBOTs, we place their hosts on the BPT diagram, typically used to separate star-forming galaxies and AGN \citep{Baldwin1981, Veilleux1987, Kewley2001, kewley2006, kauffmann2003}, in Figure~\ref{fig:BPT}. For comparison, we also show the SDSS field galaxy sample (the largest population of field galaxies that have been analyzed along the BPT diagram and the one used to determine the AGN and star-forming galaxy separation; \citealt{sdssdr7}), SLSN-I host sample from \citet{perley2016}, LGRB host sample from \citet{niino2017}, and CCSN (II, Ibc, and Ibn) host samples from \citet{qin2024}. In Figure~\ref{fig:BPT}, we highlight three variations of the BPT diagram: $\log(\textrm{OIII}/\textrm{H}\beta)$ versus $\log(\textrm{NII}/\textrm{H}\alpha)$, $\log(\textrm{SII}/\textrm{H}\alpha)$, and $\log(\textrm{OI}/\textrm{H}\alpha)$, along with the \citet{kewley2006} and \citet{cf2010} star-forming galaxy-AGN demarcation lines. For all LFBOT hosts with spectra, we present their emission line ratios in Table~\ref{tab:met}\footnote{\citet{ho2020_koala} and \citet{yao2022} also determined host galaxy emission line fluxes and/or ratios for the hosts of ZTF18abvkwl and AT2020mrf, respectively, using various line fitting techniques. Our \texttt{Prospector}-determined emission line ratios are consistent within error to their findings.}. For the host of CSS161010, we do not use the \texttt{Prospector}-determined emission line fluxes for [OIII$\lambda5007$] and H$\beta$, as the model fit determined a higher H$\beta$ flux density than [OIII$\lambda5007$], when this is clearly not the case from examining the host spectrum (see Figure~\ref{fig:appsed}). Instead, we subtract its observed spectrum from the spectral continuum model output by \texttt{Prospector} and estimate emission line flux densities by fitting the lines with a Gaussian function and integrating under the curves. We use the spectral noise to determine error. We do not find similar issues when examining other lines in its spectrum or the \texttt{Prospector} model fits to other hosts; thus, this problem is unique to the CSS161010 host fit and likely caused by the fact that these lines are so weakly detected in the host spectrum. 

In all three BPT diagrams, LFBOT hosts appear to lie slightly above the bulk of the SDSS field galaxies, but still clearly within in the star-forming phase space. We note that the host of CSS161010, which does lie in AGN region of the [SII] and [OI] BPT diagrams, is likely not being affected by an AGN, as \citet{coppejans2020} reported no AGN activity in the host. Instead, the deviation of LFBOT hosts from the SDSS field galaxy population and the placement of the host of CSS161010 probably hints that LFBOT hosts are more ionized (higher $U_\textrm{gas}$) than the normal field galaxy population, which can shift the star-forming sequence of galaxies towards higher [OIII]/H$\beta$ ratios \citep{brinchmann2008,kewley2013,kewley2019}. Indeed, we do find that the LFBOT host median gas ionization parameter $\log(U_\textrm{gas}) \approx -2.89$ leans towards the upper end of that of normal field galaxy populations, where $-3.2 \lesssim \log(U_\textrm{gas}) \lesssim -2.9$ \citep{kd2002, nakajima2014, kaasinen2018}. Higher ionization levels in galaxies are likely due to more compact, extreme star formation, a more top-heavy IMF, a larger WR star population, and/or lower metallicity \citep{Kewley2001, brinchmann2008}. As we have shown, LFBOT hosts are more metal-poor and more SFR-weighted than normal field galaxy populations, suggesting this is the most likely cause of their divergence from the field galaxy population on the BPT diagrams. Interestingly, we also find that LFBOT hosts appear distinct from CCSN hosts, which, in general, appear to trace the same phase-space as the field galaxy population. On the other hand, LFBOT hosts trace similar regions in the BPT diagrams as SLSN-I hosts and LGRBs: transients that have a strong-metallicity dependence and occur in galaxies with more extreme sSFRs than the CCSN populations.

Overall, we infer that LFBOTs do appear to have a metallicity dependence, preferring environments that are on-average more metal-poor than those of typical CCSN hosts and the general field galaxy population. However, their hosts do not appear to be as metal-poor as those of SLSNe-I and LGRBs. Once again, we do find this metallicity dependence highly indicative that LFBOTs have a stellar progenitor. Moreover, their global host properties in total (actively star-forming galaxies with recent bursts of star formation and low metallicity) are favorable conditions for massive star formation \citep{marks2012}. This may suggest that, similar to the other transients explored in this section, LFBOTs too arise from a massive core-collapse events, or, at the very least contain a stellar component in their progenitor system. We discuss further implications on LFBOT origins in Section~\ref{sec:disc}.

\section{LFBOT Locations}
\label{sec:offsets_res}
In the following section, we determine basic properties of the LFBOT local environments, including their galactocentric offsets (physical and host-normalized) and how bright the host galaxy is at the transient’s location relative to the rest of the galaxy (fractional flux measurement). To contextualize these results, we furthermore compare their offset and fractional flux measurements to the transient populations discussed in Section~\ref{sec:compare}. In Section \ref{sec:offsets}, we discuss offsets and in Section \ref{sec:frac_flux}, we measure and analyze their fractional flux.

\subsection{Galactocentric Offsets}
\label{sec:offsets}
To determine the location of our LFBOT sample within their host galaxies, we begin by collecting the localizations of their Zwicky Transient Facility (ZTF; \citealt{ZTF}) detections from ALeRCE (Automatic Learning for the Rapid Classification of Events). At the time of writing, there was no entry for AT2024wpp on ALeRCE; thus, for this event, we instead collect localizations from the discovery bots listed on the Transient Name Server (TNS). Using the collected localizations, for each LFBOT we determine a median R.A. and Decl. and the $1\sigma$ uncertainty on the localization, For CSS161010 (discovered prior to the start of ZTF), we simply use the localization and uncertainty determined in \citet{coppejans2020} through radio observations of the transient. We list all localizations and positional uncertainties ($\sigma_\textrm{pos}$) in Table~\ref{tab:lfbot_prop}. 

We determine physical galactocentric offsets ($R_\textrm{phys}$) by measuring the distance of the transient location to the center of their respective host (determined through our photometric techniques or listed in other works; Table~\ref{tab:phot}), propagating the LFBOT localization uncertainty into the offset uncertainty. We furthermore determine a host-normalized offset: $R_\textrm{hnorm} = r / r_e$, where $r$ is the angular galactocentric offset of the LFBOT and $r_e$ is the half-light radius of the host galaxy. We collect $r_e$ measurements (and uncertainties) for all LFBOT hosts from DECaLS, except for the hosts of AT2020mrf, AT2022tsd, and AT2023vth where one was not measured. Instead, we employ \texttt{SEP.flux\_radius} routine over the $g$-band  Hyper Suprime-Cam Subaru Strategic Program (HSC-SSP; \citealt{aihara2018, aihara2019}) image for AT2020mrf, Keck/LRIS $G$-band image for AT2022tsd, and PanSTARRS $g$-band image for AT2023vth to determine the radius where half the host galaxy flux is constrained. Unlike DECaLS, \texttt{SEP} does not provide an uncertainty for the $r_e$; thus we adopt an uncertainty of 1\% the measured $r_e$, which is roughly consistent with the errors obtained for the other hosts. We use perform basic error propagation, using the median and $1\sigma$ uncertainties on $r$ and $r_e$, to determine the $1\sigma$ uncertainty $R_\textrm{hnorm}$. We list the median and 1$\sigma$ uncertainty of $R_\textrm{phys}$ and $R_\textrm{hnorm}$ in Table~\ref{tab:lfbot_prop} for our LFBOT sample. 

\startlongtable
\begin{deluxetable*}{lccccccc}
\tabletypesize{\footnotesize}
\tablecolumns{7}
\tablewidth{0pc}
\tablecaption{LFBOT Localizations}
\label{tab:lfbot_prop}
\tablehead{
\colhead{LFBOT} &
\colhead{R.A.} &
\colhead{Decl.} &
\colhead{$\sigma_\textrm{pos}$ (arcsec)} &
\colhead{$z$} &
\colhead{$R_\textrm{phys}$ (kpc)} &
\colhead{$R_\textrm{hnorm}$ ($r/r_e$)} &
\colhead{Fractional Flux}
}
\startdata
AT2018cow & \ra{16}{16}{00.22} & \dec{+22}{16}{04.89} & 0.328 & 0.0141 & 1.758$\pm$0.095 & $0.86 \pm 0.05$ & $0.67^{+0.04}_{-0.05}$ \\
CSS161010 & \ra{04}{58}{34.40} & \dec{-08}{18}{03.95} & 0.03 & 0.033 & 0.258$\pm$0.02 & $0.49 \pm 0.04$ & $0.52^{+0.0}_{-0.0}$ \\
ZTF18abvkwl & \ra{02}{00}{15.20} & \dec{+16}{47}{57.32} & 0.147 & 0.272 & 2.401$\pm$0.619 & $1.11 \pm 0.29$ & $0.78^{+0.02}_{-0.3}$ \\
AT2020mrf & \ra{15}{47}{54.17} & \dec{+44}{39}{07.41} & 0.12 & 0.1353 & 1.242$\pm$0.291 & $0.9 \pm 0.21$ & $0.75^{+0.09}_{-0.2}$ \\
AT2020xnd & \ra{22}{20}{02.02} & \dec{-2}{50}{25.38} & 0.235 & 0.243 & 1.071$\pm0.911$ & $0.46 \pm 0.43$ & 0\\
AT2022tsd & \ra{03}{20}{10.86} & \dec{+08}{44}{55.96} & 0.095 & 0.256 & 6.106$\pm$0.382 & $1.67 \pm 0.1$ & $0.24^{+0.12}_{-0.23}$ \\
AT2023fhn & \ra{10}{08}{03.82} & \dec{+21}{04}{26.89} & 0.132 &  0.238 & 17.400$\pm0.503$ & $2.42 \pm 0.07$ & 0  \\
AT2023hkw & \ra{10}{42}{17.73} & \dec{+52}{29}{18.99} & 0.394 &  0.335 & 13.58$\pm$1.92 & $1.91 \pm 0.27$ & $0.0^{+0.07}_{-0.0}$  \\
AT2023vth & \ra{17}{56}{34.40} & \dec{+08}{02}{37.37} & 0.048 &  0.0747 & 2.86$\pm$0.07 & $1.57 \pm 0.04$ & $0.51^{+0.0}_{-0.0}$ \\
AT2024qfm & \ra{23}{21}{23.46} & \dec{+11}{56}{32.03} & 0.272 & 0.227 & 4.078$\pm$1.001 & $1.6 \pm 0.4$ & $0.0^{+0.23}_{-0.0}$ \\
AT2024wpp & \ra{02}{42}{05.483} & \dec{-16}{57}{22.98} & 0.206 & 0.0866 & 5.987$\pm$0.338 & $0.97 \pm 0.06$ & $0.34^{+0.09}_{-0.13}$ 
\enddata
\tablecomments{The localizations, localization uncertainties ($\sigma_\textrm{pos}$), redshifts, physical galactocentric offsets $R_\textrm{phys}$, host-normalized offsets ($R_\textrm{hnorm}$), and fractional flux for our LFBOT sample. The localization for CSS161010 was determined in \citet{coppejans2020} and the localization for AT2024wpp was determined with data available on TNS. All other localizations were determined from the listed ZTF detections of each LFBOT on ALeRCE.}
\end{deluxetable*}

\begin{figure*}[t]
\centering
\includegraphics[width=0.9\textwidth]{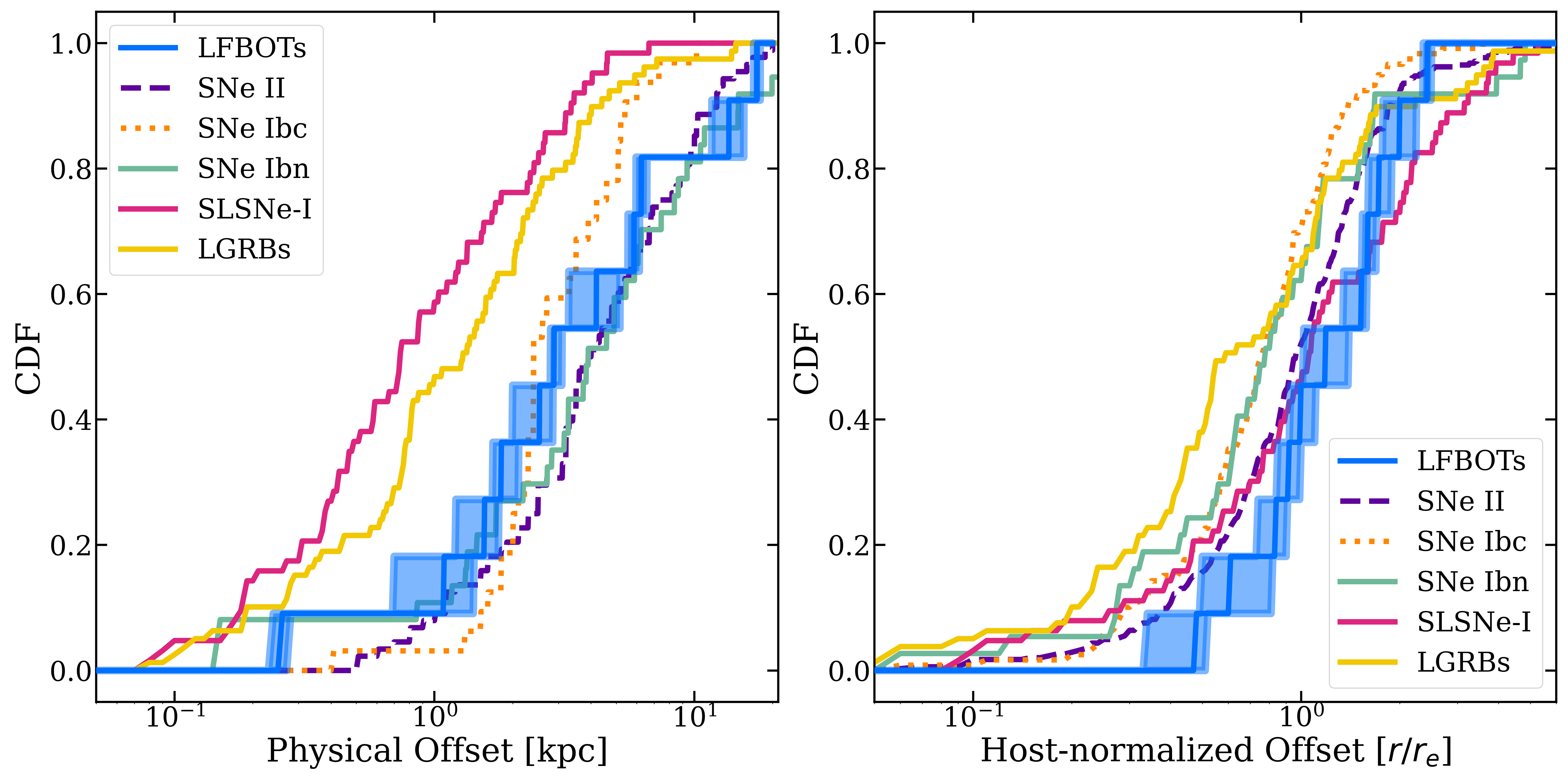}
\vspace{-0.1in}
\caption{The physical galactocentric offset (\textit{left}) and host-normalized offset (\textit{right}) distributions of LFBOTs (blue; median and 68\% interval) compared to those of SNe II (purple dashed line), Ibc (orange dotted line), Ibn (green), SLSNe-I (pink), and LGRBs (yellow). LFBOTs have similar physical offset distributions to SNe II, Ibc, and Ibn and are more offset than SLSNe-I and LGRBs. Additionally, LFBOTs have larger host-normalized offsets than LGRBs, SNe Ibc, and SN Ibn and more similar distributions to SNe II and SLSNe-I.}
\label{fig:offsets}
\end{figure*}

In Figure \ref{fig:offsets}, we present the LFBOT $R_\textrm{phys}$ and $R_\textrm{hnorm}$ CDFs and $68\%$ interval on each CDF. To properly account for uncertainty on both offset CDFs, we randomly sample 10,000 offsets for each LFBOT from a Rice Distribution using their median offset and $1\sigma$ uncertainty, following the methods outlined in \citet{blanchard2016}. Sampling from a Rice distribution is required here, as we must impose that all offsets are non-negative values. Were we to sample from a Gaussian distribution, we may obtain nonphysical negative values from events with lower offsets and/or higher uncertainties, which, in turn, would bias the resulting CDF. The Rice distribution, by contrast, guarantees that we solely sample positive, physically realistic values. We note that that when $R/\sigma R >5$, the Rice distribution reduces to a Gaussian distribution.

We find that LFBOTs have $R_\textrm{phys} = 2.89^{+9.36}_{-1.71}$~kpc (population median and 68\% interval) and $R_\textrm{hnorm} = 1.19^{+0.72}_{-0.49}$~$r/r_e$. To place these results in the context of other transient populations, we compare them to the $R_\textrm{phys}$ and $R_\textrm{norm}$ distributions of SNe II, SNe Ibc, SNe Ibn, SLSNe-I, and LGRBs. We collect offset distributions for SNe II and Ibc in \citet{prieto2008}, SNe Ibn in \citet{dong2025}, SLSNe-I in \citet{hsu2024}, and LGRBs in \citet{blanchard2016} and showcase CDFs of these distributions also in Figure~\ref{fig:offsets}. We compare the distributions with AD tests, in the same manner as described in Section~\ref{sec:mass_sf}  We are not able to reject the null hypothesis when comparing the LFBOT $R_\textrm{phys}$ distribution to SNe II, Ibc, and Ibn (0\%, 0.2\%, and 0\% of $p_{AD}<0.05$, respectively). We find that LFBOTs trace higher $R_\textrm{phys}$ than SLSNe-I and LGRBs, given that 100\% of $p_{AD}<0.05$ when comparing LFBOTs to each transient sample. We note that these results do differ from those determined in \citet{chrimes2024_offset}, which included an analysis of the offsets of 5 LFBOTs. There, they showed that LFBOTs had an $R_\textrm{phys}$ distribution more similar to those of SLSNe-I and LGRBs. It is interesting to note that with $\sim$double the population, the distribution now veers much further from the $R_\textrm{phys}$ of those transients.

Unlike the other transient populations, there appears to be a lack of LFBOTs that occur at low $R_\textrm{hnorm}$. Indeed, visually, it seems that the LFBOT $R_\textrm{hnorm}$ distribution skews towards higher offsets than SLSNe-I, LGRBs, SNe Ibc, and Ibn. When comparing these distributions with AD tests, we do find that LFBOTs trace statistically different $R_\textrm{hnorm}$ distributions than LGRBs (75\% of $p_{AD}<0.05$), SNe Ibc (93\% of $p_{AD}<0.05$), and SNe Ibn (93\% of $p_{AD}<0.05$). This strongly suggests LFBOTs occur at higher $R_\textrm{hnorm}$ than these transient populations. We do not, however, find that LFBOT $R_\textrm{hnorm}$ distributions are statistically distinct from those of SNe II and SLSNe-I, as 0\% of $p_{AD}<0.05$ in both cases.

Overall, while we do find that we can rule out similarities between offset distributions of LFBOTs and transients with likely massive star progenitors (LGRBs, SLSNe-I, and SNe II, Ibc, Ibn), LFBOTs do not appear to be extremely unique in comparison to these populations. Moreover, their $R_\textrm{phys}$ and $R_\textrm{hnorm}$ distributions lean toward significantly lower offsets than observed for short GRBs \citep{fong2022} and SNe Ia \citep{prieto2008, wang2013}: transients that derive from progenitors with long formation timescales (neutron star mergers and white dwarfs, respectively; \citealt{maoz2014, aaloc+17, gvb+17}) and large systemic velocities that allow them to migrate far from their hosts' centers \citep{zevin2020, mandhai2022, gaspari2024, pan2026}. In addition, the non-nuclearity of LFBOTs hints that they do not represent interactions with central BHs in their hosts, and thus are intrinsically different from SMBH TDEs \citep{vanvelzen2011, hung2018, hammerstein2023}. Thus, from offset distribution alone, LFBOTs are much more aligned with transients from young stellar origins, in agreement with our findings on their global host galaxy properties. We discuss implications for this result in Section~\ref{sec:disc}. 

\subsection{Fractional Flux}
\label{sec:frac_flux}

Another useful statistic for characterizing the LFBOT local environment is determining its fractional flux. As originally described in \citet{fruchter2006}, fractional flux is defined as the sum of the flux of all the pixels fainter than those at the transient location divided by the total host galaxy flux. The fractional flux essentially measures how bright the galaxy is at the location of the transient, where a fractional flux of 0 is the faintest pixel within the host (or the transient lies outside the host galaxy's light) and 1 is the brightest pixel. In star-forming galaxies, we may expect that young, massive stellar populations will trace the brightest regions of their host.

We follow the methodology in \citet{hsu2024} to determine the fractional flux and quantify its uncertainty for each LFBOT in our sample. For each LFBOT, we use the highest quality, bluest optical filter to perform the fractional flux calculation. High quality images are necessary to ensure we extract the most pixels of the host and obtain the best estimates of the fractional flux. Bluer filters, but optical filters in general, will specifically help us probe recent star formation in the hosts. We use the $g$-band image listed in Table~\ref{tab:phot} for AT2020mrf, AT2022tsd, AT2023vth, and AT2024wpp. For AT2020xnd and AT2023fhn, we use the $r$ and $V$-band images as these are the highest quality, bluest images for both of these transients. Finally, we download DECaLS $g$-band images via \texttt{FrankenBlast} for the hosts of AT2018cow, CSS161010, ZTF18abvkwl, AT2023hkw, and AT2024qfm. We determine an uncalibrated total host flux in a similar manner as described in Section~\ref{sec:phot} with \texttt{SEP}. Here, however, we measure a Kron radius using a threshold of 1$\sigma$ above sky background to ensure that we extract all potential faint pixels on the outskirts of the galaxy.

\begin{figure}[t]
\centering
\includegraphics[width=0.47\textwidth]{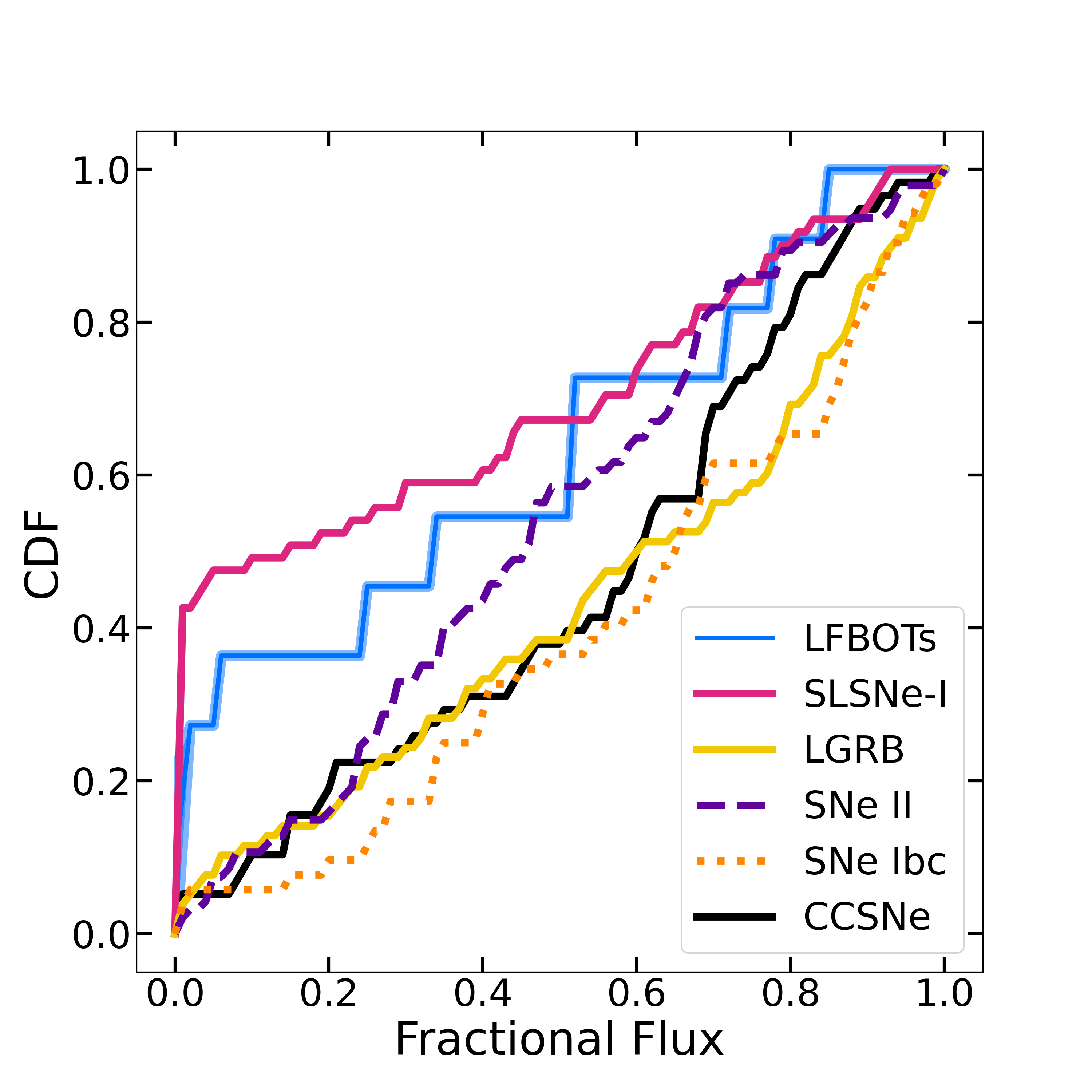}
\vspace{-0.3in}
\caption{The fractional flux distributions of LFBOTs (blue; median and 68\% interval) compared to those of SNe II (purple dashed line), Ibc (orange dotted line), SLSNe-I (pink), and LGRBs (yellow). Similar to SLSNe, a large fraction of LFBOTs occur in the faintest region of their host galaxy or outside their host galaxy's light and they clearly trend towards lower fractional fluxes than LGRBs and the CCSNe populations.}
\label{fig:ff}
\end{figure}

As done in \citet{hsu2024}, we use a Monte Carlo routine to determine the pixels that overlap with each LFBOT location, which we propagate into the fractional flux uncertainty. We sample 1000 ($x$,$y$) pixel coordinates from a 2D Gaussian distribution, assuming the pixel at the median transient localization (Table~\ref{tab:lfbot_prop}; Section~\ref{sec:offsets}) is the centroid of the Gaussian and $\sigma_\textrm{pos}$ is the standard deviation. Our images have pixel scales ranging from 0.135-0.262 arcsec/pixel and given that the hosts' light radius only encompass a few hundred to a few thousand of pixels, this number of samples should probe each LFBOT localization without oversampling. We calculate a fractional flux for each pixel and report the median and 68\% interval for each LFBOT in Table~\ref{tab:lfbot_prop}. We note that two LFBOTs (AT2020xnd and AT2023fhn) lie outside their host galaxy's Kron radius and thus they have a fractional flux of 0.

We use the 1000 fractional flux samples for each LFBOT to build 1000 realizations on the CDF, and highlight the median and 68\% interval on the CDF in Figure~\ref{fig:ff} (we note that uncertainties on the fractional flux are small, hence why the width of the CDF is also small). We contextualize these results by comparing them to fractional flux distributions of SLSNe-I (from \citealt{hsu2024}), LGRBs (from \citealt{blanchard2016}), and SNe II and Ibc (from \citealt{kelly2008}). We also include the general CCSNe fractional flux distribution from \citet{svensson2010}. These studies determined fractional fluxes at similar UV-optical rest-frame wavelengths, thus are probing similar host light distributions as our LFBOT sample. When comparing distributions with AD tests, we find that we are not able to reject the null hypothesis that LFBOTs and SLSNe-I trace the same underlying fractional flux distribution, as 0\% of $p_{AD}<0.05$. Interestingly, both SLSNe-I and LFBOTs have a high fraction of events with 0 fractional flux, implying both populations generally occur in fainter pixels in their hosts, or outside of their hosts' light. Meanwhile, we find that we can reject the null hypothesis for the LFBOT fractional flux distribution compared to LGRBs, SNe II, and SNe Ibc as 96\%, 54\%, and 98\% of $p_{AD}<0.05$, respectively. We furthermore determine that LFBOTs trace statistically different fractional flux distributions than the general CCSNe population (70\% of $p_{AD}<0.05$). If the LFBOT is of stellar origin, which is hinted at by their global host properties, these findings imply that their progenitor must receive a ``kick" to push it away from its birthsite. If LFBOTs are not of stellar origin, their fractional flux distributions suggest that their progenitors are formed at off-nuclear locations. We discuss these results in the context of possible progenitor systems in Section~\ref{sec:disc}.

\section{Possible Progenitor Systems}
\label{sec:disc}
In this section, we analyze LFBOT host stellar populations, galactocentric offset and fractional flux distributions to test the viability of various progenitor scenarios. Our goal is to determine whether proposed models such as TDEs, failed SNe, PPISNe, magnetar-powered SN, or compact object-WR mergers are physically consistent with our findings. Although LFBOTs may arise from multiple formation pathways, here we evaluate whether their environmental properties can be explained by a single progenitor channel. Finally, we emphasize that the population of LFBOTs and their hosts is currently quite small, thus it is possible that any inferences we make may change with a larger population. Nonetheless, we do find consistencies within the current LFBOT sample from which we can begin to constrain their progenitors. We organize this section by discussing each progenitor model in the context of our findings. In Table \ref{tab:prog}, we show how well each progenitor model matches with various environmental properties and the LFBOT transient emission.

\subsection{IMBH TDE?}
A TDE origin for LFBOTs was originally  proposed, motivated by the prototypical event AT2018cow, which showed a steep lightcurve decline, blue and featureless spectrum, and late-time He emission reminiscent of the tidal disruption of stars by SMBHs \citep{kwo+2019, perley2019}. In addition, the late-time UV plateau of AT2018cow is similar the late-time UV plateau observed in TDEs, and is possible further evidence of a shared progenitor connection \citep{sun2022, sun2023, inkenhaag2023, inkenhaag2025}. Unlike SMBH TDEs, LFBOT lightcurves initially rise and decay on $\approx$days timescales rather than weeks to months, which would require a less massive black hole, similar to that of IMBHs ($10^3-10^6$~$M_\odot$), to power the transient \citep{perley2019, wintergranic2025}. Early LFBOT host galaxy analyses also supported an IMBH TDE origin, as the first discovered LFBOTs were all associated with dwarf galaxies (AT2018cow, CSS161010, ZTF18abvkwl, AT2020mrf, and AT2020xnd; \citealt{perley2019, perley2021, ho2020_koala, yao2022}), which are expected to host IMBHs. 

Crucially, as we show in Section~\ref{sec:offsets}, LFBOTs are not centrally located within their hosts, 
and therefore cannot represent interactions with or accretion onto central BHs. However, their locations alone are not sufficient to completely disregard the TDE model entirely, as LFBOTs could instead be powered by off-nuclear IMBH TDEs. Recently, observational evidence has emerged for TDEs created from BHs ``wandering" from their hosts, based on the discovery of four off-nuclear TDEs with likely BH masses $>10^4$~$M_\odot$ \citep{guolo2026, guolo2026_dem, stein2026}. Although it may seem more promising that LFBOTs arise from this IMBH TDE channel, we still disfavor this scenario. Wandering IMBHs likely form through the capture of the central BHs of interacting dwarf galaxy satellites by the larger primary galaxy \citep{belllovary2010,greene2021, ricarte2021}. As galaxies grow hierarchically through mergers, larger galaxies have generally experienced more mergers and thereby should host more wandering IMBHs. Indeed, \citet{guolo2026_dem} predict that the number of wandering IMBHs, and thus the probability of off-nuclear TDEs, increases with host stellar mass. This is further supported by the properties of the current population off-nuclear TDE hosts, as all are elliptical galaxies with large stellar masses, ranging $10.7 \lesssim \log(M_*/M_\odot) \lesssim 11.2$ \citep{guolo2026_dem}. Given that LFBOTs exclusively occur in star-forming galaxies with a much lower stellar masses, we consider it improbable that they originate through a wandering IMBH channel. Thus, even with a small population of LFBOTs, we find significant evidence to suggest they do not derive from IMBH TDEs.

The IMBH TDE model is also in tension with other observed LFBOT properties. Specifically, LFBOTs are surrounded by a dense, extended CSM, inferred from their radio emission, which is difficult to justify in an IMBH TDE scenario \citep{ho2019, ho2020_koala, ho2022_xnd, margutti2019, coppejans2020, bright2022, yao2022, chrimes2024, chrimes2024_offset, nayana2025, sevilla2026}. Given the lack of support of the IMBH TDE model from both the transient and host properties, we strongly disfavor this progenitor theory for LFBOTs.

\startlongtable
\begin{deluxetable*}{lcccccc}
\tabletypesize{\footnotesize}
\tablecolumns{6}
\tablewidth{0pc}
\tablecaption{Progenitor Models of LFBOTs}
\label{tab:prog}
\tablehead{
\colhead{Progenitor Model} &
\colhead{Host $M_*$} &
\colhead{Host Star Formation} &
\colhead{Host Metallicity} &
\colhead{Local Environment} &
\colhead{Transient Emission} &
\colhead{Extended CSM} 
}
\startdata
IMBH TDE & $\times$ & $\times$ & $\times$ & $\checkmark$ & ? & $\times$ \\
Stellar Mass BH/NS TDE & $\checkmark$ & $\checkmark$ & ? & $\times$ &  $\checkmark$ & $\times$ \\
PPISN & $\checkmark$ & $\checkmark$ & $\times$ & $\times$ & ? & ?\\
Failed SN & $\checkmark$ & $\checkmark$ & $\checkmark$ & $\times$ & $\checkmark$ & ?\\
Magnetar-Powered SN & $\checkmark$ & $\checkmark$ & $\checkmark$ & $\checkmark$ & ? & $\times$ \\
BH/NS-WR Merger & $\checkmark$ & $\checkmark$ &  $\checkmark$ & $\checkmark$ & $\checkmark$ & $\checkmark$ \\
\enddata
\tablecomments{Different LFBOT progenitor theories, and how well these models align with the host global properties studied in this work ($M_*$, star formation, and metallicity), their local environments (offsets and fractional fluxes), and transient emission. We find the most support for the \citet{metzger2022} BH/NS-WR merger. }
\end{deluxetable*}

\subsection{Stellar Mass Compact Object TDE?}
We next turn our attention to the stellar mass compact object TDE progenitor channel discussed in \citet{tsuna2025}. In this scenario, a massive star in a tight field binary undergoes a CCSN, forming a remnant NS or BH. The natal kick imparted on the compact object forces it to collide with its (typically, main-sequence) stellar companion. The tidal forces on the newly-formed compact object then disrupt the stellar companion, which leads to the formation of an accretion disk around the compact object. \citet{tsuna2025} predict that the transient emission from this collision and accretion will reach the peak luminosities observed for LFBOTs, with similarly rapid lightcurve evolution when modeling scenarios with low ejecta masses. In addition, their model predicts late-time H emission ($\approx$20~days post-peak, similar to AT2018cow; \citealt{prentice2018, perley2019}) observed in some LFBOTs, which is difficult to reproduce in other progenitor scenarios. This late-time emission arises from CCSN ejecta becoming optically thin, allowing detection of H lines from the disrupted companion star.

Our findings on LFBOT environments are also in tension with this progenitor model. First, from our analysis on LFBOT fractional fluxes, we find clear evidence that the LFBOT progenitor has been forced out of the star-forming region within its host and birthsite. This is likely caused by either massive star that has received a significant kick or a progenitor with a longer formation timescale than a core-collapse event. This BH/NS TDE model satisfies neither of these scenarios, given that the TDE occurs almost immediately ($\approx$~weeks) after the first star undergoes core-collapse: an insignificant amount of time for the binary to migrate away from its these regions. Moreover, to the extent that most massive stars are in binaries at birth, there is nothing particularly different about this progenitor set-up than normal CCSN models, except that the orientation of the SN natal kick forced it to collide with, and subsequently tidally disrupt, its companion star. Therefore, if LFBOTs were to arise from this channel, their environments should largely be similar those of the normal CCSN population. In Section~\ref{sec:compare}, we clearly show that LFBOT hosts are quite distinct from CCSN hosts in both metallicity and star formation, which disfavors most LFBOTs being connected to this progenitor channel. Thus, it is highly unlikely that these stellar mass compact object TDEs are the progenitors of LFBOTs, from environmental properties alone.

\subsection{Stellar Death or Supernovae?}
An alternative progenitor theory postulates that LFBOTs arise from stellar deaths, with possible origins including  PPISNe \citep{leung2020}, failed SNe \citep{margutti2019, perley2019, qlc2019, antoni2022}, and magnetar-powered SNe \citep{prentice2018, vurm2021}. Central to each of these scenarios is the formation of a compact object that is either rapidly-rotating or accreting at super-Eddington rates to heat the supernova ejecta and power the LFBOT lightcurve. The classification of many lower-luminosity FBOTs \citep{drout2014} as CCSNe \citep{ho2023_fbot} highlights a possible connection between LFBOTs and the deaths of massive stars. As shown in Sections \ref{sec:compare} and \ref{sec:offsets_res}, we find substantial evidence from LFBOT global host properties (star formation, metallicities) and their localizations that their progenitor comprises a young, massive stellar component and, thus, may originate from the final stages of massive star evolution, which we explore below.

\subsubsection{PPISNe?}
Under the assumption that LFBOTs originate from stellar deaths of massive stars, their comparisons to the environments of SLSNe-I, LGRBs, and CCSNe does impose some constraints on the nature of their progenitor. For one, we show in Section~\ref{sec:metallicity} that LFBOTs occur in more metal-poor environments than SNe Ibc and SNe II, whose progenitors are thought to be stripped-envelope WR stars and red supergiants (RSGs), respectively \citep{smartt2009, smith2011, smith2014, smartt2015}. At face value, this may suggest that the LFBOT progenitors are more massive stars, which more easily form at low metallicities. However, we also find that LFBOT hosts are neither as metal-poor nor as intensely star-forming as those of SLSNe-I and LGRBs. Both of these transients are thought to derive from the core-collapse of very massive, highly metallicity-dependent stars (e.g., \citealt{heger2003, woosley2006, leloudas2015, nicholl2017, blanchard2020}). Thus, we infer that while LFBOT progenitors may come from more massive stars than SNe Ibc and II, they are likely not as massive or metallicity-dependent as SLSNe-I and LGRB progenitors. We, therefore, strongly disfavor PPISNe as an LFBOT progenitor, as it is assumed that they originate from more massive and metal-poor stars than SLSNe-I and LGRBs \citep{heger2003, woosley2017} and would reside in hosts with lower metallicities than either population. Furthermore, there is some question as to whether this model can produce the dense, extended CSMs observed for LFBOTs \citep{metzger2022}. Taken together, we reject this model.

\subsubsection{Failed SNe?}
On the other hand, the failed SN progenitor model could explain why LFBOTs have a stronger metallicity dependence than CCSNe, but weaker than SLSNe-I and LGRBs. In this scenario, a massive star (possibly a blue supergiant; \citealt{margutti2019}) collapses  directly into a BH after neutrino-induced mass loss, failing to produce a SN, and accretes the remaining stellar material. This accretion creates luminous emission \citep{lovegrove2017,qlc2019,antoni2022}, which may be similar to that of LFBOTs. \citet{chrimes2026} further showed that this model can reproduce the long-lived UV emission observed for AT2018cow. It is thought that high angular momentum of the progenitor star is crucial in forming the accretion disk, although this has been debated \citep{qlc2019}. Metallicity is thought to play a large role in the intrinsic angular momentum of stars, since at higher metallicity, stars will lose angular momentum due to mass loss from metal-driven winds \citep{heger2003}. There is indeed observational support for this, as metal-driven mass loss likely leads to the formation of more stripped-envelope SNe Ibc, which are found in more metal-rich hosts than SNe II \citep{prieto2008, anderson2010, modjaz2011, kelly2012, galbany2016, schulze2021, qin2024}. Thus, we would expect failed SN to preferentially occur in lower metallicity environments than at least SNe Ibc, which we do find for LFBOTs.

If LFBOTs are powered by failed SNe, however, it is not clear why their underlying host light distribution significantly deviates from that of CCSNe. The high fraction of LFBOTs at in the faintest pixel of their host or outside their host's light radius signals that their progenitor has been ejected from their birthsite and is traveling at high systemic velocities (higher than assumed for CCSN). Yet, it is unclear why certain core-collapse events would receive much stronger kicks than others. A possible way to reconcile these differences is if the LFBOT failed SN progenitor is in a binary system, since binary interactions can displace SNe in their hosts \citep{blaauw1961, ams+2017, renzo2019, wagg2025}. Binary interactions may also spin up stars, establishing the necessary conditions for a star to undergo a failed SN. However, considering that most massive star are in binary or multiple stellar systems \citep{chini2012,sana2012,dunstall2015,md2017}, we can presume that a large number of SNe Ibc and II are also influenced by binary physics. Consequently, to fully validate the failed SN scenario for LFBOTs, we must determine why this specific model would impart significantly larger natal kicks than those assumed for standard CCSNe. This may be done, for instance, via binary modeling (e.g., \citealt{wagg2025_cogsworth}). Thus, while we do not completely rule out the failed-SN model, we believe it is difficult to justify why this would be the LFBOT progenitor from host properties alone.

\subsubsection{Magnetar-powered SNe?}
It was also suggested that LFBOTs originate from magnetar-powered SNe, in which a successful CCSN produces a rapidly spinning, millisecond magnetar remnant (e.g., \citealt{prentice2018}). In this case, the LFBOT lightcurve is powered by the magnetar spin-down that heats the SN ejecta, keeping the transient emission blue for an extended period. Interestingly, this is also the most prevailing progenitor theory for SLSNe-I \citep{kasen2010, leloudas2015, metzger2015, nicholl2015, nicholl2017, blanchard2020}. Similar to LFBOTs, SLSNe-I have high peak luminosities and lightcurves that cannot be solely explained by $^{56}$Ni decay, suggesting they too have a central engine. However, SLSNe-I are generally much more slowly evolving transients than LFBOTs. The disparity can potentially be reconciled by progenitor mass; for instance, LFBOTs may involve less massive progenitors that produce lower ejecta masses  than those of SLSNe-I \citep{margutti2019}. Our environmental analysis clearly reveals that LFBOTs and SLSNe-I do not occur in similar host galaxies, as SLSN-I hosts are more metal-poor, more star-forming, and less massive than LFBOT hosts. The only apparent similarity in SLSNe-I and LFBOT environments are their fractional flux distributions, where a large fraction of both transient populations occur in fainter regions of their hosts. Thus, nominally, the differences between their environments do not indicate that SLSNe-I and LFBOTs share the same progenitor. Moreover, as the magnetar-powered SN model is a much more popular progenitor theory for SLSNe-I, this would suggest that LFBOTs likely do not also arise from those same progenitors.

A possible way to connect SLSNe-I and LFBOTs, assuming they both originate from magnetar-powered SNe, is by invoking binary versus solitary star progenitor channels. To form a millisecond magnetar, the progenitor star must have high angular momentum, which is either intrinsic to the star or achieved through a binary companion. As just discussed in the case of failed SNe, retaining high angular momentum throughout a stellar lifetime can be a natural consequence of low metallicity. On the other hand, low metallicity is not a strict requirement for a star that inherits angular momentum through its binary companion. With this in mind, we may infer that SLSNe-I originate from a single star channel, as they do occur in lower metallicity environments, whereas the LFBOT progenitor is in a binary system. Yet, this dichotomy breaks down when considering the local environments of these transients. The fractional flux distributions of both SLSNe-I and LFBOTs highlight that both of their progenitors have received a kick to migrate away from star-forming regions within their hosts, which is incommensurate with a solitary star channel, but a possible outcome in the binary scenario. Indeed, \citet{hsu2024} speculates that SLSNe-I arise from a disrupted binary system to achieve estimated systemic velocities on the order of 100~km~s$^{-1}$. Thus, while we don't completely rule out the possibility that LFBOTs represent a millisecond magnetar channel, we struggle to rationalize this progenitor scenario if SLSNe-I are indeed magnetar-powered SNe. 

Finally, there is also some question as to whether the millisecond magnetar channel can produce the dense, extended CSM and asymmetric ejecta observed in several LFBOTs \citep{margutti2019, vurm2021, metzger2022} or the large accretion disk inferred by modeling the late time UV and X-ray spectrum \citep{Chen+23,Migliori+24,Inkenhaag+25,wintergranic2025}. Moreover, \citet{sun2022} shows that magnetar emission alone cannot power the early and late-time emission of AT2018cow, thereby requiring an additional energy source to explain LFBOT emission. Therefore, we are very skeptical if LFBOTs originate from magnetar-powered SNe, given that their environments are so disparate from those of SLSNe-I and the transient emission does not seem fully described by this model.

\subsection{Compact Object and Wolf-Rayet Star Merger?}
We lastly consider the possibility that LFBOTs arise from the merger between a stellar mass BH or NS and the evolved He core of a binary companion (WR star), which has been proposed in  \citet{metzger2022} and  \citet{klencki2025}. In this model, the merger is predicted to occur after a binary system, initially comprised of a compact object and main-sequence star, undergoes a common-envelope or stable mass transfer phase. During this phase, the Hydrogen envelope of the star is accreted by the compact object, tightening the binary and leading to a BH/NS-WR star pair. The system eventually merges due to angular momentum loss, causing the compact object to tidally disrupt the WR star. Luminous transient emission is produced from accretion of the stellar material at super-Eddington rates and shocks with the CSM, created from the earlier WR star mass-loss episode. Since explosive nucleosynthesis is minimal, this merger model is also consistent with the low $^{56}$Ni yields inferred in LFBOTs \citep{metzger2022, neights2026}.

Our findings on LFBOT environments are compatible with this progenitor scenario, especially in comparison the aforementioned progenitor models. First, \citet{klencki2025} determines that these mergers will preferentially occur in environments with subsolar, but not extremely low, metallicities, given that (i) massive compact objects more easily form at low metallicities, and (ii) binary interactions with a WR star are more frequent at low metallicity \citep{klencki2020}.  Thus, this model naturally explains the slight low-metallicity bias we find in the LFBOT host population. It is furthermore expected that there will be a delay ($\approx$few Myr) between formation of the compact object (via CCSNe) and the merger. This allows enough time for the natal kick from the SN to propel the system to the outskirts of the host galaxy and provides clear rationale for the observed LFBOT fractional flux distribution. Lastly, the total time delay between the birth of the binary system until the merger is not so long ($\approx$~tens of Myr) that we would expect their hosts to evolve significantly off the SFMS or grow substantially in stellar mass, which takes on-order hundreds to thousands of Myr \citep{peng2010, torrey2018, iyer2020, sampaio2024}. In turn, we would expect these mergers to favor actively star-forming hosts, possibly with recent bursts of star formation in which the binary was formed, that are not exceptionally massive: exactly what we currently observe with the LFBOT host population. 

It is further interesting to note that \citet{metzger2022} predicts that some SNe Ibn/Icn may also come from this same progenitor channel. Specifically, they hypothesize that these SNe originate from systems with longer delays (by $\sim 10^{3}-10^{5}$ yr) between the first common-envelope phase and the merger than assumed for the LFBOT channel. Our results may indicate a common origin between LFBOTs and SNe Ibn, though, we note both of their host populations are small and thus any observed similarity will be influenced by small-number statistics. We show that SNe Ibn occur in hosts that are slightly less star-forming, but have similar metallicities to LFBOT hosts. From the similar metallicity dependence of both transient classes, we may infer that the LFBOT and SN Ibn arise from similar stellar progenitors. The slight difference in star-formation activity in their hosts may solely be an indication of the longer delay times of the SNe Ibn progenitor channel, where we would expect the hosts to be more evolved, as the SN Ibn progenitor formation timescale \citet{metzger2022} predicts is more aligned with galactic evolutionary timescales \citep{peng2010, torrey2018, iyer2020, sampaio2024}. While no fractional flux distribution currently exists for SNe Ibn, several events occur well outside host galaxy's light \citep{hosseinzadeh2019, dong2025}, including one event within the small but growing population of SNe Icn \citep{aster2026, hu2026, shi2026}. This suggests, similar to LFBOTs, that their progenitors have been impacted by kicks that eject them from their birthsite. Future studies on larger populations of SNe Ibn/Icn and LFBOTs should consider whether similar progenitors can power both of their lightcurves. Stronger correlations between their global and local host properties may also indicate that there is indeed a shared progenitor channel. 

A potential caveat to the BH/NS-WR star merger progenitor scenario for LFBOTs is that a similar progenitor system was proposed to power ultra-long GRB jet ($\approx 10^3-10^5$~s) during the merger \citep{fryer1998, zhang2001, klencki2025}, though perhaps only in the subset of events in which the BH is rapidly spinning. On-axis GRB jets have been confidently ruled-out within the current population of LFBOTs \citep{margutti2019, coppejans2020, ho2020_koala, yao2022, ho2023, nayana2025}. However, it remains plausible for several events to be associated with (undetected) off-axis GRB jets \citep{margutti2019, ho2020_koala, ho2023, nayana2025}. If GRB jets do precede LFBOTs, it would intrinsically connect them to the population of long and ultra-long GRBs. Typical LGRBs are thought to arise from massive collapsar events, in which a massive, rapidly rotating star collapses into a BH, forming an accretion disk that powers the relativistic jet \citep{woosley1993}. To gain enough angular momentum to create the accretion disk, several studies advocate that the LGRB progenitor must be in a tight binary system \citep{fryer1999, fryer2025}, including a BH/NS-WR star system. Alternatively, \citet{fryer1998} and \citet{zhang2001} suggest that a BH-WR star merger, with a similar formation channel as discussed in \citet{metzger2022}, could be responsible for some LGRBs. This, furthermore, was the proposed progenitor model for ultra-long GRBs 101225A \citep{thone2011} and 250702B \citep{neights2026}. Notably, ultra-long GRBs are also expected to have lower $^{56}$Ni yields than typical long GRBs (or, rather, the SNe Ic-BL following the long GRB), which does match the expectations for LFBOTs \citep{neights2026}. While we do compare LFBOTs to LGRBs, we note that the population of ultra-long GRBs and their hosts is too small for a meaningful comparison. Lastly, no study has yet determined if LFBOT emission contributes to long or ultra-long GRB lightcurves, but this may be informative to confirm if they do arise from the same sources.

If LGRBs and LFBOTs arise from similar tight binary evolutionary pathways, this may be hinted at in their environments. Here, we show that LFBOTs occur in more metal-rich hosts, have higher galactocentric offsets, and are associated with fainter regions of their hosts than LGRBs. In addition, it appears that LGRB hosts lie above the SFMS, whereas LFBOTs mostly track the SFMS (although we find no statistical difference between their 1D-sSFR and $M_*$ distributions). The metallicity difference between LFBOTs and LGRBs, as well as the possible distinction between the hosts on the sSFR-$M_*$ plane is easily understood by the subtle differences in their progenitors. First, while it is not a requirement in the LFBOT merger progenitor scenario that the WR star is rapidly rotating, this is a requirement with the LGRB scenario (whether they come from the same binary mergers or simply represent core-collapse events in tight binary systems). As such, the LGRBs should occur in lower-metallicity environments than LFBOTs, where the stellar companion can more easily retain high angular momentum. Moreover, if ultra-long GRBs precede at least some LFBOTs, as suggested in \citealt{klencki2025}, their progenitors should systematically arise from less massive stars than LGRBs, in either the merger or core-collapse scenario \citep{zhang2001, fryer2025}. The formation of less massive compact objects and stellar companions inherently does not require as low metallicities as forming a more massive system. Thus, the discrepancy in the host metallicities of LFBOTs and LGRBs can be well-understood if LFBOTs are formed from less massive binaries than those of LGRBs.

The distinction between LGRB and LFBOT offsets and fractional fluxes may also be related to the total mass of the progenitor. More massive binaries likely travel with lower systemic velocities than less massive binaries. Furthermore, if we suppose that most LGRB progenitors undergo core-collapse before their tight-binary system can merge, they may also have slightly shorter formation timescales. Their progenitors thereby have a higher likelihood of staying closer to their birthsite, rather than getting kicked out as we see for LFBOTs. We believe that this comparison between LGRB and LFBOT hosts, although speculative on both of their progenitors, gives further support to the LFBOT BH/NS merger origin. 

In total, we find the highest support for this compact object-WR star merger from LFBOT environments. All other progenitor models discussed have major caveats that contradict either the observed environmental properties of LFBOTs or their emission. While we don't completely rule out a stellar origin for LFBOTs, we implore future studies on LFBOTs and their progenitors to consider the effect that these models will have on their transients' observed fractional flux distribution, metallicity dependence, and comparison to other well-studied core-collapse events.

\section{Conclusion}
\label{sec:conc}
In this paper, we conducted an analysis of the environments of 11 LFBOTs, comprising the largest uniform analysis of LFBOT hosts to-date. To determine their host galaxy stellar population properties (stellar mass, amount of active star formation, metallicity, etc.), we modeled their host photometry and/or spectroscopy with \texttt{Prospector}. We additionally performed a simplistic analysis of LFBOT local environments by determining their physical and host-normalized  galactocentric offset distributions and measuring their fractional fluxes. To contextualize these results, we compared them to those derived for several CCSNe populations (SNe II, Ibc, and Ibn), LGRBs, and SLSNe-I: transients that LFBOTs have been frequently compared to in the literature. We list our main conclusions below:
\begin{itemize}
    \item LFBOT hosts have $\log(M_*/M_\odot) = 9.61^{+0.74}_{-1.61}$ (population median and 68\% interval). LFBOT hosts have statistically similar stellar mass distributions to those of SN II, Ibc, Ibn, and LGRB hosts, but trace higher stellar masses than SLSNe-I.
    \item We determine that LFBOT hosts have present-day SFR = $0.99^{+14.85}_{-0.95}$~$M_\odot$~yr$^{-1}$ and log(sSFR) = $-9.51^{+0.71}_{-0.64}$. Compared to SLSN-I hosts, LFBOT hosts have lower sSFR. Meanwhile, they have higher sSFR than CCSN hosts and similar sSFR distributions to LGRB hosts.
    \item LFBOT hosts are all actively star-forming, but have a variety of SFH shapes. The majority of hosts are either currently in their most star-forming period or have had a star formation burst in the past $\approx$100~Myr. 
    \item LFBOT hosts prefer SFR-weighted solutions in the 2D SFR-$M_*$ plane, and possibly more SFR-weighted distributions than SNe Ibc and II hosts. They additionally occupy a different region in the 2D SFR-$M_*$ plane than SLSNe-I and LGRBs, which appear to lean toward a higher-sSFR and lower-$M_*$ region than LFBOT hosts.
    \item LFBOT hosts have a gas-phase oxygen abundance $12+\log(\textrm{O}/\textrm{H}) =8.59^{+0.18}_{-0.22}$. This is lower metallicity than the general field galaxy population and CCSN hosts, but higher metallicity than SLSN-I and LGRB hosts.
    \item All LFBOT hosts lie in the star-forming galaxy phase space in the BPT diagrams, but have higher $\log(\textrm{OIII}/\textrm{H}\beta)$ ratios in comparison to the general field galaxy population and CCSN hosts. These elevated ratios are also observed for LGRB and SLSN-I hosts and are likely caused by their hosts having low metallicities and high sSFRs.
    \item LFBOTs have physical galacotocentric offsets $=2.89^{+9.36}_{-1.71}$~kpc and host-normalized galactocentric offsets $1.19^{+0.72}_{-0.49}$~$r/r_e$. LFBOTs are more physically offset from their hosts than SLSNe-I and LGRBs but have similar physical offset distributions to CCSNe. In addition, they have larger host-normalized offsets than LGRBs, SNe Ibc, and SNe Ibn, but similar host-normalized offsets to SLSNe-I and SNe II.
    \item LFBOTs have a high fraction of events that occur in the faintest region of their host or outside of their hosts' light radii, which is similar to SLSNe-I, but distinct from LGRBs and CCSNe.
\end{itemize}
Our results indicate that the LFBOT progenitor includes a young, massive star, has a short formation timescale, and has received a large kick to eject it from star-forming regions within the host. Based on our findings, we strongly disfavor IMBH and stellar mass BH/NS TDEs progenitor channels for LFBOTs. Unlike TDEs, which tend to occur in more evolved, quiescent galaxies, LFBOTs exhibit a strict dependence on active, ongoing star formation that is incompatible with typical TDE host demographics. While we are skeptical of stellar death and CCSN progenitor scenarios, we are not able to completely rule-them out. In particular, we believe that there is a possibility that LFBOTs could comprise a failed SN or magnetar-powered CCSNe channel. However, the failed SN channel does not yet sufficiently explain the LFBOT fractional flux distribution, as it is unclear why this channel in particular would force them to have higher offsets from star-forming regions in the host than typical CCSNe. It is furthermore difficult to reconcile the magnetar-powered CCSNe channel if SLSNe-I also arise from the same progenitors, given the differences in their environments. In addition, it is not clear that this magnetar model can explain all aspects of observed LFBOT transient emission, including their dense CSM, asymmetric ejecta, and early and late-time lightcurve evolution.

We do find sufficient evidence that LFBOTs originate from BH/NS-WR star mergers: the star formation and metallicity dependence of LFBOTs, as well as their offset and fractional flux distributions can all be easily explained within this progenitor model. It is assumed that that these binaries will preferentially form in low metallicity environments (although, not as extreme as those of SLSNe-I and LGRBs, similar to LFBOTs) and will have short enough merger timescales to occur soon ($\approx$tens of Myr) after the star formation episode that produced the binary. Thus, we would expect the majority of their hosts to still be actively star-forming at the time of merger, as we have seen for LFBOTs. Moreover, the formation of the compact object may impart a natal kick on the system, allowing it to travel away from its birthsite, and naturally explaining the observed LFBOT fractional flux distribution. This scenario also may connect the LFBOT progenitor to a variety of other astrophysical transients, including SNe Ibn/Icn, LGRBs, and ultra-long GRBs. A connection between their progenitors is also hinted at in their environments. For instance, some SNe Ibn/Icn are observed to occur well outside their host galaxies' light radius, similar to LFBOTs, suggesting all three transients may be influenced by kicks. LGRBs are observed to occur in lower metallicity hosts than LFBOTs, embedded within star-forming regions of the host. These differences can be explained if LGRBs arise from more massive BH/NS-WR binaries with higher angular momenta than those of LFBOTs, which would more easily form at lower metallicities than LFBOTs and have weaker natal kicks causing them to more frequently stay within their birthsite.

Our study establishes a crucial framework for constraining LFBOT progenitor channels through their host galaxy properties. We emphasize that the current sample of LFBOTs is quite small; as such, our findings should be viewed as a preliminary foundation for future population studies. Luckily, with the onset of the Vera C. Rubin Observatory's Legacy Survey of Space and Time (Rubin; \citealt{LSST, lsst2019}), we may soon achieve a statistically robust population of LFBOTs and their hosts. Indeed, given current LFBOT rate estimates \citep{coppejans2020, ho2023_fbot}, Rubin could discover tens to hundreds of LFBOTs each year. This large population will undoubtedly lend crucial insight into the breadth of LFBOT host properties, offsets, and their connection to recent star formation within their hosts. Moreover, with an increased sample sizes of all transients, we will better discern if LFBOTs share similar formation pathways to other events, including CCSNe, LGRBs, SLSNe-I, and SNe Ibn. Ultimately, the next few years promises to bring a much deeper understanding of these enigmatic, fascinating transients.

\section*{Acknowledgments}
We thank the anonymous referee for their insightful comments that greatly improved this manuscript. We are grateful to Peter Blanchard, Lisa Kewley, Sarah Biddle, and Daichi Tsuna, who all provided extremely helpful guidance and suggestions on several parts of this study. The Villar Astro Time Lab acknowledges support through the David and Lucile Packard Foundation, the Research Corporation for Scientific Advancement (through a Cottrell Fellowship), the National Science Foundation under AST-2433718, AST-2407922 and AST-2406110, as well as an Aramont Fellowship for Emerging Science Research. This work is supported by the National Science Foundation under Cooperative Agreement PHY-2019786 (the NSF AI Institute for Artificial Intelligence and Fundamental Interactions).  BDM acknowledges support from NASA ATP (grant number 80NSSC22K0807), the Fermi Guest Investigator Program (grant number 80NSSC24K0408) and the Simons Foundation (grant number 727700). The Flatiron Institute is supported by the Simons Foundation. DF's contribution to this material is based upon work supported by the National Science Foundation under Award No. AST-2401779.

The computations in this paper were run on the FASRC Cannon cluster supported by the FAS Division of Science Research Computing Group at Harvard University.

Observations reported here were obtained at the MMT Observatory, a joint facility of the Smithsonian Astrophysical Observatory and the University of Arizona.

This research has made use of the Keck Observatory Archive (KOA), which is operated by the W. M. Keck Observatory and the NASA Exoplanet Science Institute (NExScI), under contract with the National Aeronautics and Space Administration.

Based in part on data collected at the Subaru Telescope and retrieved from the HSC data archive system, which is operated by Subaru Telescope and Astronomy Data Center at National Astronomical Observatory of Japan. The Hyper Suprime-Cam (HSC) collaboration includes the astronomical communities of Japan and Taiwan, and Princeton University. The HSC instrumentation and software were developed by the National Astronomical Observatory of Japan (NAOJ), the Kavli Institute for the Physics and Mathematics of the Universe (Kavli IPMU), the University of Tokyo, the High Energy Accelerator Research Organization (KEK), the Academia Sinica Institute for Astronomy and Astrophysics in Taiwan (ASIAA), and Princeton University. Funding was contributed by the FIRST program from Japanese Cabinet Office, the Ministry of Education, Culture, Sports, Science and Technology (MEXT), the Japan Society for the Promotion of Science (JSPS), Japan Science and Technology Agency (JST), the Toray Science Foundation, NAOJ, Kavli IPMU, KEK, ASIAA, and Princeton University. 

The Pan-STARRS1 Surveys (PS1) have been made possible through contributions of the Institute for Astronomy, the University of Hawaii, the Pan-STARRS Project Office, the Max-Planck Society and its participating institutes, the Max Planck Institute for Astronomy, Heidelberg and the Max Planck Institute for Extraterrestrial Physics, Garching, The Johns Hopkins University, Durham University, the University of Edinburgh, Queen’s University Belfast, the Harvard-Smithsonian Center for Astrophysics, the Las Cumbres Observatory Global Telescope Network Incorporated, the National Central University of Taiwan, the Space Telescope Science Institute, the National Aeronautics and Space Administration under Grant No. NNX08AR22G issued through the Planetary Science Division of the NASA Science Mission Directorate, the National Science Foundation under Grant No. AST-1238877, the University of Maryland, and Eotvos Lorand University (ELTE) and the Los Alamos National Laboratory.

The Legacy Surveys consist of three individual and complementary projects: the Dark Energy Camera Legacy Survey (DECaLS; Proposal ID \#2014B-0404; PIs: David Schlegel and Arjun Dey), the Beijing-Arizona Sky Survey (BASS; NOAO Prop. ID \#2015A-0801; PIs: Zhou Xu and Xiaohui Fan), and the Mayall z-band Legacy Survey (MzLS; Prop. ID \#2016A-0453; PI: Arjun Dey). DECaLS, BASS and MzLS together include data obtained, respectively, at the Blanco telescope, Cerro Tololo Inter-American Observatory, NSF’s NOIRLab; the Bok telescope, Steward Observatory, University of Arizona; and the Mayall telescope, Kitt Peak National Observatory, NOIRLab. Pipeline processing and analyses of the data were supported by NOIRLab and the Lawrence Berkeley National Laboratory (LBNL). The Legacy Surveys project is honored to be permitted to conduct astronomical research on Iolkam Du’ag (Kitt Peak), a mountain with particular significance to the Tohono O’odham Nation.

NOIRLab is operated by the Association of Universities for Research in Astronomy (AURA) under a cooperative agreement with the National Science Foundation. LBNL is managed by the Regents of the University of California under contract to the U.S. Department of Energy.

This project used data obtained with the Dark Energy Camera (DECam), which was constructed by the Dark Energy Survey (DES) collaboration. Funding for the DES Projects has been provided by the U.S. Department of Energy, the U.S. National Science Foundation, the Ministry of Science and Education of Spain, the Science and Technology Facilities Council of the United Kingdom, the Higher Education Funding Council for England, the National Center for Supercomputing Applications at the University of Illinois at Urbana-Champaign, the Kavli Institute of Cosmological Physics at the University of Chicago, Center for Cosmology and Astro-Particle Physics at the Ohio State University, the Mitchell Institute for Fundamental Physics and Astronomy at Texas A\&M University, Financiadora de Estudos e Projetos, Fundacao Carlos Chagas Filho de Amparo, Financiadora de Estudos e Projetos, Fundacao Carlos Chagas Filho de Amparo a Pesquisa do Estado do Rio de Janeiro, Conselho Nacional de Desenvolvimento Cientifico e Tecnologico and the Ministerio da Ciencia, Tecnologia e Inovacao, the Deutsche Forschungsgemeinschaft and the Collaborating Institutions in the Dark Energy Survey. The Collaborating Institutions are Argonne National Laboratory, the University of California at Santa Cruz, the University of Cambridge, Centro de Investigaciones Energeticas, Medioambientales y Tecnologicas-Madrid, the University of Chicago, University College London, the DES-Brazil Consortium, the University of Edinburgh, the Eidgenossische Technische Hochschule (ETH) Zurich, Fermi National Accelerator Laboratory, the University of Illinois at Urbana-Champaign, the Institut de Ciencies de l’Espai (IEEC/CSIC), the Institut de Fisica d’Altes Energies, Lawrence Berkeley National Laboratory, the Ludwig Maximilians Universitat Munchen and the associated Excellence Cluster Universe, the University of Michigan, NSF’s NOIRLab, the University of Nottingham, the Ohio State University, the University of Pennsylvania, the University of Portsmouth, SLAC National Accelerator Laboratory, Stanford University, the University of Sussex, and Texas A\&M University.

BASS is a key project of the Telescope Access Program (TAP), which has been funded by the National Astronomical Observatories of China, the Chinese Academy of Sciences (the Strategic Priority Research Program “The Emergence of Cosmological Structures” Grant \# XDB09000000), and the Special Fund for Astronomy from the Ministry of Finance. The BASS is also supported by the External Cooperation Program of Chinese Academy of Sciences (Grant \# 114A11KYSB20160057), and Chinese National Natural Science Foundation (Grant \# 12120101003, \# 11433005).

The Legacy Survey team makes use of data products from the Near-Earth Object Wide-field Infrared Survey Explorer (NEOWISE), which is a project of the Jet Propulsion Laboratory/California Institute of Technology. NEOWISE is funded by the National Aeronautics and Space Administration.

The Legacy Surveys imaging of the DESI footprint is supported by the Director, Office of Science, Office of High Energy Physics of the U.S. Department of Energy under Contract No. DE-AC02-05CH1123, by the National Energy Research Scientific Computing Center, a DOE Office of Science User Facility under the same contract; and by the U.S. National Science Foundation, Division of Astronomical Sciences under Contract No. AST-0950945 to NOAO.

\vspace{5mm}
\facilities{Keck (LRIS); MMT (Binospec, MMIRS); VLT (FORS2)}

\software{\texttt{SEP} \citep{SEP}; \texttt{POTPyRI} \citep{POTPyRI}; \texttt{FrankenBlast} \citep{frankenblast};  \texttt{PypeIt} \citep{prochaska2020}; \texttt{Prospector} \citep{jlc+2021}; \texttt{FSPS} and \texttt{python-FSPS} \citep{FSPS_2009, FSPS_2010}; \texttt{dynesty} \citep{Dynesty}}

\bibliography{refs}

\appendix 
\restartappendixnumbering

\section{Host Galaxy Observations}
\label{app:host_obs}
Here, we highlight LFBOT host galaxy photometric (Table~\ref{tab:phot}) and spectroscopic (Table~\ref{tab:spec}) observations.
\startlongtable
\begin{deluxetable*}{l|cccccc}
\tabletypesize{\footnotesize}
\tablecolumns{7}
\tablewidth{0pc}
\tablecaption{Host Galaxy Photometry}
\label{tab:phot}
\tablehead{
\colhead{LFBOT} &
\colhead{Host R.A.} &
\colhead{Host Decl.} &
\colhead{Instrument or Survey} &
\colhead{Filter} &
\colhead{AB Mag} &
\colhead{Ref.}}
\startdata
AT2018cow & \ra{16}{16}{00.58} & \dec{+22}{16}{08.300} & GALEX & FUV & 18.38 $\pm$ 0.21 & \citet{perley2019} \\
 &  &    & GALEX & NUV & 17.88$\pm$ 0.04 &  \citet{perley2019} \\
 &  &   & SDSS & $u$ & 16.76 $\pm$ 0.04 &  \citet{perley2019} \\
 &  &   & SDSS & $g$ & 15.58 $\pm$ 0.003 &  \citet{perley2019} \\
 &  &    & SDSS & $r$ & 15.02 $\pm$ 0.002 &  \citet{perley2019} \\
 &  &    & SDSS & $i$ & 14.73 $\pm$ 0.01 &  \citet{perley2019} \\
 &  &   & SDSS & $z$ & 14.53 $\pm$ 0.02 &  \citet{perley2019} \\
 &  &   & Pan-STARRS & $y$ & 14.48$\pm$ 0.05 &  \citet{perley2019} \\
 &  &   & 2MASS & $J$ & 14.15 $\pm$ 0.05 & \citet{perley2019} \\
 &  &    & 2MASS & $H$ & 14.07 $\pm$ 0.08 &  \citet{perley2019} \\
 &  &    & 2MASS & $K$ & 14.32 $\pm$ 0.11 &  \citet{perley2019} \\
 &  &    & WISE & $w1$ &15.37$\pm$ 0.01 &  \citet{perley2019} \\
 &  &    & WISE & $w2$ & 16.01$\pm$ 0.02 &  \citet{perley2019} \\
 &  &    & WISE & $w3$ & 14.99 $\pm$ 0.03 &  \citet{perley2019} \\
 &  &    & WISE & $w4$ & 14.67$\pm$ 0.24 &  \citet{perley2019} \\ 
 \hline
CSS161010 & \ra{04}{48}{34.398} & \dec{-08}{18}{04.337} & Pan-STARRS & $g$ & 21.35 $\pm$ 0.08 &  This work \\
 &  &    & Pan-STARRS & $r$ & 21.99 $\pm$ 0.10 &  This work \\
 &  &   & Pan-STARRS & $i$ & 21.00 $\pm$ 0.05 &  This work \\
 &  &   & Pan-STARRS & $z$ & 20.36 $\pm$ 0.10 &  This work \\
 &  &   & Pan-STARRS & $y$ & 20.22 $\pm$ 0.30 &  This work \\
 \hline
 ZTF18abvkwl & \ra{02}{00}{15.238} & \dec{+16}{47}{57.151}  & SDSS & $u$ & 21.95 $\pm$ 0.20 &  \citet{ho2020_koala} \\
 &  &   & Pan-STARRS & $g$ & 21.33 $\pm$ 0.07 &  This work \\
 &  &    & Pan-STARRS & $r$ &  21.09 $\pm$ 0.05 &  This work \\
 &  &    & Pan-STARRS & $i$ & 21.03 $\pm$ 0.05 &  This work \\
 &  &    & Pan-STARRS & $z$ & 20.27 $\pm$ 0.07 &  This work \\
 &  &    & Pan-STARRS & $y$ & 19.87 $\pm$ 0.15 &  This work \\
  &  &   & 2MASS & $J$ & 18.88 $\pm$ 0.04 & This work \\
 &  &   & 2MASS & $K$ & 17.70 $\pm$ 0.04 &  This work \\
 \hline
AT2020mrf & \ra{15}{47}{54.20} & \dec{+44}{39}{07.01} &  GALEX & FUV & $> 23.28$ & \citet{yao2022} \\
 &  &   & GALEX & NUV & $> 23.56$ & \citet{yao2022}  \\
 &  &   & HSC & $g$ & 23.28 $\pm$ 0.03 &  \citet{yao2022} \\
 &  &   & HSC & $r$ & 23.15 $\pm$ 0.05 &  \citet{yao2022} \\
 &  &   & HSC & $i$ & 22.64 $\pm$ 0.04 &  \citet{yao2022} \\
 &  &   & HSC & $z$ & 22.72 $\pm$ 0.08 & \citet{yao2022} \\
 &  &   & HSC & $y$ & 22.36 $\pm$ 0.13 &  \citet{yao2022} \\
 \hline
 AT2020xnd & \ra{22}{20}{02.04} & \dec{-2}{50}{25.44} & VLT/FORS2 & $u$ & 25.6 $\pm$ 0.3 & \citet{perley2021} \\
 &  &   & VLT/FORS2 & $g$ & 24.75 $\pm$ 0.10 &  \citet{perley2021}\\
 &  &   & MMT/Binospec & $r$ & 24.03 $\pm$ 0.11 &  This work \\
 &  &   & VLT/FORS2 & $I$ & 24.56 $\pm$ 0.25 & \citet{perley2021} \\
 \hline
 AT2022tsd & \ra{03}{20}{10.80} & \dec{+8}{44}{57.15} &  Keck/LRIS & $G$ & 21.41 $\pm$ 0.02 & This work \\
 &  &   & MMT/Binospec & $r$ & 20.61 $\pm$ 0.01 &  This work\\
 &  &   & Pan-STARRS & $i$ & 20.61 $\pm$ 0.06 &  This work \\
 &  &   & MMT/Binospec & $z$ & 20.13 $\pm$ 0.01 & This work \\
 \hline
 AT2023fhn & \ra{10}{08}{03.744} & \dec{+21}{04}{22.44} & SDSS & $u$ & 20.44 $\pm$ 0.01 & This work \\
 &  &   & DECaLS & $g$ & 19.46 $\pm$ 0.01 &   This work \\
 &  &   & Keck/LRIS & $V$ & 19.36 $\pm$ 0.002 &   This work\\
 &  &   & DECaLS & $r$ & 18.91 $\pm$ 0.01 &   This work\\
 &  &   & DECaLS & $i$ & 18.60 $\pm$ 0.01 &  This work \\
 &  &   & DECaLS & $z$ & 18.54 $\pm$ 0.01 & This work\\
 \hline 
 AT2023hkw & \ra{10}{42}{17.64} & \dec{+52}{29}{16.32} & GALEX & $NUV$ & $>$21.44 & This work \\
  &  &   & Pan-STARRS & $g$ & 20.82 $\pm$ 0.09 &  This work \\
 &  &    & Pan-STARRS & $r$ &  19.74 $\pm$ 0.03 &  This work \\
 &  &    & Pan-STARRS & $i$ & 19.47 $\pm$ 0.02 &  This work \\
 &  &    & Pan-STARRS & $z$ & 19.23 $\pm$ 0.04 &  This work \\
 &  &    & Pan-STARRS & $y$ & 19.11 $\pm$ 0.09 &  This work \\
  &  &   & 2MASS & $J$ & 18.09 $\pm$ 0.03 & This work \\
 &  &   & 2MASS & $H$ & 17.49 $\pm$ 0.02 &  This work \\
 \hline
 AT2023vth & \ra{17}{56}{34.50} & \dec{+08}{02}{35.04} & GALEX & $NUV$ & 21.19 $\pm$ 0.23 & This work \\
 &  &   & Pan-STARRS & $g$ & 19.60 $\pm$ 0.02 &  This work \\
 &  &    & Pan-STARRS & $r$ &  18.94 $\pm$ 0.01 &  This work \\
 &  &    & Pan-STARRS & $i$ & 18.56 $\pm$ 0.01 &  This work \\
 &  &    & Pan-STARRS & $z$ & 18.42 $\pm$ 0.02 &  This work \\
 &  &    & Pan-STARRS & $y$ & 18.37 $\pm$ 0.04 &  This work \\
  &  &   & 2MASS & $J$ & 17.56 $\pm$ 0.02 & This work \\
  &  &   & 2MASS & $H$ & 16.64 $\pm$ 0.01 & This work \\
&  &   & 2MASS & $K$ & 16.76 $\pm$ 0.01 &  This work \\
 \hline
 AT2024qfm & \ra{23}{21}{23.40} & \dec{+11}{56}{32.64} & Pan-STARRS & $g$ & 20.58 $\pm$ 0.05 & This work \\
 &  &  & Pan-STARRS  & $r$ & 19.78 $\pm$ 0.02 & This work\\
 &  &  & Pan-STARRS & $i$ & 19.45 $\pm$ 0.02 &  This work\\
 &  &  & Pan-STARRS  & $z$ & 19.21 $\pm$ 0.02 &  This work\\
 &  &  & MMT/MMIRS & $Y$ & 18.89 $\pm$ 0.02 &  This work\\
 &  &  & MMT/MMIRS & $J$ & 18.77 $\pm$ 0.01 & This work\\
  &  &  & MMT/MMIRS & $H$ & 18.29 $\pm$ 0.03 &  This work\\
 &  &   & MMT/MMIRS & $K$ & 17.66 $\pm$ 0.03 & This work\\
 \hline
AT2024wpp & \ra{02}{42}{05.592} & \dec{-16}{57}{26.280} & MMT/Binospec & $g$ & 20.19 $\pm$ 0.01 & This work \\
 &  &  & Pan-STARRS  & $r$ & 19.83 $\pm$ 0.02 &  This work\\
 &  &  & MMT/Binospec & $i$ & 19.58 $\pm$ 0.02 &  This work\\
 &  &  & MMT/Binospec  & $z$ & 19.34 $\pm$ 0.02 &  This work\\
 &  &   & MMT/MMIRS & $Y$ & 19.53 $\pm$ 0.04 & This work\\
 &  &   & MMT/MMIRS & $J$ & 19.64 $\pm$ 0.04 & This work\\
  &  &   & MMT/MMIRS & $H$ & 19.72 $\pm$ 0.06 &  This work\\
 &  &  & MMT/MMIRS & $K$ & 20.87 $\pm$ 0.12 & This work\\
 \hline
 \enddata
\tablecomments{The localizations and photometric observations for our sample of LFBOT host galaxies. Magnitudes are not corrected for Galactic extinction.}
\end{deluxetable*}
\startlongtable
\begin{deluxetable*}{lccccc}
\tabletypesize{\footnotesize}
\tablecolumns{7}
\tablewidth{0pc}
\tablecaption{Host Galaxy Spectroscopy}
\label{tab:spec}
\tablehead{
\colhead{LFBOT} &
\colhead{$z$} &
\colhead{Facility/Instrument} &
\colhead{Exposures} &
\colhead{Lines Identified} &
\colhead{Ref.}}
\startdata
AT2018cow & 0.0141 & MMT/Binospec & 4$\times$200~s & Ca II H+K,  H$\beta$, [O~III]$\lambda 5007$, Mg$\lambda5175$, NaD$\lambda5892$, & This work\\
& & &  & H$\alpha$, [N~II]$\lambda \lambda 6549,6584$, [S~II]$\lambda \lambda 6717, 6731$ \\
CSS161010 & 0.033 & Keck/LRIS & 3300~s & [O~II]$\lambda 3727$, He~I$\lambda3889$, Ca II H+K, [OIII]$\lambda5007$ H$\alpha$, & \citet{coppejans2020}\\
& & &  & [N~II]$\lambda 6584$, [S~II]$\lambda \lambda 6717, 6731$ \\
ZTF18abvkwl$^*$ & 0.272 & Keck/LRIS & 1$\times$900~s & [O~II]$\lambda 3727$, He~I$\lambda3889$, H$\gamma$, H$\beta$, H$\gamma$, [O~III]$\lambda\lambda 4959, 5007$, & \citet{ho2020_koala}\\
& & &  & OI$\lambda6300$, H$\alpha$, [N~II]$\lambda \lambda 6549,6584$, [S~II]$\lambda \lambda 6717, 6731$ \\ 
AT2020mrf &  0.1353 & Keck/LRIS & 4$\times$850~s & [O~II]$\lambda 3727$, H$\alpha$, [N~II]$ \lambda 6584$ & \citet{yao2022}\\
AT2020xnd & 0.243 & MMT/Binospec & 4$\times$900~s & None & This work\\
AT2022tsd & 0.256 & MMT/Binospec & 4$\times$900~s & [O~II]$\lambda 3727$, He~I$\lambda3889$, H$\gamma$, H$\beta$, [O~III]$\lambda\lambda 4959, 5007$, & This work\\
& & &  & OI$\lambda6300$, H$\alpha$, [N~II]$\lambda \lambda 6549,6584$, [S~II]$\lambda \lambda 6717, 6731$ \\ 
AT2023fhn & 0.238 & MMT/Binospec & 6$\times$900~s & [O~II]$\lambda 3727$, Ca II H+K,  H$\beta$, [O~III]$\lambda\lambda 4959, 5007$,  & This work\\
& & &  & OI$\lambda6300$, H$\alpha$, [N~II]$\lambda \lambda 6549,6584$, [S~II]$\lambda \lambda 6717, 6731$ \\
AT2023hkw & 0.335 & MMT/Binospec & 4$\times$900~s & [O~II]$\lambda 3727$, Ca II H+K,  & This work\\
& & &  & H$\alpha$, [N~II]$\lambda \lambda 6549,6584$, [S~II]$\lambda \lambda 6717, 6731$ \\
AT2023vth & 0.0747 & MMT/Binospec & 3$\times$900~s & [O~II]$\lambda 3727$, Ca II H+K, H$\beta$, [O~III]$\lambda\lambda 4959, 5007$,  & This work\\
& & &  & OI$\lambda6300$, H$\alpha$, [N~II]$\lambda \lambda 6549,6584$, [S~II]$\lambda \lambda 6717, 6731$ \\
AT2024qfm &  0.227 & MMT/Binospec & 4$\times$900~s & [O~II]$\lambda 3727$, H$\beta$, [O~III]$\lambda 5007$, & This work \\
 & & &  & H$\alpha$, [N~II]$\lambda \lambda 6549,6584$, [S~II]$\lambda \lambda 6717, 6731$ \\ 
AT2024wpp  & 0.0866 & MMT/Binospec & 3$\times$900~s & [O~II]$\lambda 3727$, Ca II H+K, H$\beta$, [O~III]$\lambda\lambda 4959, 5007$, & This work\\
  & & &  & OI$\lambda$6300, H$\alpha$, [N~II]$\lambda \lambda 6549,6584$, [S~II]$\lambda \lambda 6717, 6731$ \\
 \hline
 \enddata
\tablecomments{Spectroscopic observations of LFBOT host galaxies. We reduce all available spectroscopic observations, including those that were previously published except for that of ZTF18abvkwl. We also list spectral lines we detect in each of the spectra.\\
$^*$Not re-reduced.}
\end{deluxetable*}

\section{Host Galaxy SED Fits}
\label{app:host_fit}
\begin{figure*}[t]
\centering
\includegraphics[width=0.5\textwidth]{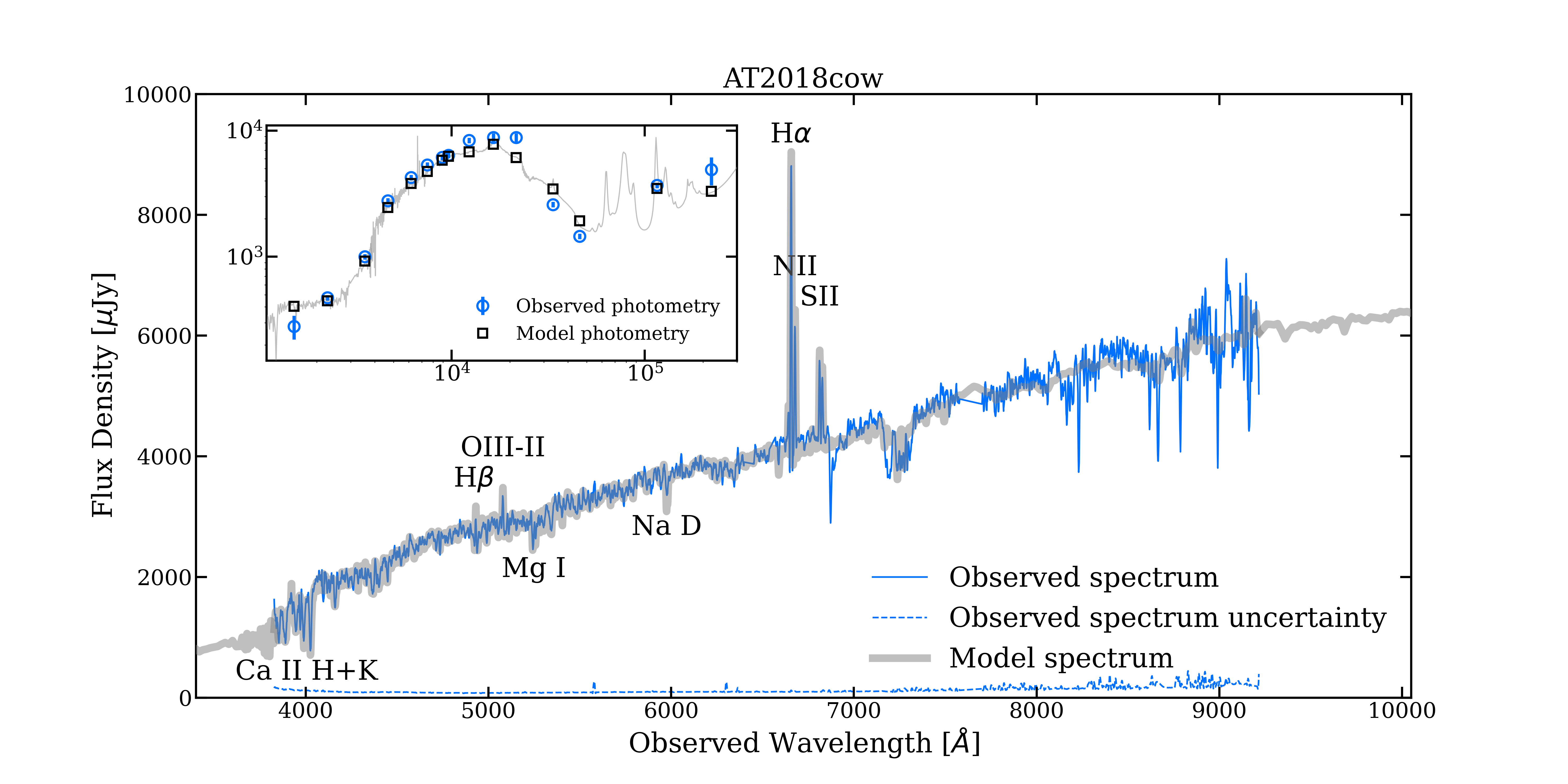} \hspace{-0.3in}
\includegraphics[width=0.5\textwidth]{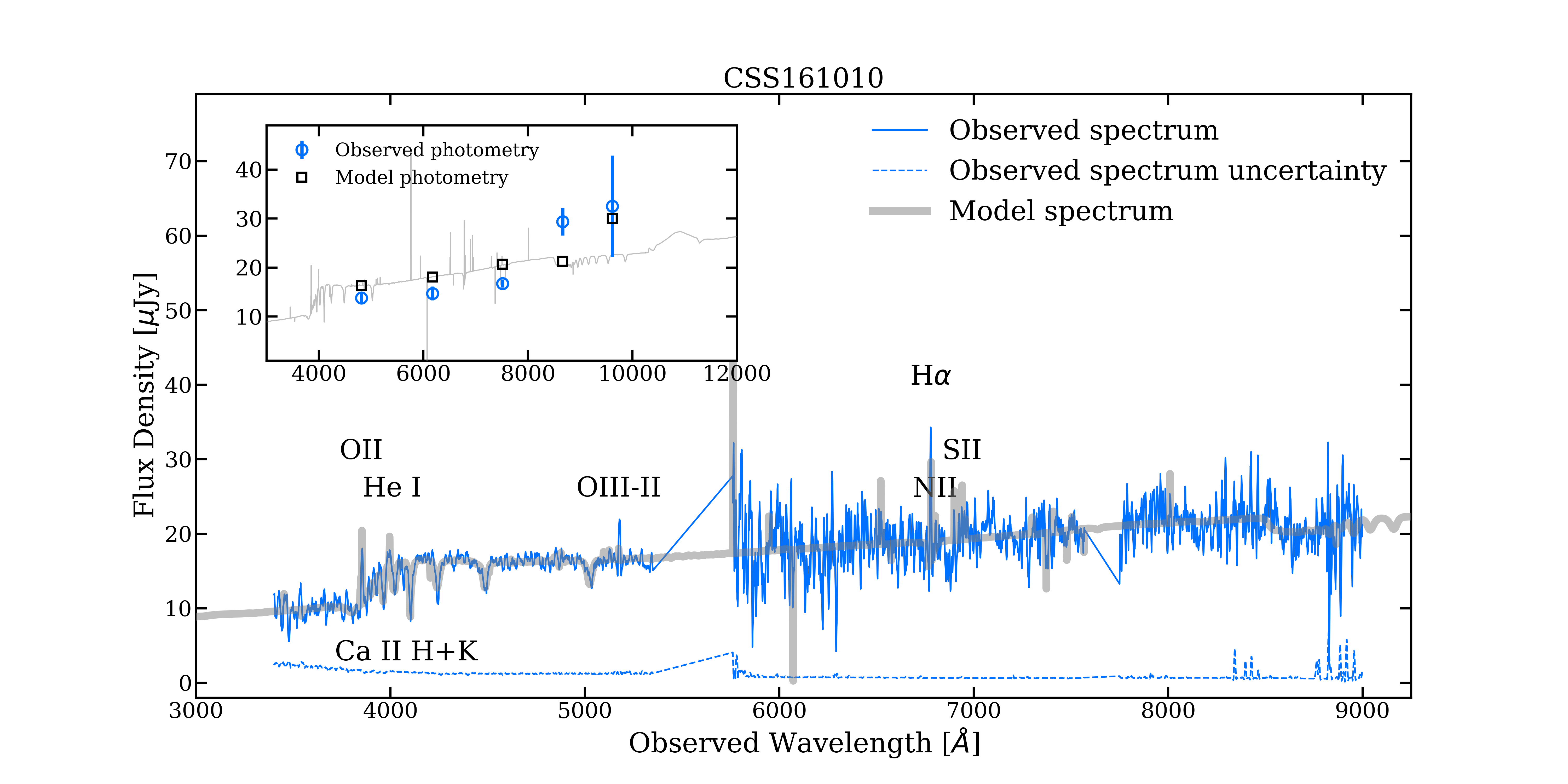}
\includegraphics[width=0.5\textwidth]{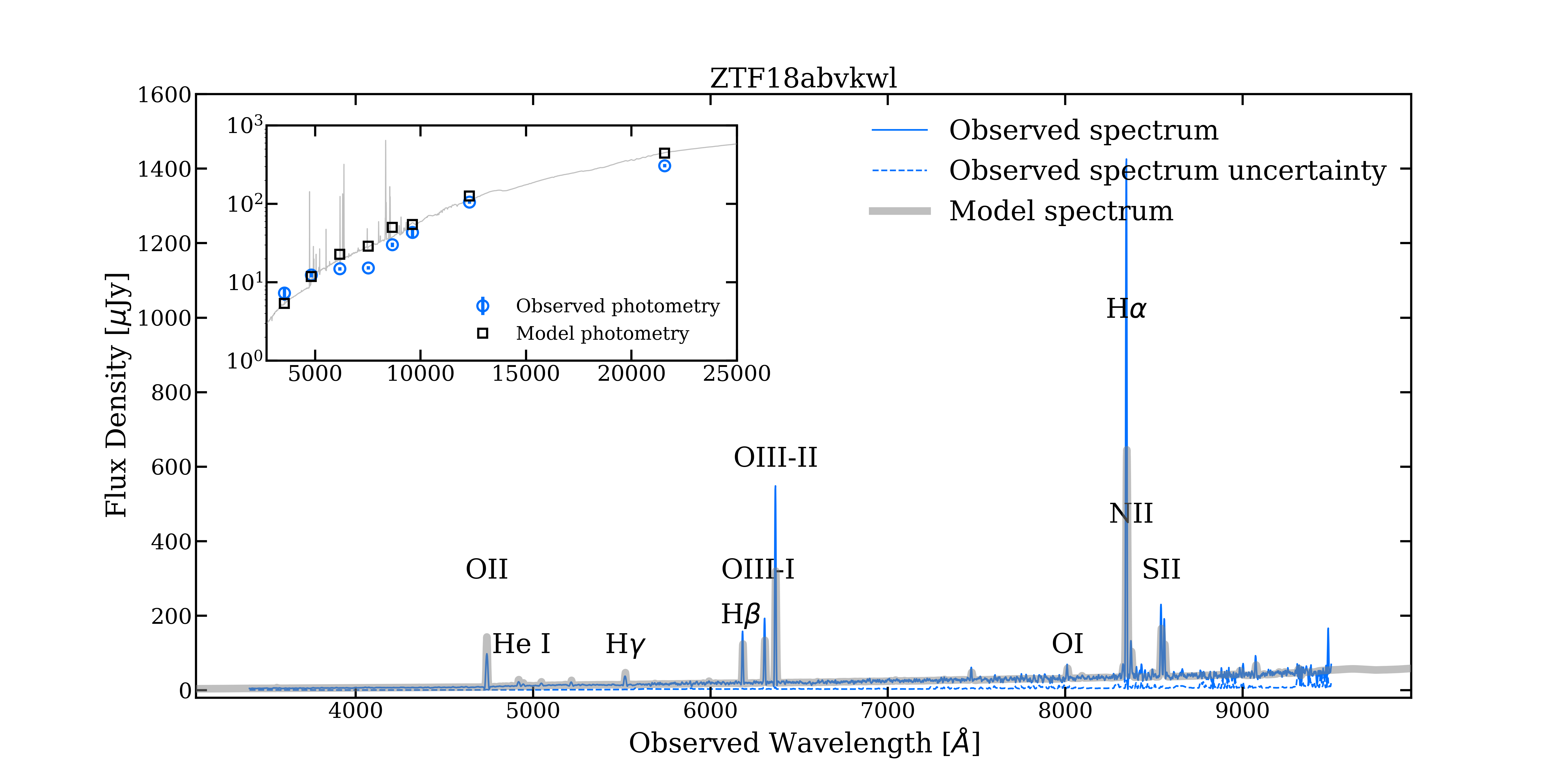} \hspace{-0.3in}
\includegraphics[width=0.5\textwidth]{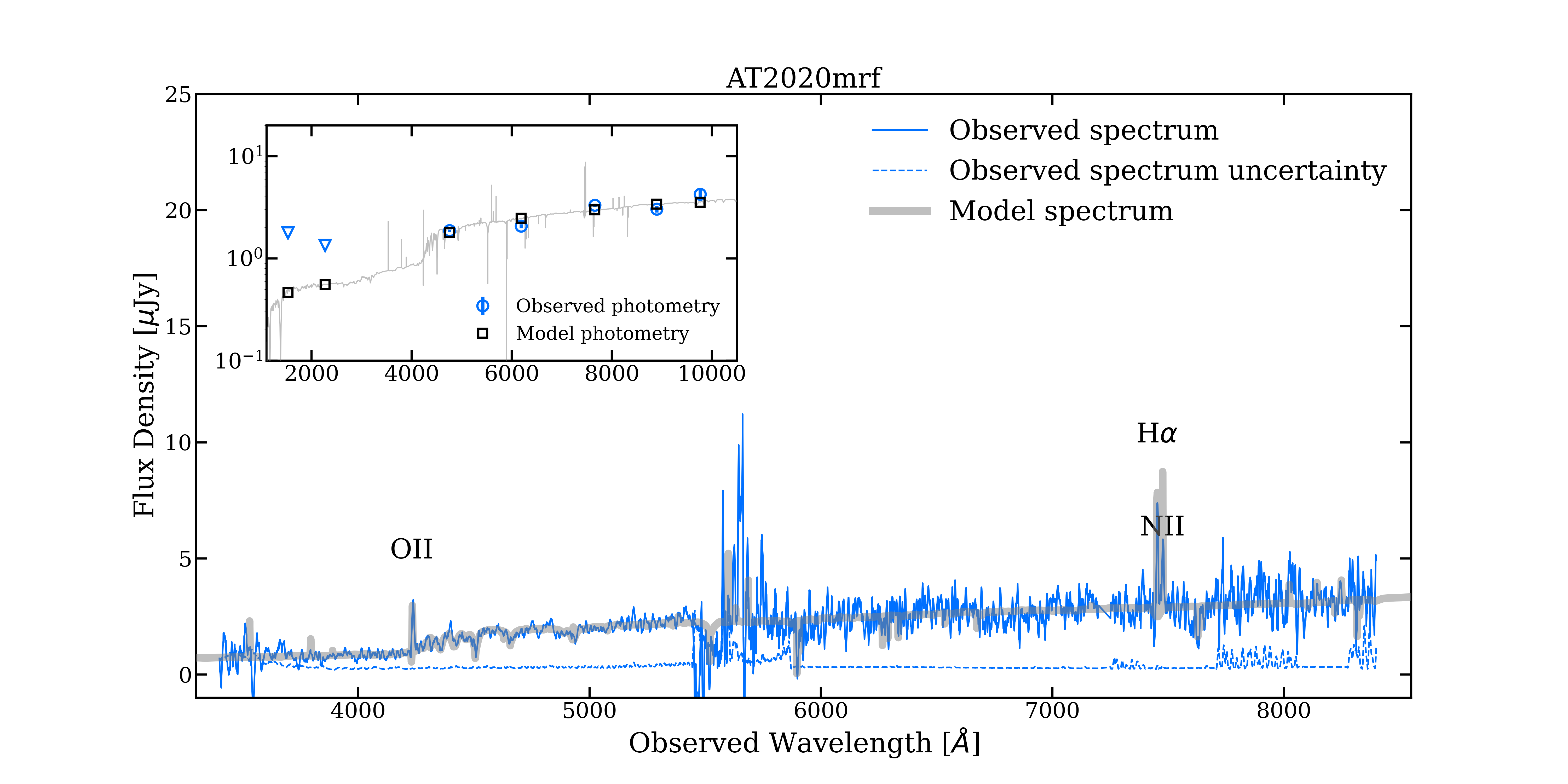}
\includegraphics[width=0.5\textwidth]{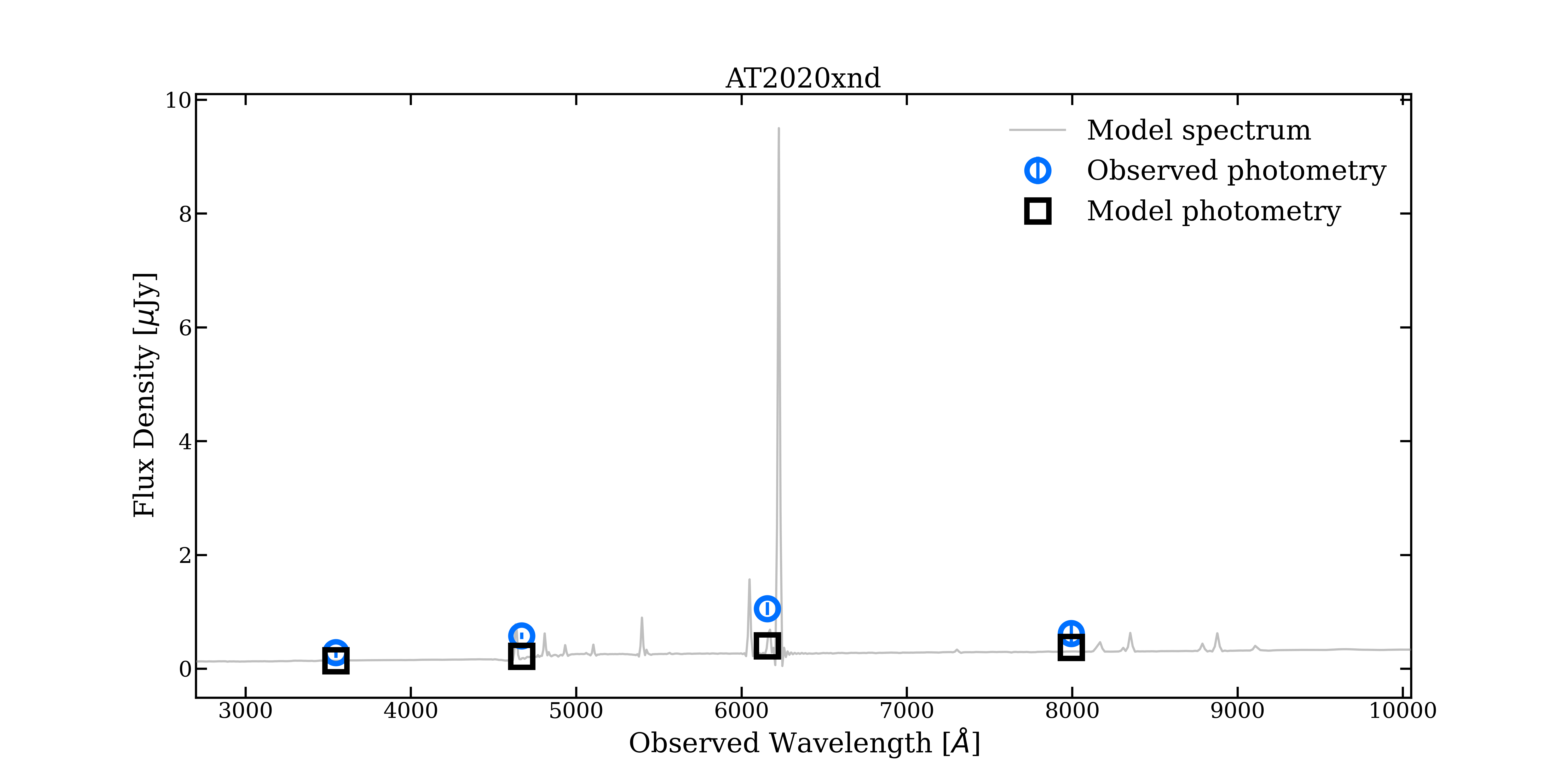} \hspace{-0.3in}
\includegraphics[width=0.5\textwidth]{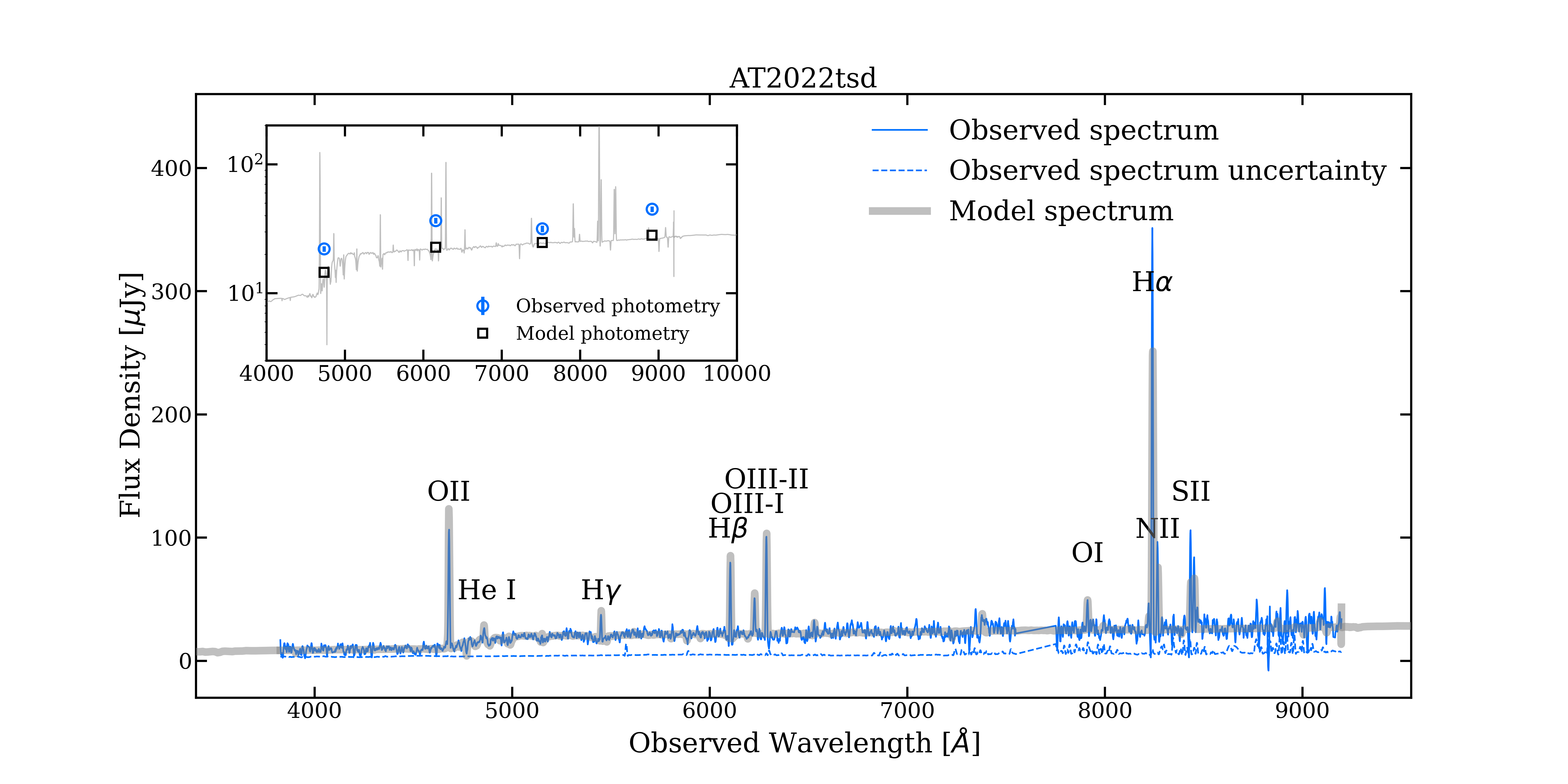}
\caption{The observed photometry (blue circles) and observed spectrum (blue line) compared to the \texttt{Prospector} model photometry (black squares) and spectra (grey lines) for our LFBOT hosts derived at the median of their stellar population property posteriors. We highlight high S/N emission and absorption lines observed in each host spectrum.}
\label{fig:appsed}
\end{figure*}

In Figure~\ref{fig:appsed}, we highlight the \texttt{Prospector}-produced model SEDs compared to the observed SEDs for each LFBOT host in our sample. For all fits, we find good agreement between the model and observed data.

\clearpage 
\begingroup
\addtocounter{figure}{-1}
\renewcommand{\thefigure}{B\arabic{figure} (Cont.)}
\renewcommand{\theHfigure}{Cont\arabic{figure}} 

\begin{figure*}
\centering
\includegraphics[width=0.5\textwidth]{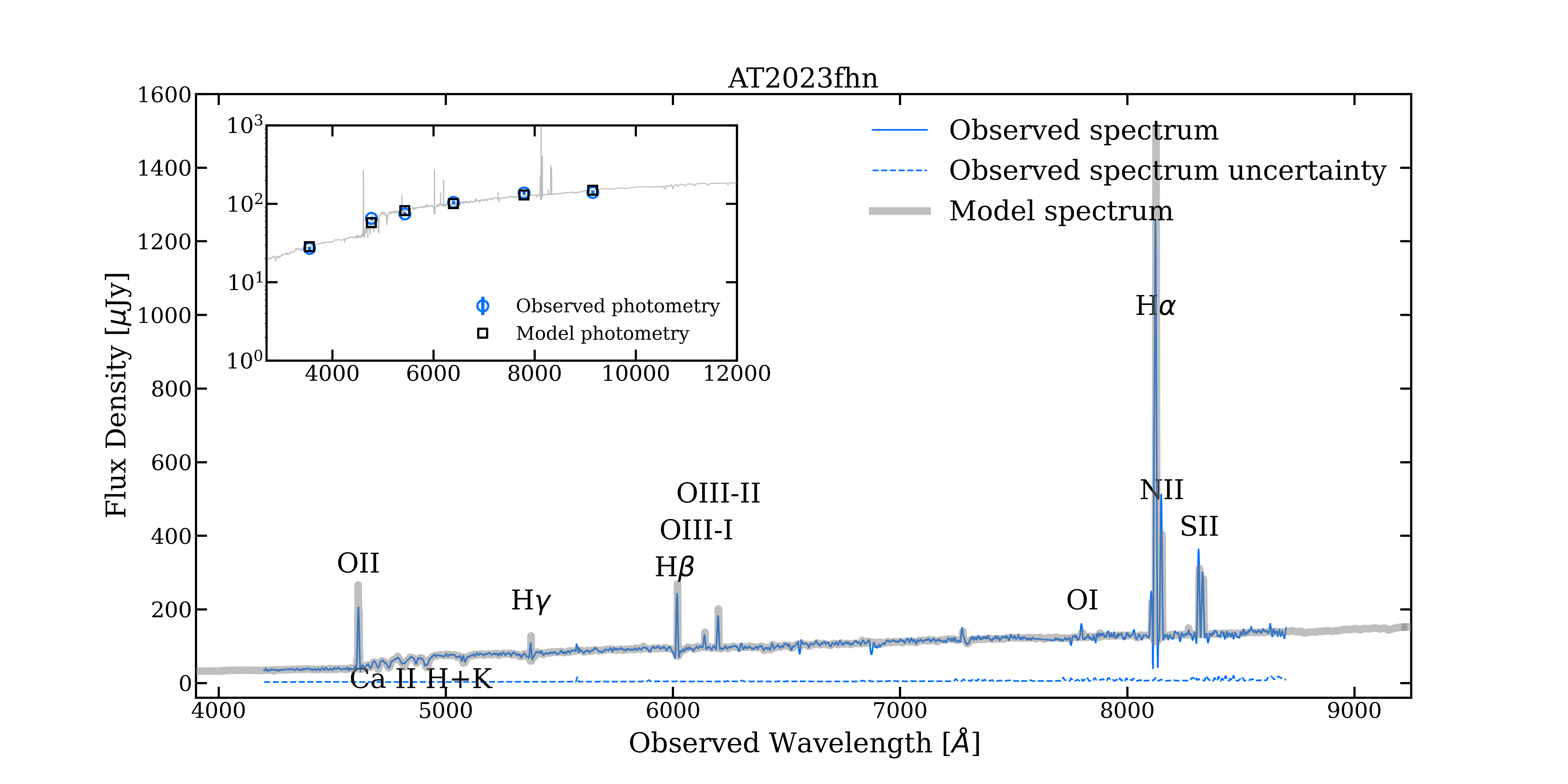} \hspace{-0.3in}
\includegraphics[width=0.5\textwidth]{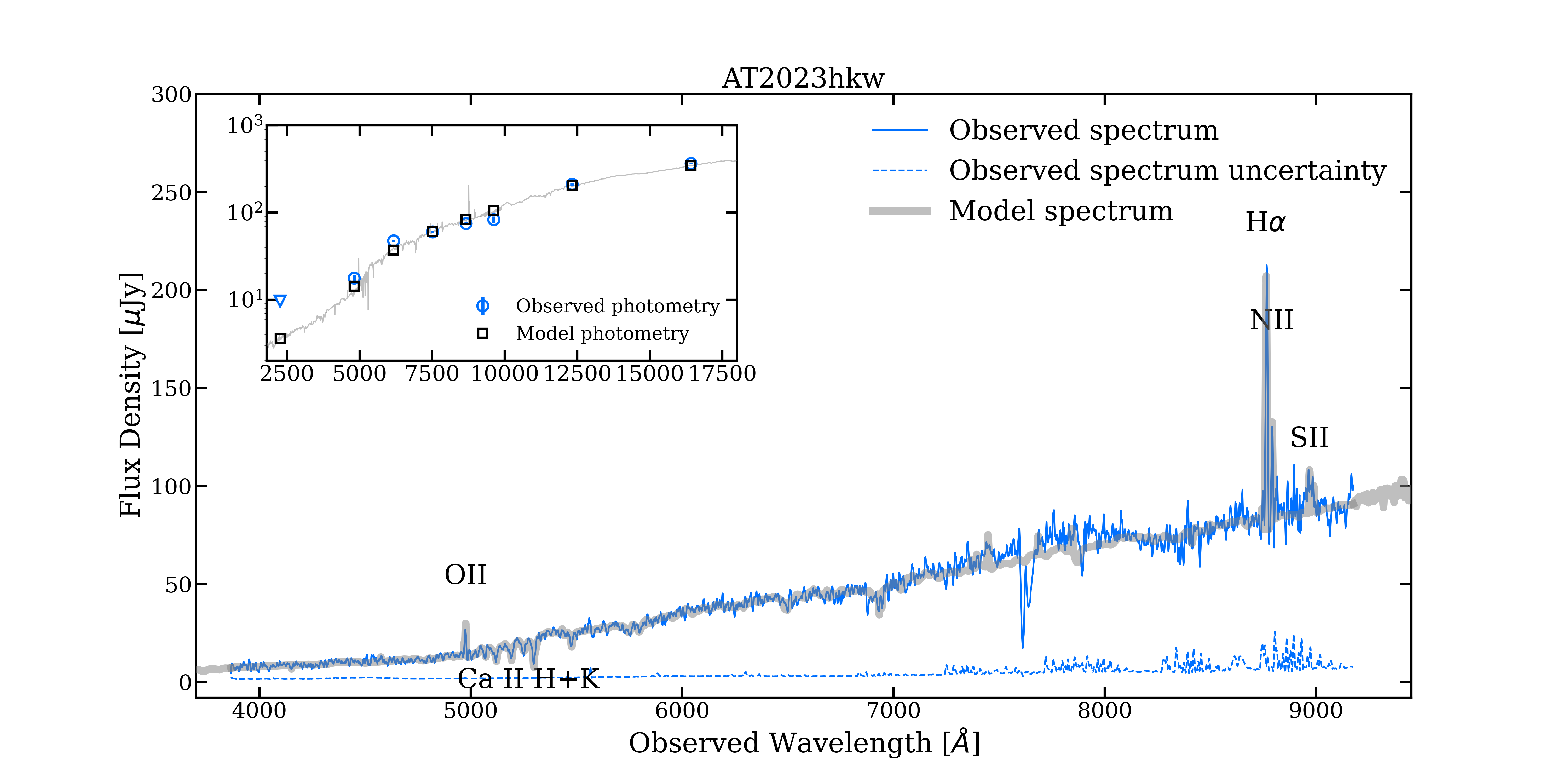} 
\includegraphics[width=0.5\textwidth]{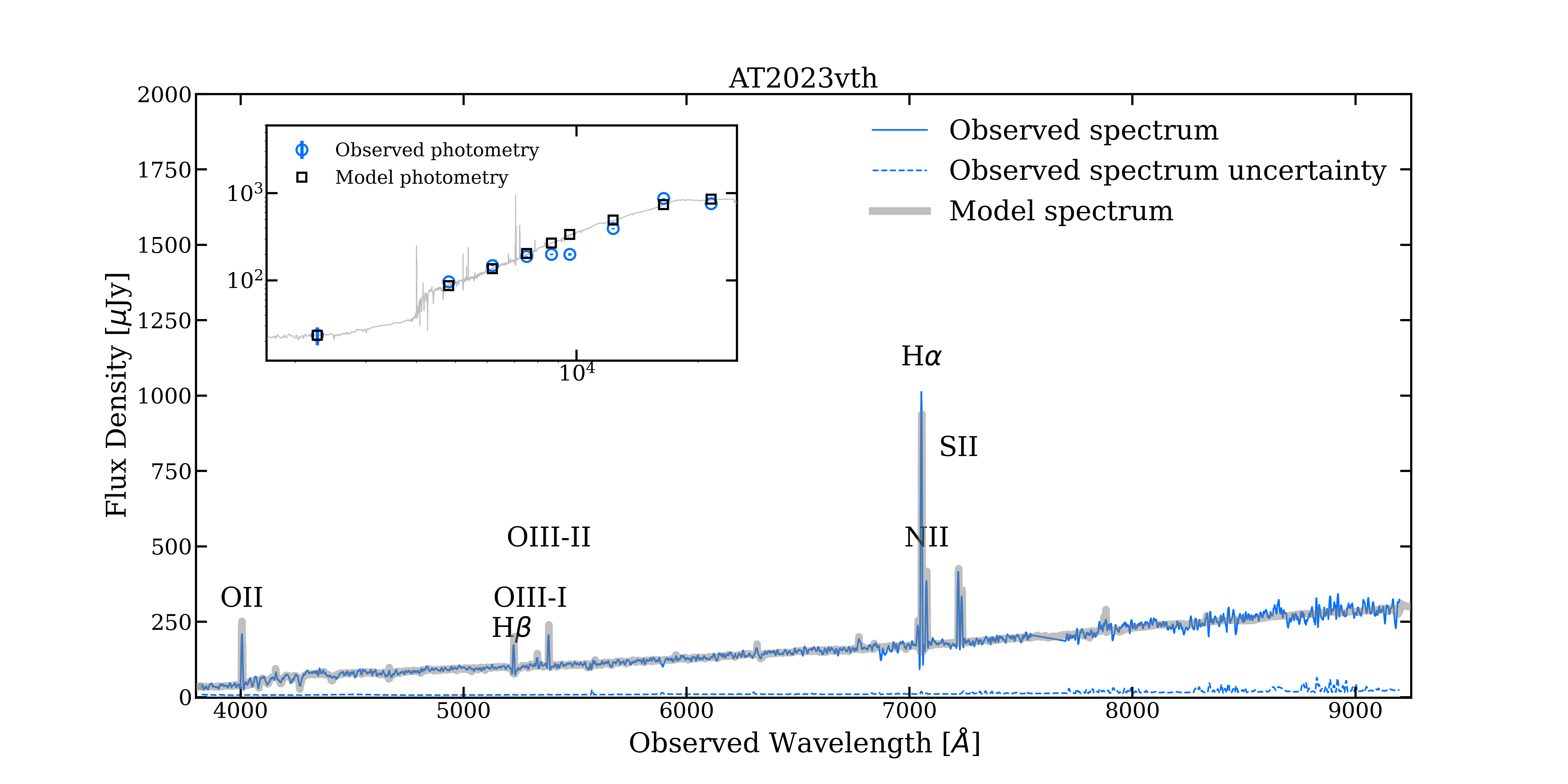} \hspace{-0.3in}
\includegraphics[width=0.5\textwidth]{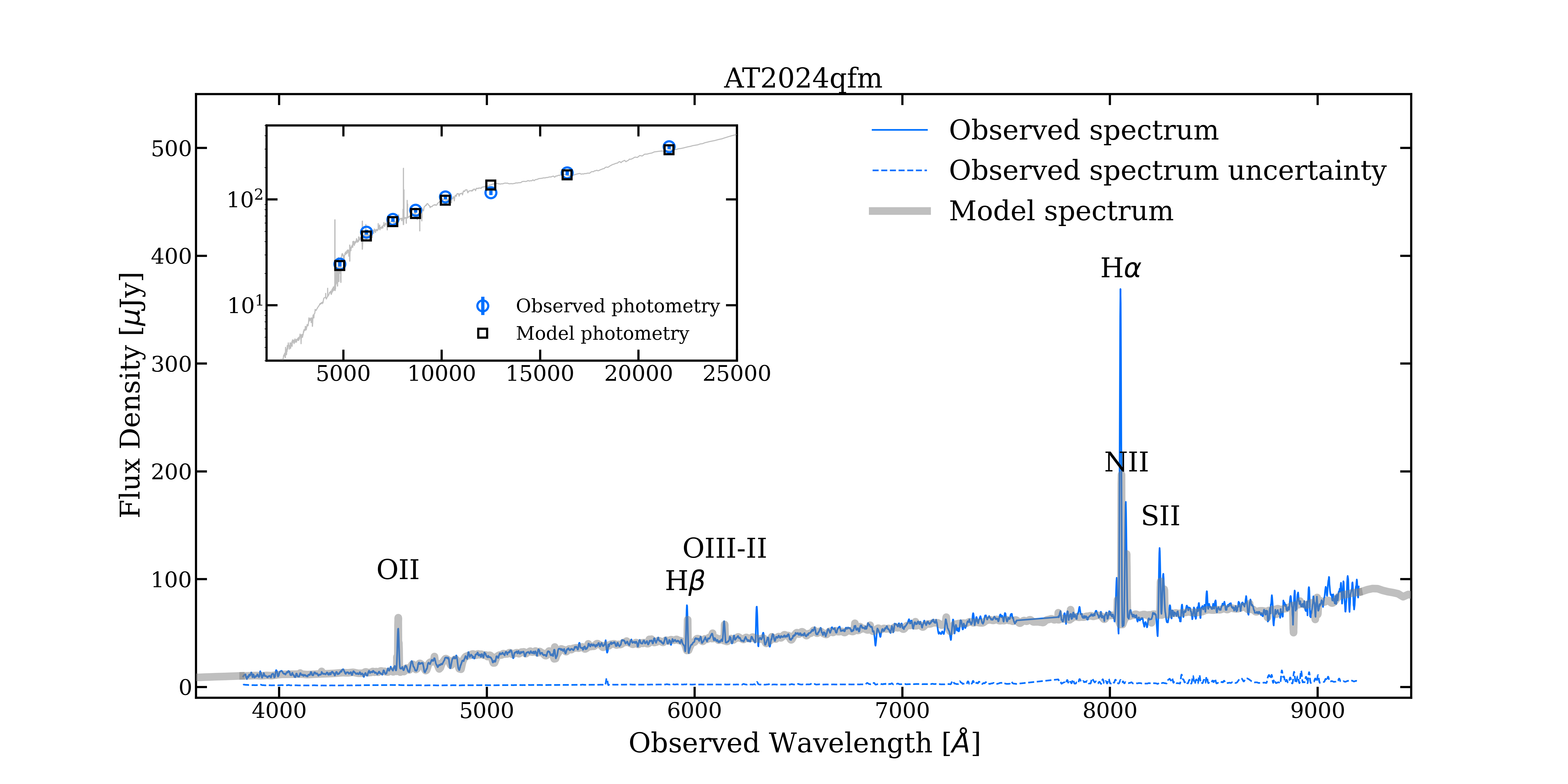}
\includegraphics[width=0.5\textwidth]{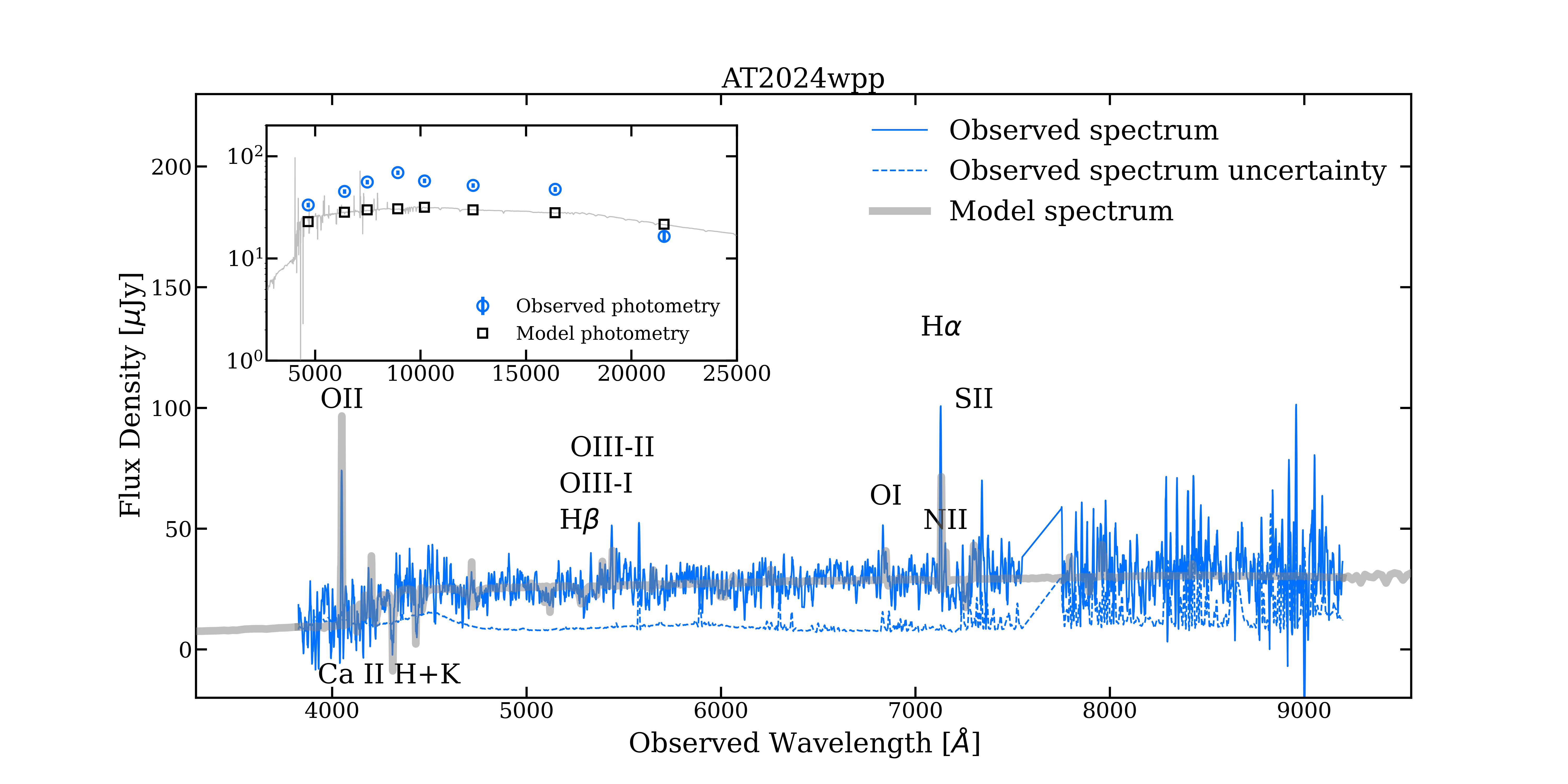}
\caption{The observed photometry (blue circles) and observed spectrum (blue line) compared to the \texttt{Prospector} model photometry (black squares) and spectra (grey lines) for our LFBOT hosts derived at the median of their stellar population property posteriors. We highlight high S/N emission and absorption lines observed in each host spectrum.}
\end{figure*}
\endgroup

\section{Literature Samples}
\label{app:lit}
In Table~\ref{tab:lit}, we cite the literature from which we collected samples of transient hosts that we compared to those of LFBOTs. 

\begin{deluxetable*}{lcccc}
\tabletypesize{\footnotesize}
\tablecolumns{5}
\tablewidth{0pc}
\tablecaption{Literature Comparison and References for Transient Environment Properties}
\label{tab:lit}
\tablehead{
\colhead{Transient Sample} &
\colhead{Mass and SFR} &
\colhead{12 + log(O/H)} &
\colhead{Galactocentric Offsets} &
\colhead{Fractional Flux} 
}
\startdata
SNe Ibc & \citet{frankenblast} & \citet{qin2024} & \citet{prieto2008} & \citet{kelly2008} \\
SNe Ibn & \citet{frankenblast} & \citet{qin2024} & \citet{dong2025} & $\cdots$ \\
SNe II  & \citet{frankenblast} & \citet{qin2024} & \citet{prieto2008} & \citet{kelly2008} \\
LGRBs   & \citet{svensson2010}, &  \citet{wang2014} & \citet{blanchard2016} & \citet{blanchard2016} \\
 &  \citet{perley2013}, & & \\
 & \citet{wang2014}, & & \\
 & \citet{niino2017} & & \\
SLSNe-I &  \citet{schulze2021} &  \citet{perley2016} & \citet{hsu2024} & \citet{hsu2024} \\
\enddata
\tablecomments{Literature from which we collected host properties of SNe Ibc, Ibn, II, LGRBs, and SLSNe-I.}
\end{deluxetable*}

\end{document}